\RequirePackage{fix-cm}
\documentclass[natbib,smallextended]{svjour3}       
\smartqed  
\usepackage{graphicx}
\usepackage{amssymb}
\usepackage{comment}
\usepackage{deluxetable}
\usepackage{color}
\usepackage{lscape}

\makeatletter
\def\@to{to}
\makeatother
\usepackage{graphicx}
 \journalname{ISSI Book on TDEs}
\newcommand{\xraychap}{X-ray Chapter}
\newcommand{\gammachap}{Gamma-ray Chapter}

\newcommand{\radiochap}{Radio Chapter}
\newcommand{\echochap}{Echo Chapter}

\newcommand{\impostchap}{One-off Events and Imposters Chapter}
\newcommand{\hostchap}{Host Galaxies Chapter}

\newcommand{\ratechap}{Rates Chapter}

\newcommand{\emischap}{Emission Mechanisms Chapter}

\newcommand{\optuv}{optical-ultraviolet}
\newcommand{\bb}{blackbody}

\usepackage{ref_macros} 

\begin{document}

\title{Optical-Ultraviolet Tidal Disruption Events}




\author{Sjoert~van~Velzen               \and
        \mbox{Thomas~W.-S.~Holoien}     \and
        Francesca~Onori                 \and
        Tiara~Hung                      \and
        Iair~Arcavi
}


\institute{S. van Velzen \at
              Center for Cosmology and Particle Physics, New York University, NY 10003 \\
              Department of Astronomy, University of Maryland, College Park, MD 20742
           \and
           T. Holoien \at
              The Observatories of the Carnegie Institution for Science, 813 Santa Barbara St., Pasadena, CA 91101, USA \\
              Carnegie Fellow
          \and
          F. Onori \at
              Istituto di Astrofisica e Planetologia Spaziali (INAF), via del Fosso del Cavaliere 100, Roma, I-00133, Italy
          \and
          T. Hung \at
          Department of Astronomy and Astrophysics, University of California, Santa Cruz, California, CA 95064, USA
          \and
          I. Arcavi (chapter coordinator) \at
          The School of Physics and Astronomy, Tel Aviv University, Tel Aviv 69978, Israel \\
          CIFAR Azrieli Global Scholars program, CIFAR, Toronto, Canada\\
          \email{arcavi@tauex.tau.ac.il}
}

\date{Received: date / Accepted: date}

\maketitle

\begin{abstract}
The existence of \optuv\ Tidal Disruption Events (TDEs) could be considered surprising because their electromagnetic output was originally predicted to be dominated by X-ray emission from an accretion disk. Yet over the last decade, the growth of optical transient surveys  has led to the identification of a new class of optical transients occurring exclusively in galaxy centers, many of which are considered to be TDEs. Here we review the observed properties of these events, identified based on a shared set of both photometric and spectroscopic properties. We present a homogeneous analysis of 33 sources that we classify as robust TDEs, and which we divide into classes. The criteria used here to classify TDEs will possibly get updated as new samples are collected and potential additional diversity of TDEs is revealed. We also summarize current measurements of the \optuv\ TDE rate, as well as the mass function and luminosity function. Many open questions exist regarding the current sample of events. We anticipate that the search for answers will unlock new insights in a variety of fields, from accretion physics to galaxy evolution.

\keywords{Tidal Disruption Events: Optical \and Tidal Disruption Events: Ultraviolet, Bowen Fluorescence}
\end{abstract}

\section{Introduction}
\label{sec:intro}

The first predictions for TDE observables considered direct emission from an accretion disk. Therefore, it was expected that TDEs be seen mainly in the X-rays. Indeed, the first TDEs were detected during the {\it ROSAT} all sky survey as luminous, soft X-ray flares from the nuclei of otherwise quiescent galaxies \citep{Bade96,KomossaGreiner99,Grupe99, Greiner00}. 
Subsequently, similar X-ray events were discovered through dedicated searches or serendipitous discoveries with {\it Chandra} and {\it XMM-Newton} \cite[][]{Esquej07, Esquej08, Lin11, Saxton12}. {\it Swift} has been extremely useful in providing rapid follow-up observations for many of these events, and for discovering a new class of higher energy gamma-ray events. A detailed description of TDE characteristics in the X-rays and gamma-rays can be found in the {\xraychap} and the \gammachap\ in this book, respectively.

In the last decade, several transients discovered in optical and ultraviolet wavelengths have been attributed to TDEs. The source of TDE emission at these wavelengths is currently debated to be either reprocessing of the X-ray emission from the accretion disk by optically thick material surrounding the disk \citep[e.g.][]{Guillochon13,Roth16}, or as emission from outer shocks between the debris streams as they collide \citep[e.g.][]{Piran15}, or perhaps some combination of both \citep[e.g.][]{Jiang16,Lu20}. A detailed discussion of these emission mechanisms is presented in the \emischap\ in this book. Here we focus on the observational properties of the main classes of optical and ultraviolet transients suggested in the literature to be TDEs. Additional one-of-a-kind events, whose nature is less clear, are discussed in the \impostchap.

The main class of \optuv\ TDEs contains approximately 30 events (Table \ref{tab:prop}) and is identified mainly by broad ($\sim10,000$\,km\,s$^{-1}$) H and/or He lines in their spectra (with some events showing additional features), blue-continuum ($\sim$few$\times10^4$\,K) emission lasting for several months, and a $\sim$month-long rise to a peak optical absolute magnitude of approximately $-20$. A diversity within this class does exist, and a full unbiased view of the population is not yet available. This class of events show a surprising and still unexplained strong preference for rare post-starburst (or E+A) host galaxies (see \hostchap), while almost none have been found in active galactic nuclei (AGN)\footnote{A notable candidate exception is PS16dtm \citep{Blanchard17} which is discussed in the \impostchap.}. There are obvious observational difficulties in finding TDE flares embedded in AGN light, and also in discerning TDEs from rare AGN flares \citep[e.g.][]{Trakhtenbrot2019_CLAGN, Trakhtenbrot2019_Bowen}. \cite{Chan2019} recently suggested that TDEs in AGN will look different than those in non-active galaxies due to the interaction of the stellar debris with the pre-existing accretion disk. Distinguishing TDEs from AGN is discussed in more detail in the \impostchap.

Here we present a current picture of the main class of established events. However, it's important to keep in mind that with new events being discovered regularly, the map of \optuv\ TDEs is still being drawn. While we refer to all of the events discussed here as TDEs, some might still be considered ``TDE candidates" until a better picture of the population of real events and their imposters emerges. 

In the next section, we begin with a brief review of the history of \optuv\ TDE discoveries in order to provide a picture of the different biases affecting the current sample of events. We then focus on the photometric (Section~\ref{sec:lightcurves}) and spectroscopic (Section~\ref{sec:spectra}) properties of this sample and discuss how a robust sample of \optuv\ TDEs can be constructed (Section~\ref{sec:obsparams}). We then discuss rate measurements and the luminosity function of \optuv\ TDEs (Section~\ref{sec:rates}), and end with some concluding remarks (Section~\ref{sec:conclusions}).

\section{A Brief History of TDE Detections in Optical and Ultraviolet Surveys}\label{sec:intro_hist}

The first \optuv\ TDE candidate to be discovered was GALEX-D3-13 \citep{Gezari06}; two more events (D1-9 and D23-H1) were reported in \cite{Gezari08,Gezari09}. These three events were all found using multi-epoch ultraviolet images from the Galaxy Evolution Explorer \citep[GALEX;][]{Martin05} which were selected to have a rich set of archival data (e.g., the Groth field). Using this archival data, recurring AGN flares could be removed. 

Simultaneous optical imaging is available for two of the GALEX flares, allowing for an accurate estimate of the \bb\ temperature. All three TDEs found in GALEX data have relatively high \bb\ temperatures ($T \approx 5\times {10^4}$\,K). Since GALEX imaging is relatively deep (limiting magnitude 23 in the near-ultraviolet band), the GALEX search for TDEs has the greatest depth of all imaging searches to date \citep[see Table~3 in][]{vanVelzen18}. However this search was not done in real-time, hence spectroscopy of the flares was never obtained. 

The GALEX transient search \citep{Gezari13} remains the only ulraviolet-based TDE search to date.
Below we provide an overview of the optical surveys that have led to the discovery of TDEs  that can be classified based on their spectral properties (as listed in Table~\ref{tab:prop}).

\subsection{SDSS}
The Sloan Digital Sky Survey \citep[SDSS;][]{york02} provided two types of TDE candidates. First, the SDSS catalog of galaxy spectra \citep{strauss02} enabled the discovery of coronal-line TDEs \citep[e.g.][]{Komossa2008,Wang2011,Wang2012,Yang2013}. For these sources, the TDE is not observed directly, but rather transient narrow high-ionization emission lines\footnote{such as a strong Fe X $\lambda$6376 to O III $\lambda$5007 ratio, and strong Fe X $\lambda$6376, Fe XI $\lambda$7894, Fe XIV $\lambda$5304, Ar XIV $\lambda$4414, and S XII $\lambda$7612 lines.} are seen, and are interpreted as an ``echo" of a soft X-ray flare (required to produce the coronal lines) originating in a TDE. These events, as well as echoes showing broad emission lines \citep[e.g.][]{Komossa2008}, are discussed in more detail the \echochap\ of this book. Here we note that due to the lack of multi-epoch imaging observations that cover the flare itself, our knowledge of the photometric properties of these events (e.g., timescales, location in the host galaxy, and \bb\ temperature) is limited.

Second, SDSS provided the first TDEs selected directly using optical imaging observations \citep{vanVelzen10}. These were discovered in Stripe~82 \citep{frieman08}, a region of $\approx 300$~deg$^2$ that was observed with a cadence of $\sim$few days for three months form 2005 to 2007 (lower cadence observation date back to 2000). A systematic search for nuclear flares from all galaxies in Stripe~82 yielded two events (dubbed SDSS-TDE1 and SDSS-TDE2) from otherwise quiescent galaxies. The key property distinguishing these flares from normal supernovae and AGN is a high \bb\ temperature (as measured using SDSS $ugri$ photometry) that remained constant for at least three months (this property later became a basis for the photometric selection of TDEs). Both events also show transient ultraviolet emission, detected in GALEX more than one year after peak. For one of the two flares, SDSS-TDE2, an optical spectrum was obtained near peak. The spectrum displays broad emission features around H$\alpha$, and H$\beta$ and He II $\lambda$4686.

\subsection{Pan-STARRS}
Two TDEs (from non-AGN galaxies) were discovered by the Pan-STARRS medium-deep survey \citep{Chambers07}: PS1-10jh \citep{Gezari12} and PS1-11af \citep{Chornock14}. Both were identified via spectroscopic follow-up of photometric transients in the medium-deep fields \citep{Chambers16}, a set of fields that were targeted with a three-day cadence to a typical single-epoch depth of $m=23$ (though the effective flux limit of the TDE search is likely to be shallower due to the requirement for spectroscopic classification). 

Both PS1-10jh and PS1-11af show transient ultraviolet emission (detected by GALEX) and long-term high \bb\ temperatures, but display different spectral properties. PS1-10jh displays a broad He II $\lambda4686$ emission feature in its optical spectrum, while PS1-11af shows two transient absorption features in the rest-frame ultraviolet, possibly attributed to Mg II. PS1-10jh is the first TDE with multiple pre-peak detections and multiple spectroscopic observations, making it the archetype for the main \optuv\ class of TDEs.

While the medium-deep survey has concluded, the Pan-STARSS telescope continued to search for transients \citep{Huber15} and later discovered the TDEs PS17dhz \citep[AT\,2017eqx;][]{Nicholl2019} and PS18kh \citep[AT\,2018zr;][]{Holoien18a}. Another TDE candidate from the Pan-STARRS survey was discovered in a narrow-line Seyfert~1 galaxy \citep[PS16dtm;][]{Blanchard17}. This event, and others like it, for which an association with a TDE or enhanced AGN activity is not decisive, are discussed in the \impostchap.

\subsection{PTF}
Three TDEs were discovered in Palomar Transient Factory \citep[PTF:][]{Law09} data by searching for spectroscopically-observed nuclear flares with an $R$-band absolute magnitude in the range $-21<M_R<-19$ \citep{Arcavi14}. The three events discovered (PTF09ge, PTF09djl, and PTF09axc) all display
relatively high \bb\ temperatures, and have similar spectroscopic signatures as PS1-10jh and SDSS-TDE2, namely broad He and/or H emission lines. The discovery of the events established the spectroscopic sequence of TDEs (see Section \ref{sec:spectra}), and also identified the peculiar nature of the host galaxies of such events (see \hostchap). For the PTF TDEs, no ultraviolet observations were obtained near peak, but two sources were detected in {\it HST} far-ultraviolet imaging observations, five years after peak \citep{vanVelzen18_FUV}. 

\subsection{iPTF}
The iPTF search for TDEs \citep{Hung18} used the properties of previous TDEs to identify new candidates in real time, hence more follow-up data (in particular with {\it HST} and {\it Swift}) of promising TDEs could be obtained in real time. The typical single-epoch magnitude limit of iPTF data is $\approx 20.5$. No selection on the absolute magnitude was made, but instead a blue color for the flare ($g-R<-0.1$) was required to reduce the background of supernovae. Follow-up spectroscopy and {\it Swift}-UVOT observations were used to confirm the TDE nature. Three TDEs were found: iPTF15af \citep{Blagorodnova19}, iPTF16fnl \citep{Blagorodnova17}, and  iPTF16axa \citep{Hung17}. iPTF16fnl is an outlier in terms of its low luminosity and rapid decline rate. It raises the question of how many more such TDEs might have been missed due to an observational bias against finding rapidly evolving events.

\subsection{ASAS-SN} 
To date, the All Sky Automated Survey for supernovae \citep[ASAS-SN;][]{Shappee14} discovered eight TDEs: ASASSN-14ae \citep{Holoien14}, ASASSN-14li \citep{Holoien16a}, ASASSN-15oi \citep{Holoien16b}, ASASSN-18pg \citep[AT 2018dyb;][]{leloudas19,Holoien2020_18pg}, ASASSN-18ul (AT\,2018fyk), ASASSN-18zj \citep[AT\,2018hyz;][]{Short2020,gomez20,Hung2020}, ASASSN-19bt \citep[AT\,2019ahk;][]{Holoien19_bt}, and ASASSN-19dj (AT\,2019azh). The ASAS-SN discoveries come from an all-sky search using single-band observations (either $V$- or $g$-band) with a typical limiting magnitude of 17. The TDE identification is established via spectroscopic follow-up. This shallow and wide nature of ASAS-SN means that the survey is sensitive to the lowest-redshift TDEs ($z\sim 0.03$), making them ideal for intensive followup observations. Ultraviolet detections from {\it Swift} have been obtained for all ASAS-SN TDEs. 

ASASSN-14li is the first optical TDE with a well-sampled X-ray light curve \citep{Holoien16a,Miller15}, and (at the time of writing) the only published optical TDE detected in the radio \citep[see the  \radiochap;][]{vanVelzen16,Alexander16}. 

The nature of an additional ASAS-SN event, ASASSN-15lh, is controversial. It has been interpreted as either an extreme superluminous supernova \citep{Dong16,Godoy-Rivera17} or an extreme TDE \citep{Leloudas16,Margutti17,Kruhler18}.

\subsection{ATLAS}
The Asteroid Terrestrial-impact Last Alert System \cite[ATLAS;][]{Tonry2018} is a survey primarily designed for discovering potentially hazardous asteroids. However, it regularly discovers transients from its two-day cadence search down to a limiting magnitude of 19--20. ATLAS has discovered three TDEs so far \citep{vanvelzen2020ZTF}: ATLAS18way (AT\,2018hco), ATLAS18yzs (AT\,2018iih) and ATLAS19qqu (AT\,2019mha).

\subsection{OGLE}
The Optical Gravitational Lensing Experiment \citep[OGLE;][]{Udalski2015} also runs a transient search \citep{Wyrzykowski2014}. The survey is focused on an area of the sky around the Magellanic Clouds, but this includes many background galaxies. Each field is observed roughly every four days with a limiting magnitude of $\sim20$ in the $I$-band. OGLE has discovered one TDE \citep[OGLE16aaa,][]{Wyrzykowski17,Kajava20}.

\subsection{ZTF}
The Zwicky Transient Facility \citep[ZTF;][]{Bellm19} started in early 2018 and is still running. The single-epoch depth and filters are similar to iPTF, but the field of view is eight times larger. The source PS18kh was detected during the ZTF commissioning phase \citep{vanVelzen18_NedStark}, and sixteen more TDEs have been detected in ZTF data since \citep{vanvelzen2020ZTF}, signaling a large increase of the TDE discovery rate.    

\section{Light Curves}
\label{sec:lightcurves}

Optical-ultraviolet TDEs typically rise to a peak \bb\ luminosity of $L_{\rm bb}\sim$10$^{43.5-44.5}$ erg s$^{-1}$ on timescales of weeks-months with a decay broadly consistent with t$^{-5/3}$. This is the expected decline rate of the post-disruption mass return flow (\citealt{Rees88,Phinney89}), though the density profile of the disrupted star may influence this rate \citep{Lodato09,Gezari12}, as will relativistic effects \citep{Kesden12}. It is also important to keep in mind that fitting a power law to TDE light curves introduces a large uncertainty on the inferred power-law index, due to the freedom in setting the time of disruption.

\begin{figure}
\includegraphics[width=\textwidth]{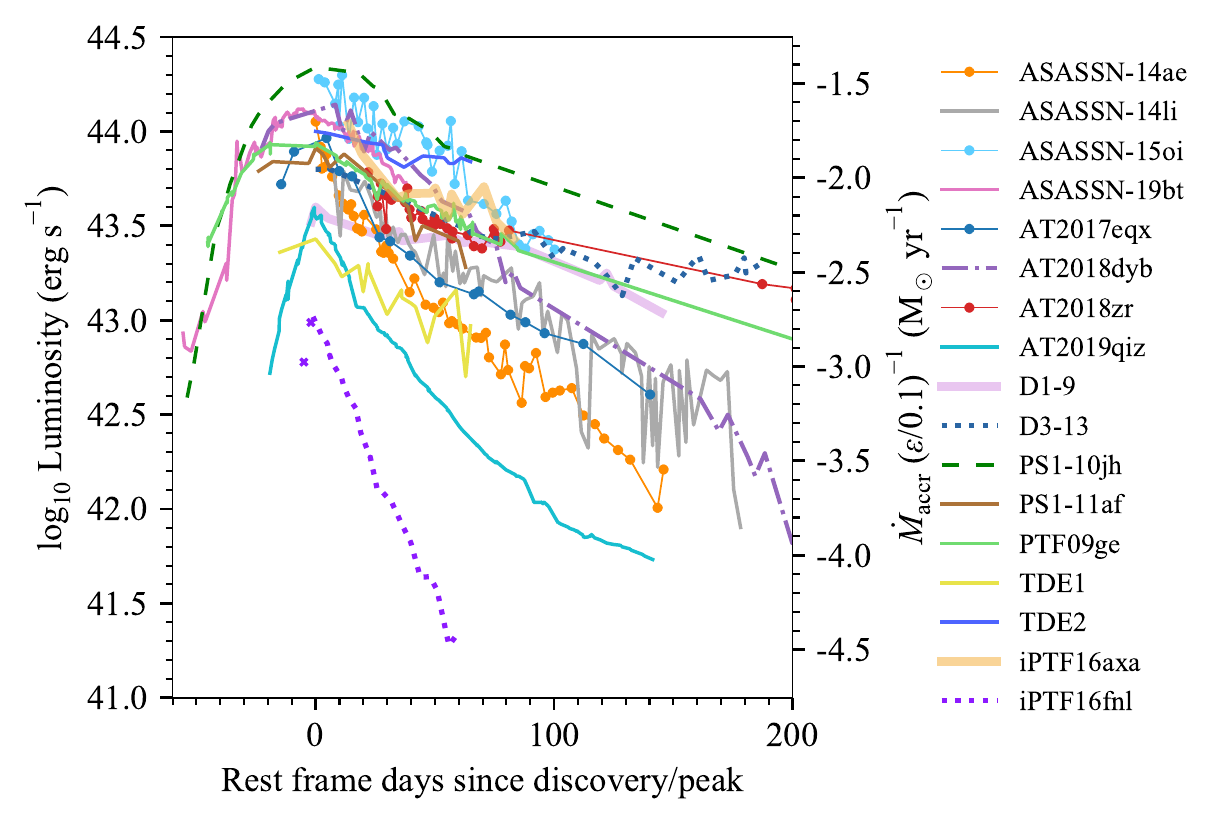}
\caption{A compilation of TDE light curves. The integrated \optuv\ luminosity is shown as inferred from spectral energy distribution fitting. The x-axis shows the time elapsed since peak ֿfor PTF09ge, PS1-10jh, iPTF16fnl, AT2017eqx, AT2018dyb, AT2018zr, and ASASSN-19bt and the time elapsed since discovery for ASASSN-14ae, ASASSN-14li, ASASSN-15oi, and iPTF16axa. The two crosses in purple are derived from pre-peak $g$-band data of iPTF16fnl assuming a \bb\ temperature of $2\times10^4$\,K. The mass accretion rate on the right hand side is normalized to an efficiency $\varepsilon$ of 0.1. Figure adapted from \cite{Hung17}, data from \cite{vanVelzen10,Gezari12,Chornock14,Arcavi14,Holoien14,Holoien16a,Holoien16b,Blagorodnova17,Hung17,vanVelzen18_NedStark,Nicholl2019,leloudas19,Holoien19_bt,Nicholl2020}
\label{fig:optical_lc}}       
\end{figure}

\begin{figure}
\includegraphics[trim={0 0 0 65mm},clip, width=\textwidth]{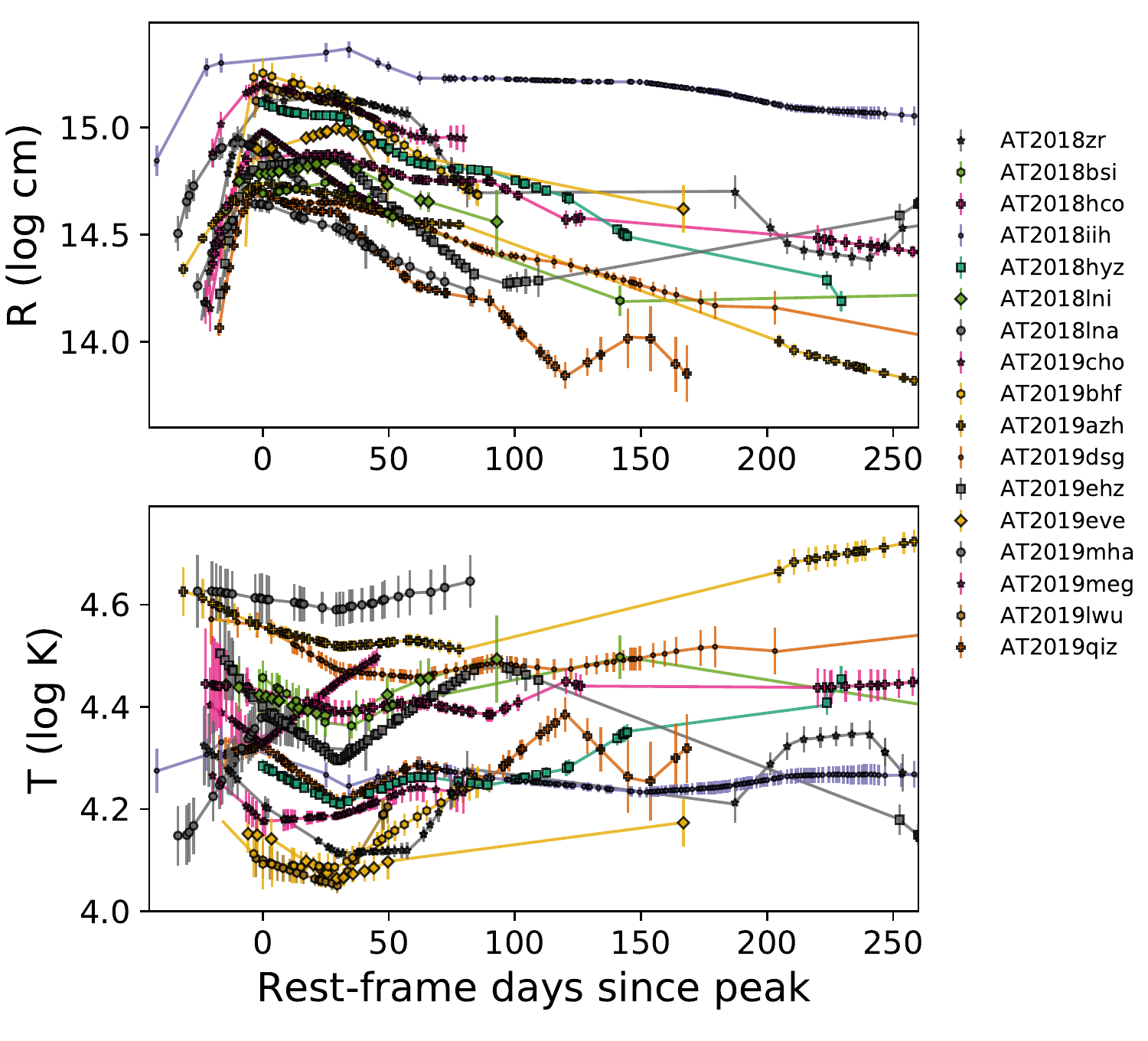}
\caption{A compilation of \bb\ radii and temperatures obtained from \optuv\ photometric observations of seventeen TDEs from \cite{vanvelzen2020ZTF}. All sources show a decrease of the \bb\ radius after maximum light. The majority of the sources show a near-constant \bb\ temperature. While some TDEs show modest temperature variability near peak or a significant temperature increase at late-time, none of the known TDEs show a monotonic decrease of the \bb\ temperature. Since most of the sources in the ZTF sample were identified as TDEs within a few weeks of maximum light and the temperature is inferred from ultraviolet follow-up observations, this lack of cooling cannot be solely explained as a selection effect.}
\label{fig:ZTF_LTR}       
\end{figure}

Peak luminosities and time scales vary between events (Fig. \ref{fig:optical_lc}). What does not seem to vary between events is the lack of cooling of the post-peak \bb\ temperature  (Fig. \ref{fig:ZTF_LTR}), though some of this homogeneity may be artificially enhanced by the use of the non-evolving temperature as a selection criterion to discriminate TDEs from supernovae. 

Perhaps the first notable outlier in its photometric properties (being still ``normal'' spectroscopically) is iPTF16fnl, displaying a fainter and faster light-curve evolution than the rest of the class \citep[Fig. \ref{fig:optical_lc};][]{Blagorodnova17,Onori2019}. This raised the question as to whether many more such events exist but are missed due to observational biases. Indeed, recently, more rapidly-decaying events have been found in the ZTF survey (see Table~\ref{tab:lc} and the analysis of AT\,2019qiz by \citealt{Nicholl2020}). 

Another photometric outlier is ASASSN-15lh \citep{Leloudas16}, which displayed a double peak in the bluer bands of its light curve. This event is an outlier also spectroscopically, and its identification as a TDE is contended \citep{Dong16,Godoy-Rivera2017}. For this reason we do not include it in Table \ref{tab:prop}. ASASSN-15lh is discussed in more detail in the \impostchap.

Recently, \cite{vanvelzen2020ZTF} doubled the sample of known TDE light curves, and found a correlation between photometric properties and the spectroscopic types (this spectroscopic classification is introduced in the next section). Namely, that TDEs showing Bowen features have smaller \bb\ radii compared to the rest of the population, while events showing only He in their spectra have the longest rise times. For the overall population, the rise timescale and fade timescale show no clear correlation, while a strong correlation would be expected if rise and fade timescales were both determined by the fallback rate. \citet{vanvelzen2020ZTF} conclude that this might be an indication that the rise time of TDE light curves is not related to the fallback time but rather to a diffusion time through some extended material (which may be the source of the spectral differences as well). 

\begin{figure}
\includegraphics[trim={0 0 0 33}, clip, width=\textwidth]{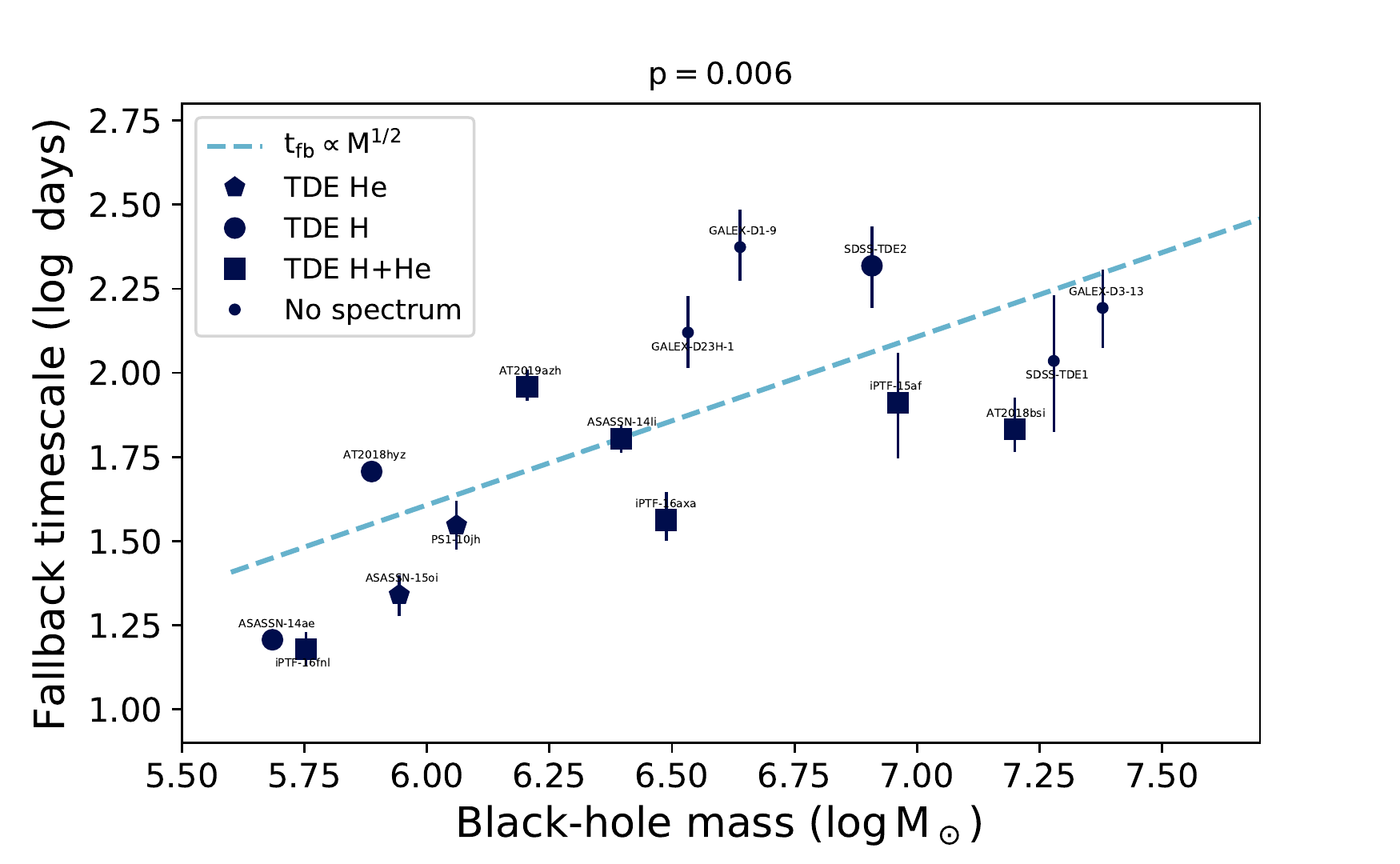}
\caption{Black-hole mass versus fallback time. The black-hole mass is estimated from the \mbox{$M$-$\sigma$} relation and the fallback time follows from fitting a $L\propto t^{-5/3}$ power-law decay to the \optuv\ \bb\ lightcurve. The dashed line indicates the predicted fallback rate for a star of one solar mass \citep{Stone13}.}
\label{fig:mass-t0}       
\end{figure}

Several authors have reported a (potential) correlation between the decay rate of the light curve and black-hole mass \citep{Blagorodnova17,Wevers17} or total host galaxy mass \citep{vanvelzen2020ZTF}. In Figure~\ref{fig:mass-t0} we reproduce this result using our homogeneous lightcurve analysis (see Section~\ref{sec:obsparams}). Here we estimate the black-hole mass from the $M$-$\sigma$ relation \citep{Gultekin09}, using the velocity dispersion measurement of \citet{Wevers17,Wevers19} or SDSS \citep{Thomas13}. The fallback timescale is estimated by fitting a power-law decay to the observed \optuv\ \bb\ emission, with the power-law index fixed to $-5/3$. We find a statistically significant correlation between the black-hole mass and decay rate ($p=0.006$ for a Kendall's Tau test). This correlation could indicate that the early-time decay rate of the \optuv\ emission is indeed determined by the fallback rate.

\begin{figure}
\includegraphics[width=\textwidth]{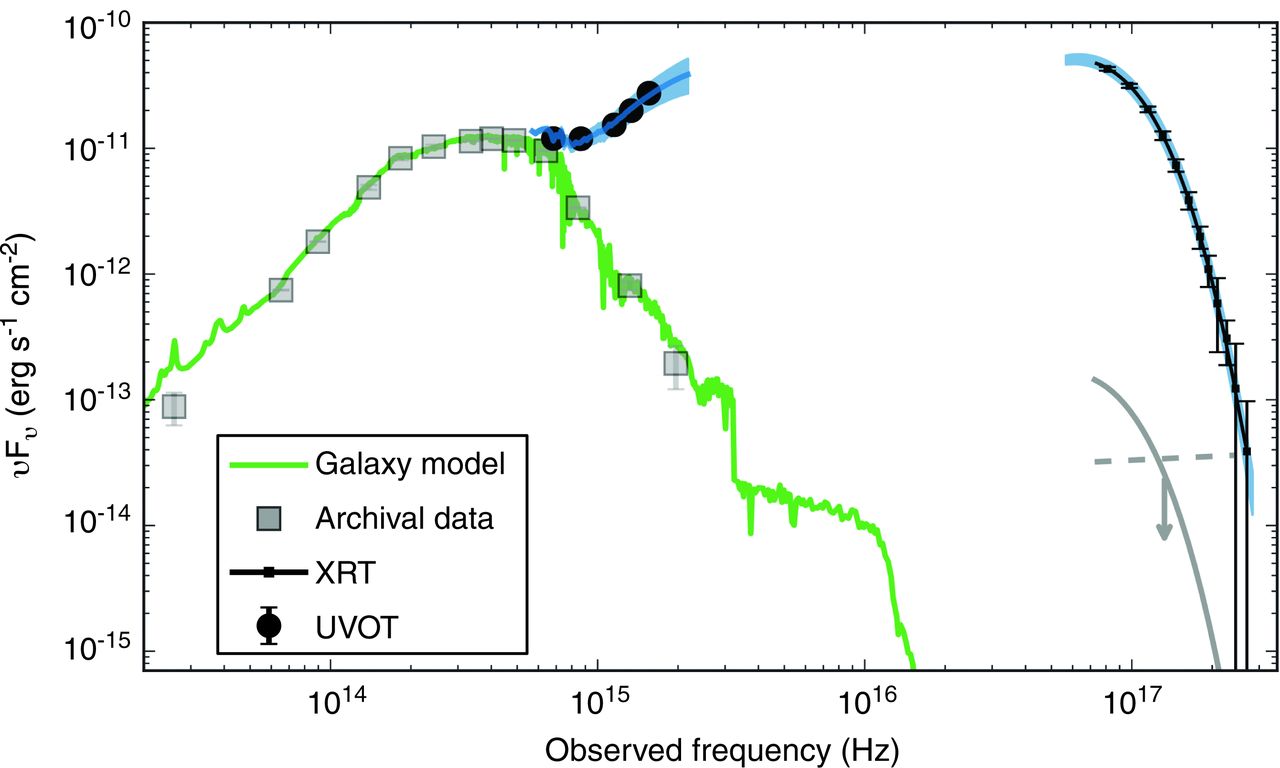}
\caption{{\it Swift} X-ray and broadband \optuv\ observations of ASASSN-14li. A double-\bb\ fit with temperatures of T=7.7$\times$10$^{5}$ K and T=3.5$\times$10$^{4}$ K is shown in blue. The pre-flare X-ray limit for the \bb\ spectrum is show in grey. The SED of the host galaxy (grey squares) together with a best fit synthetic galaxy spectrum (green) are shown for comparison. Figure from \citet{vanVelzen16}.}
\label{fig:sed}       
\end{figure}

X-ray detections are rare in the case of optically-discovered TDEs, and there are only a few cases showing both \optuv\ and X-ray emission: ASASSN-14li \cite[][Fig. \ref{fig:sed}]{Miller_14li,Holoien16,vanVelzen16}, AT\,2018fyk \cite[][]{Wevers19a} AT\,2018zr,  \cite[][]{vanVelzen18_NedStark}, and most recently, AT\,2019azh, AT\,2019dsg, and AT\,2019ehz \citep{vanvelzen2020ZTF}.

The existence of an observational dichotomy between \optuv\ and X-ray TDEs has been a topic of study. \cite{LoebUlmer97}, \cite{strubbe_quataert09}, \cite{guillochon14}, \cite{Roth16} and \cite{Auchettl16} invoke material surrounding the accretion disk as responsible for reprocessing of the X-ray photons from the disk into optical and ultraviolet photons. \cite{Dai18} expand on this picture by proposing that the reprocessing material is concentrated in the equatorial direction, concluding that the observed dichotomy is a viewing angle effect: the optical and ultraviolet emission are seen from the equatorial direction, while a polar viewing angle will not cross the reprocessing material and provide a line of sight directly into the X-ray emitting accretion disk. These models are further discussed in the \emischap\ of this book.

Many \optuv\ TDEs with observations more than 100 days post-peak show a flattening of the light curve to almost constant luminosity. A clear example is ASASSN-14li \citep{Brown17a}. Using both {\it HST} and {\it Swift}/UVOT observations, \citet{vanVelzen18_FUV} find a flattening of the ultraviolet light curve for 10 out of 12 TDEs with late-time observations. They show that this plateau phase can be explained by emission from an accretion disk that forms after the disruption \citep{Cannizzo90}. \citet{Mummery20} show that both the late-time ultraviolet plateau emission and the early-time X-ray emission of ASASSN-14li can be explained using an evolving accretion disk.

\section{Spectra}
\label{sec:spectra}

Spectra are crucial for distinguishing TDEs from other nuclear activity due to AGN variability, or other transients such as supernovae. They can also provide insights into the sources of TDE emission and dynamics. The increasing sample of TDEs monitored by dense spectroscopic follow-up campaigns has revealed a set of spectroscopic classes differing in their line species, profiles and evolution.  

\subsection{Optical}
\begin{figure}
\includegraphics[width=\textwidth]{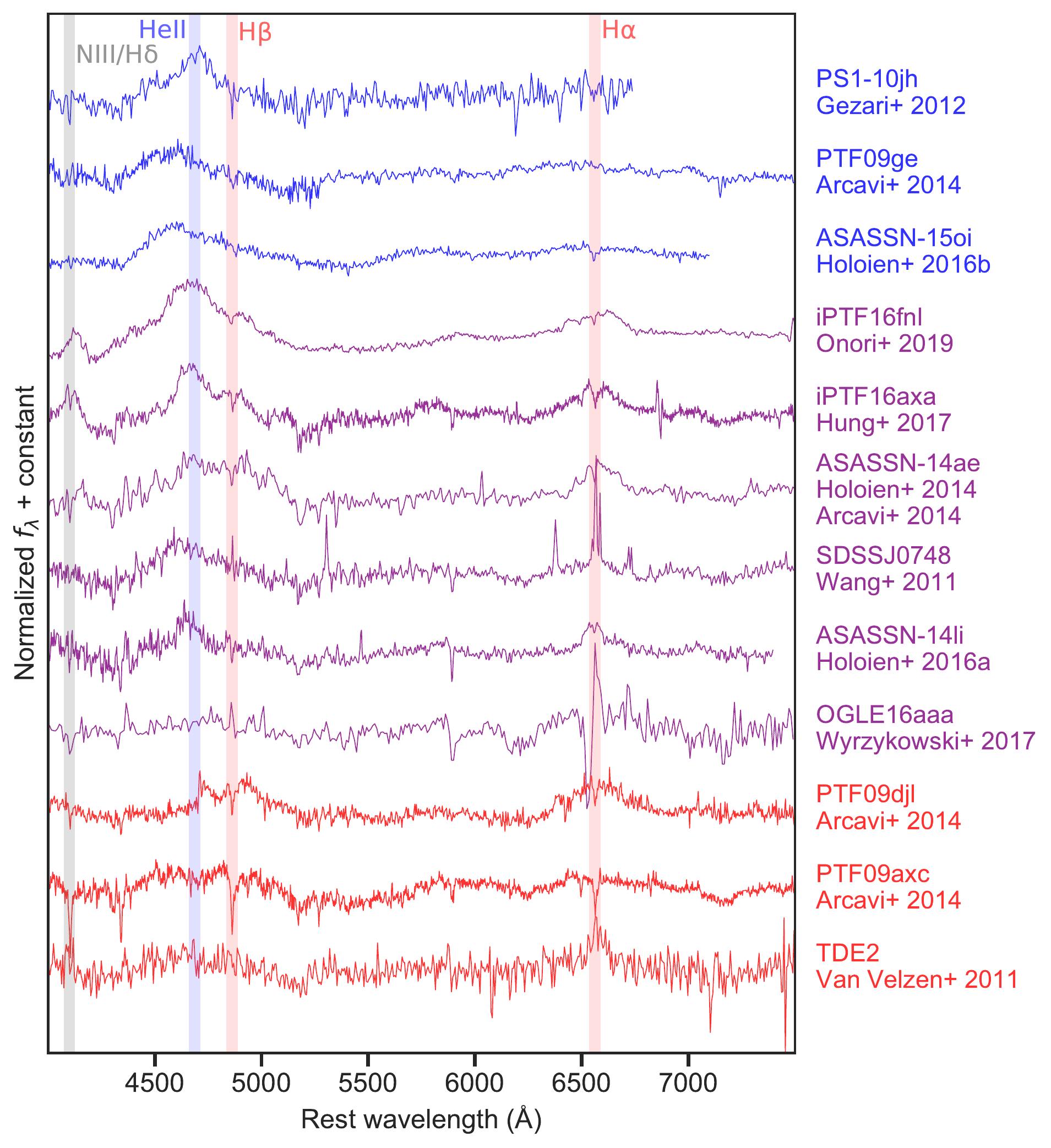}
\caption{Continuum-subtracted optical spectra of optically-discovered TDEs. Events in blue are dominated by a broad He II $\lambda$4686 feature with no signatures of H$\alpha$ or H$\beta$ (we denote these ``TDE\,He''). Events in purple are characterized by the presence of both broad He II $\lambda$4686 and H$\alpha$, with variation in the intensity of these lines from one event to another, and possible broad H$\beta$ seen in some events (we denote these``TDE\,H+He''). Events in red are H-dominated (we denote these ``TDE\,H''). Additional N III features (perhaps blended with H$\delta$) are seen for some events in the ``H+He" class. Figure adapted from \cite{Arcavi14}, data from \cite{Wang2011,vanVelzen10,Gezari12,Arcavi14,Holoien14,Holoien16b,Hung17,Onori2019}.}
\label{fig:optical_spec}       
\end{figure}

Spectra of the first \optuv\ TDEs identified were dominated by a strong and consistent blue continuum with broad (FWHM$\sim$10$^{4}$\,km\,s$^{-1}$) emission lines superimposed on it. The most common and prominent spectral feature is a broad and intense He II $\lambda$4686 line \citep[e.g.][]{Gezari12,Gezari15}, which is not seen at these widths, intensities and durations in spectra of other known transients. In some cases, broad H$\alpha$ and H$\beta$ lines are also identified \citep[e.g.][]{Arcavi14}. This led to an initial map of TDE spectral diversity as a continuum of H-rich to He-rich events (Fig. \ref{fig:optical_spec}). \citet{vanVelzen20} divided the current TDE population into three spectroscopic types and showed that TDEs in these classes have significantly different photometric properties. Following their approach, in Table \ref{tab:prop} we label events showing only He II lines as ``TDE\,He'', those showing only hydrogen lines as ``TDE\,H'' and those showing both He II and H lines as ``TDE\,H+He''.

The lack of any hydrogen signatures in the spectra of PS1-10jh was initially interpreted as evidence for the disruption of a rare helium star \citep{Gezari12}. However, with the discovery of additional events with hydrogen-free spectra, this hypothesis became harder to justify. \cite{guillochon14}, using hydrodynamical simulations, was able to explain the PS1-10jh spectra as produced by the disruption of a main sequence star. In this scenario, an optically thick reprocessing envelope determines the location of the line-forming region, and the lack of hydrogen lines is attributed to the conditions for forming the lines in the envelope, rather than the presence or absence of hydrogen. \cite{Roth16} showed that varying the radius of the reprocessing layer (given some composition and ionizing flux) can suppress the hydrogen lines while keeping the He II lines strong, thus reproducing the observed spectral sequence displayed in Figure \ref{fig:optical_spec}.

\begin{figure}
\includegraphics[width=\textwidth]{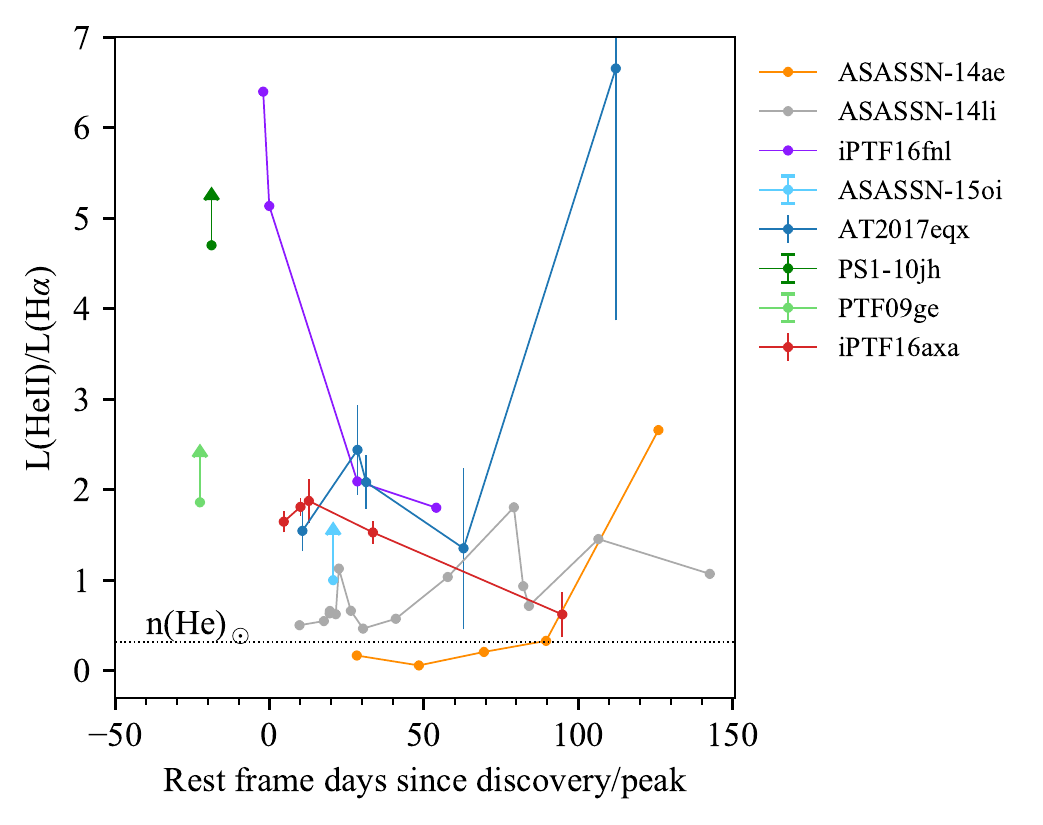}
\caption{He II $\lambda$4686 to H$\alpha$ line luminosity ratio and its evolution with time for a sample of \optuv\ TDEs with well-sampled spectroscopic sequences. The ratio changes with time in a different way for each event, yet most events show a ratio higher than the expected value for a nebular environment with solar abundance (dotted line). Figure from \cite{Hung17}, data from \cite{Gezari12,Arcavi14,Chornock14, Holoien14, Holoien16a, Holoien16b,Blagorodnova17}.}
\label{fig:lineratio}       
\end{figure}

In some events, the relative hydrogen to helium line intensity is seen to change with time. For instance, the early optical spectra of ASASSN-14ae \citep{Holoien14} are characterized by an intense and broad H$\alpha$ emission line and a weak signature of a broad component in the He II $\lambda$4686 line, which developed only later. An even more extreme He II to H line-ratio evolution is seen in AT\,2017eqx (Figure~\ref{fig:lineratio}) and interpreted, following \cite{Roth16}, as evidence for contraction of the reprocessing layer \citep{Nicholl2019}.

In Figure \ref{fig:lineratio} we show the time evolution of the He II / H$\alpha$ luminosity ratio for a number of \optuv\ TDEs with well-sampled spectroscopic sequences. The time evolution of this line ratio varies from event to event. Yet, in most, the He II / H$\alpha$ luminosity ratio is above the expected value for a nebular environment with solar abundance \cite[dotted line in Figure \ref{fig:lineratio}]{Hung17}. This helium enhancement has been explained as the result of the suppression of the Balmer lines (which are optically thick) by high gas densities ($>$10$^{10}$\,cm$^{-3}$) in the emitting region. This scenario was first proposed by \cite{bogdanovic04} and further investigated using \texttt{CLOUDY} calculations by \cite{Gaskell14} and \cite{strubbe15}, as well as with radiative transfer calculations by \cite{Roth16}.

\begin{figure}
\includegraphics[width=\textwidth]{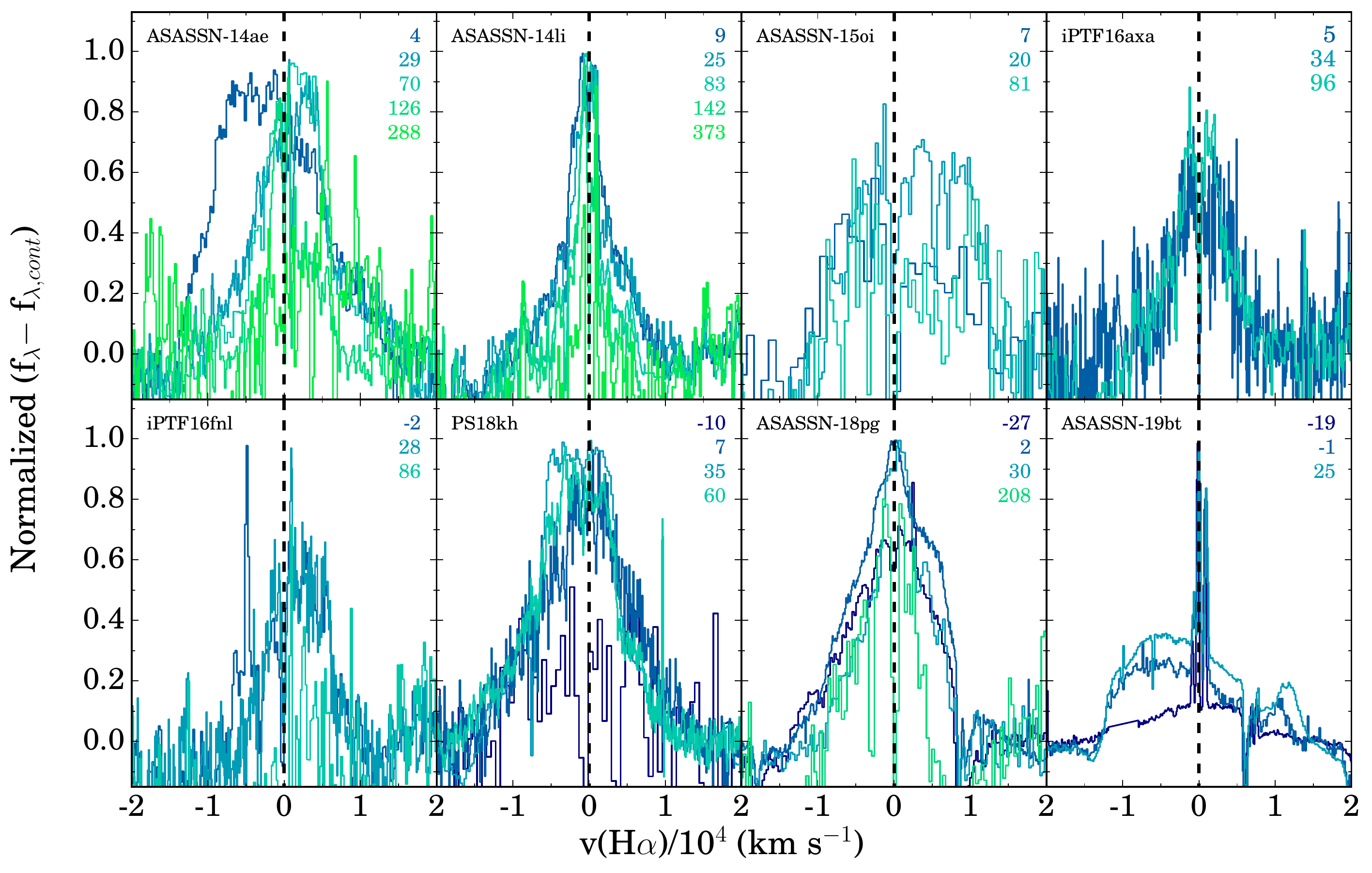}
\caption{Continuum-subtracted line profiles of H$\alpha$ (rest-frame zero velocity marked by black dashed line)} for eight \optuv\ TDEs. Colors correspond to rest-frame days from peak or discovery as indicated. ASASSN-14ae can be seen to have an asymmetric profile at early times which later becomes symmetric. Data from \cite{Holoien14,Holoien16a,Holoien16b,Brown16,Brown16b,Hung17,Blagorodnova17,Holoien19_bt,Holoien19_kh,Holoien2020_18pg}.
\label{fig:optical_spec_Ha}       
\end{figure}

\begin{figure}
\includegraphics[width=\textwidth]{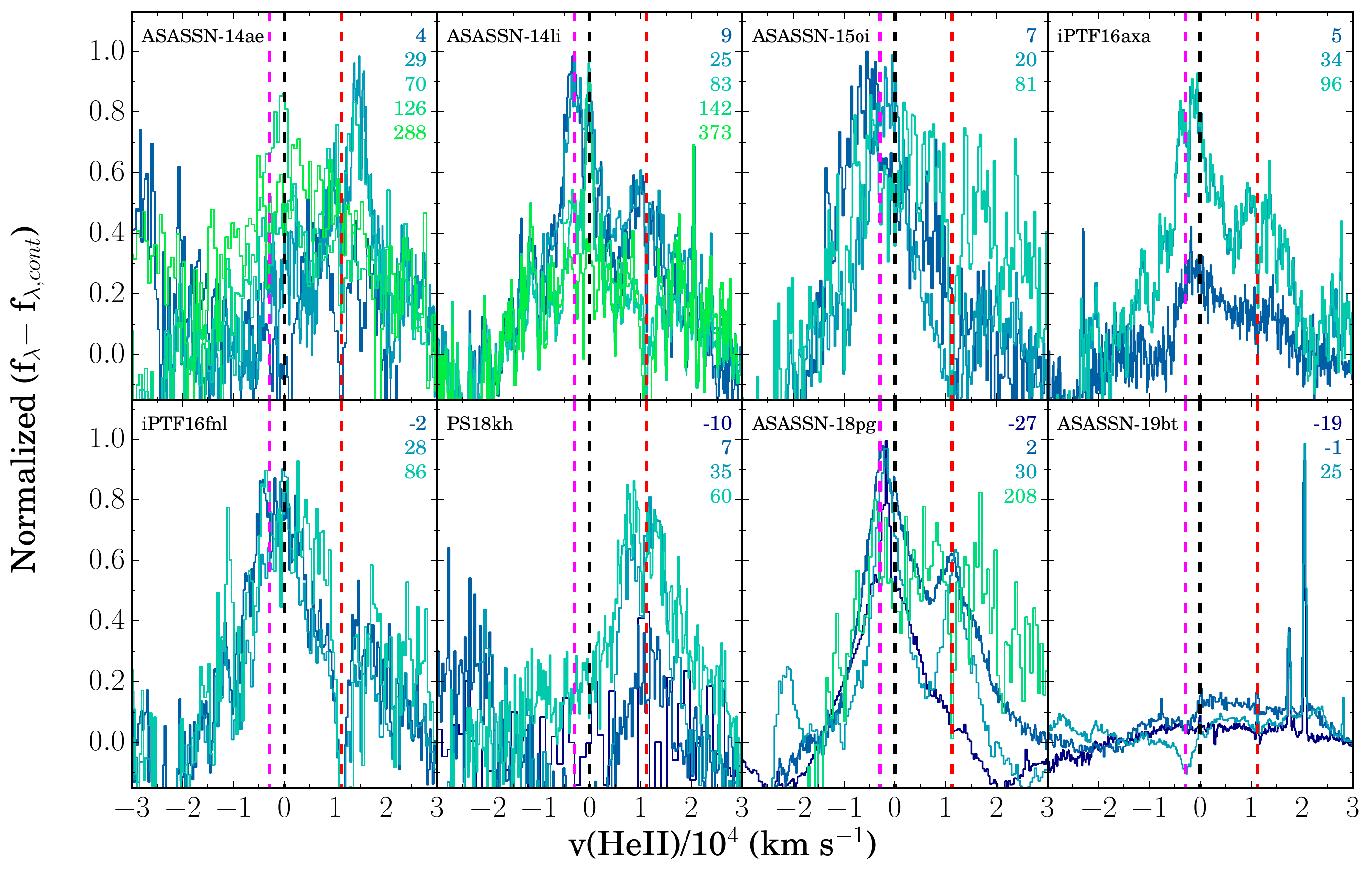}
\caption{Continuum-subtracted line profiles of He II (rest-frame zero velocity marked by black dashed line), N III (magenta dashed line),} and H$\beta$ (red dashed line) for eight \optuv\ TDEs. Colors correspond to rest-frame days from peak or discovery as indicated. ASASSN-15oi can be seen to have an asymmetric He II line profile at early times which later becomes symmetric. Data sources are the same as in Figure \ref{fig:optical_spec_Ha}.
\label{fig:optical_spec_He}       
\end{figure}

The broad features observed in TDEs are, in general, pure emission line profiles, with no signs of absorption or P-cygni features. 
We present the line-profile evolution of H$\alpha$ and He II for several TDEs in Figures \ref{fig:optical_spec_Ha} and \ref{fig:optical_spec_He} respectively. Some events display evolving asymmetric profiles. For example, ASASSN-14ae \cite[]{Holoien14} shows a broad and asymmetric H$\alpha$ line, with a prominent red wing and a blue-shifted centroid at early times. Later, the H$\alpha$ line becomes symmetric and centered around the host rest-frame wavelength. Similar evolution is seen in the He II line of ASASSN--15oi \cite[]{Holoien16b}. 

\begin{figure}
\includegraphics[trim={0 108mm 0 0},clip,width=\textwidth]{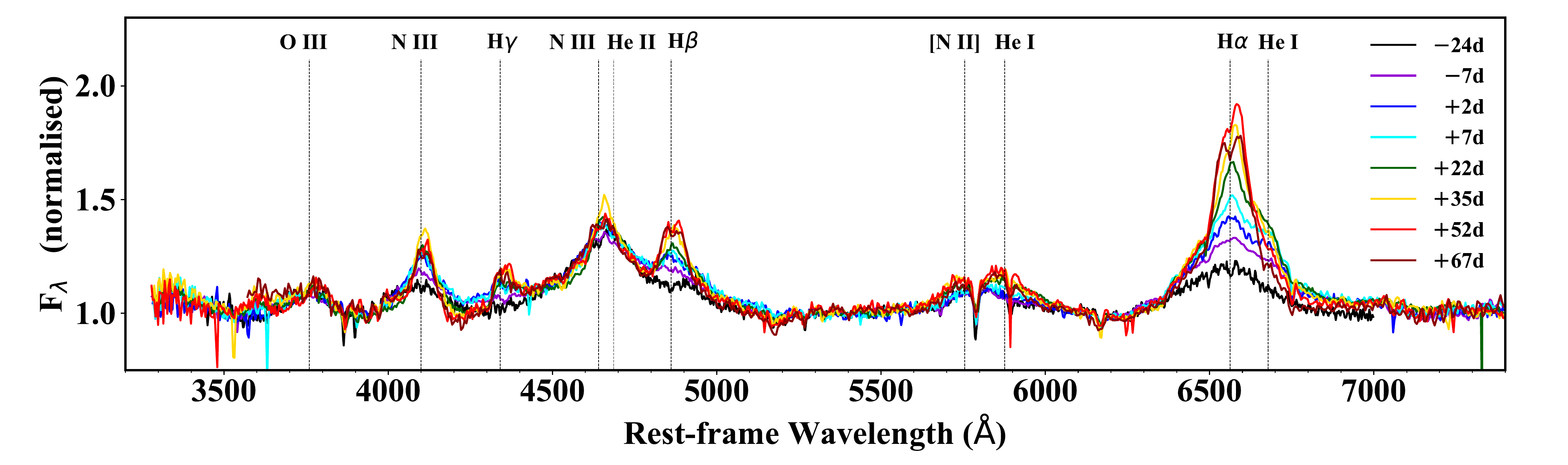}
\caption{
Spectra of the Bowen (N-rich) TDE AT\,2018dyb after normalising to the continuum. The different colors indicate different phases in rest-frame days relative to peak luminosity. A larger richness in line species compared to the initial hydrogen to helium mapping of TDE spectra (Fig. \ref{fig:optical_spec}) can be seen. Figure from \cite{leloudas19}.
}
\label{fig:n_rich_tde}       
\end{figure}

Optically-thick outflows have been proposed to explain such line profiles and their evolution, especially in the case of H$\alpha$ \cite[]{Roth18}. However, in some cases, the blue-shifted component in the He II $\lambda$4686 line has been attributed to a blend with possible N III/C III lines \cite[]{Gezari15, Brown18}. \cite{leloudas19} found a rich set of line species in the spectra of the TDE AT\,2018dyb (Fig. \ref{fig:n_rich_tde}). In addition to the Balmer series and He II, they identify also He I, O III $\lambda$3760, and, indeed, N III $\lambda\lambda$4100, 4640 (the latter blended with He II). This event demonstrates the richness of \optuv\ TDE spectral diversity beyond the hydrogen to helium sequence seen initially \citep[see also][]{Holoien2020_18pg}. \cite{leloudas19} identify possible N III lines also in the TDEs ASASSN-14li and iPTF16axa. Together with iPTF15af \citep{Blagorodnova19} and iPTF16fnl \citep{Onori2019}, they form a N-rich class of TDEs. 

\begin{figure}
\includegraphics[width=\textwidth]{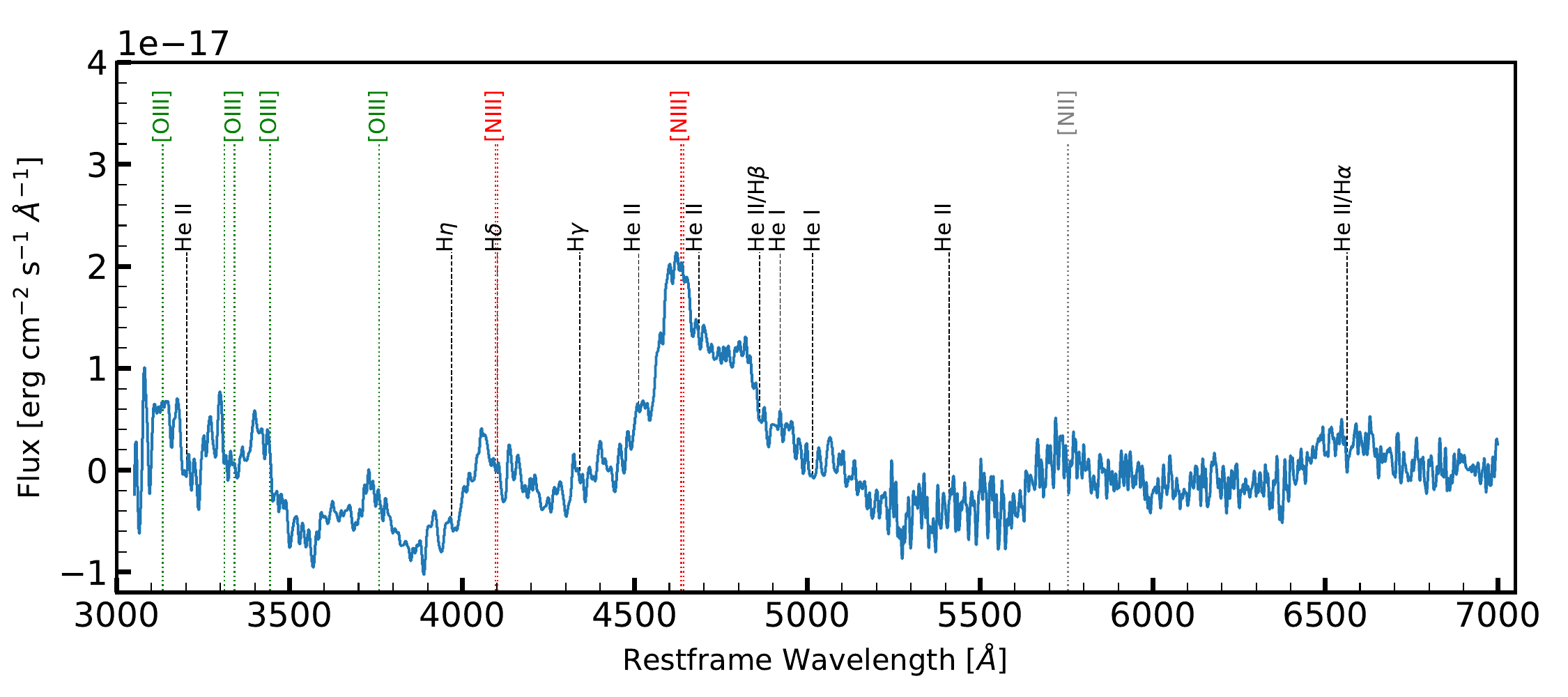}
\caption{
Spectrum of the Bowen TDE iPTF15af. Bowen fluorescence lines can be seen ([N III] lines are marked in red and [O III] lines in green - the centroids of these forbidden lines appear shifted by $−3000$\,km\,s$^{−1}$). This event identifies yet an additional possible layer of \optuv\ TDE spectral diversity. Figure from \cite{Blagorodnova19}.
}
\label{fig:bowen_tde}       
\end{figure}

In addition to N III, \cite{Blagorodnova19} identify several Bowen fluorescence lines in the optical and ultraviolet spectra of iPTF15af (Fig. \ref{fig:bowen_tde}). Bowen fluorescence \citep{Bowen1928,Weyman1969} is a process by which extreme-ultraviolet and X-ray photons ionize He II, which later recombines, emitting Ly-$\alpha$ photons. These then excite certain O III and N III transitions with similar wavelengths, launching a cascade of transitions observed in the optical and ultraviolet regimes. In particular, strong O III $\lambda$3133, N III $\lambda$4640, and He II $\lambda$4686 have been observed in some planetary nebulae and X-ray binaries \citep{Schachter1989,Kastner1996}. Such emission was predicted to occur around active supermassive black holes decades ago \citep{Netzer1985}, but was only recently observed for the first time in an extreme outburst of an AGN \citep{Trakhtenbrot2019_Bowen}. Its detection also in TDEs \citep{Blagorodnova19} may indicate the presence of similar conditions involving extreme-ultraviolet photons hitting high-density and high-optical-depth material, in both TDEs and some AGN outbursts. In Table \ref{tab:prop} we denote both the N-rich TDEs and those with Bowen lines with a ``Bowen'' label (it is not yet clear whether these two spectral classes are distinct). We adopt the criterion of \cite{leloudas19} that use the presence of either N III $\lambda$4640 or $\lambda$4100, or O III $\lambda$3760 to identify an event as ``Bowen''. As first pointed out by \citet{leloudas19}, evidence for these Bowen lines is (so far) detected only in spectra of TDEs that show both He and Balmer emission lines (i.e. TDE\,H+He's). In fact, almost all TDEs typed as ``H+He" show one or more emission lines that might be associated with the Bowen mechanism \citep{vanVelzen20}.

\cite{wever19} identified the Fe II 37,38 multiplet in narrow emission lines in the the late-time spectra of AT\,2018fyk (we add a "Fe" label to this event in Table \ref{tab:prop}) and suggest similar lines might be present in SDSS J0748, PTF09ge, and ASASSN–15oi. These low-ionization lines are also found in narrow-line Seyfert 1 AGN, in some coronal-line emitters, and in flares or TDEs in active galaxies such as PS16dtm \cite[][]{Blanchard17}, AT\,2018bcb \citep{Neustadt20}, and AT\,2018dyk \citep{Frederick19}.

The combination of the TDE spectral features (species, line profiles and intensity) discussed so far are distinct from those of all known transients, including all types of supernovae.

In many TDEs, the widths of the lines evolve with time, starting at $\sim$10$^{4}$\,km\,s$^{-1}$ at early times and becoming narrower on time scales of months.
The narrowing of the lines with time, as the luminosity of the event declines, is the opposite behaviour observed in the case of reverberation mapping in AGN \cite[e.g. as noted   by][]{Holoien16b}. While the width of broad spectral lines is usually associated with the kinematics of the emitting gas, in some cases, it can be strongly affected by electron scattering instead. This can lead to an overestimation of the emitting region bulk velocity, and a mistaken interpretation of its evolution. 

\begin{figure}
\includegraphics[width=\textwidth]{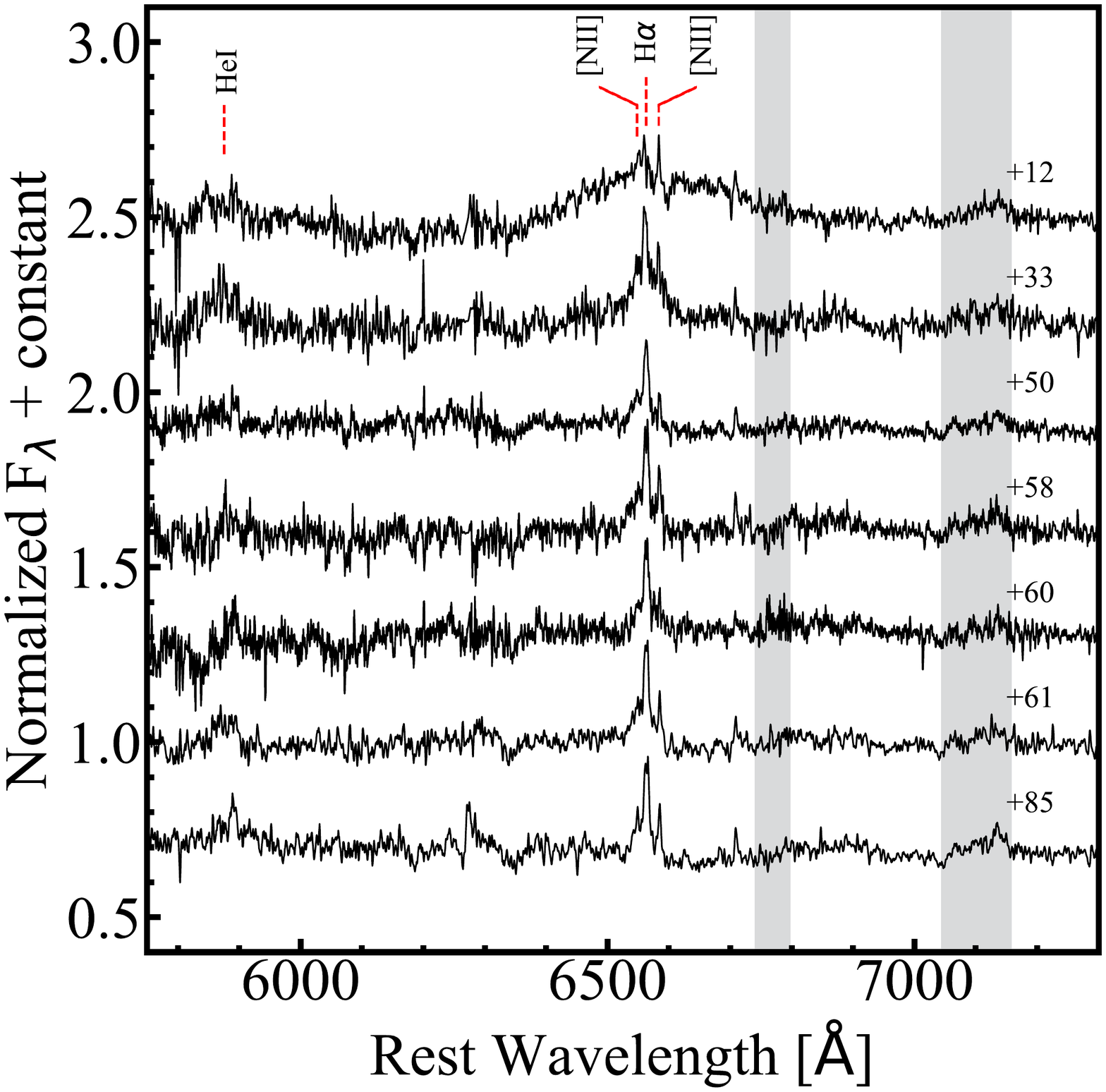}
\caption{A sequence of host-subtracted spectra of iPTF16fnl. The evolution of the H$\alpha$ line profile is characterized by the narrowing of the broad component together with the development of a narrow core at late times. This may be an indication of decreasing optical depth, rather than of bulk velocities of the line-emitting material \citep{Roth18}. The main emission lines are indicated with vertical dashed red lines, while the gray  areas indicate the wavelength ranges affected by telluric lines. Days relative to peak luminosity are shown on the right. Figure adapted from \cite{Onori2019}.}
\label{fig:optical_spec_Ha_xsh}       
\end{figure}

\cite{Roth18} investigate spectral line formation due to electron scattering in an outflowing optically-thick gas. Using radiative transfer calculations they show that, indeed, non-coherent scattering of hot electrons can play a primary role in the broadening of the lines in a high-density emitting region.
In this case, the line profiles are characterized by the presence of a narrow line core superimposed on a broad component at the base of the line. The broad line `wings' are produced by photons scattered off of high-velocity electrons in the gas. The contribution of the wing to the line profiles becomes more prominent with increasing optical depth. In some cases, the broad wings may dominate the line profile completely, with the narrow core becoming visible only later as the optical depth decreases. This is seen very clearly in the case of the H$\alpha$ line of iPTF16fnl \citep[Fig. \ref{fig:optical_spec_Ha_xsh};][]{Onori2019}.
In this scenario, the observed line profile evolution, in which a broad component becomes narrower and a central, narrow core emerges, reflects a decline in the emitting region's density, rather than a drop in its bulk velocity.

Among the current sample of \optuv\ TDEs, four events exhibit broad emission lines with boxy or double-peaked profiles, which have been associated with a line-emitting disk: PTF09ge \citep{Arcavi14}, ASASSN-14li \citep{Liu17}, AT\,2018zr \citep{Holoien18a}, and AT\,2018hyz \citep{Short2020,Hung2020}.
In particular, for AT\,2018zr and AT\,2018hyz, the double-peaked line profile was seen to develop gradually with time, and has been interpreted as the revealing of an accretion disk through the thinning reprocessing material. \cite{Hung2020} deduce a disk-formation time scale of about a month for AT\,2018hyz, implying rapid circularization of the stellar debris after disruption. \cite{Hung2020} further model the H$\alpha$ emission line in AT\,2018hyz as a combination of a double-peaked disk profile and a Gaussian component, which is likely produced in the outflowing gas. The Gaussian component becomes stronger over time and dominates the H$\alpha$ line shape at late times. This competition with a non-disk component, and the requirement of a high inclination angle (close to edge-on) to see a disk profile may explain why double-peaked line profiles are rare in TDEs.

\subsection{Ultraviolet}

\begin{figure}
\includegraphics[width=\textwidth]{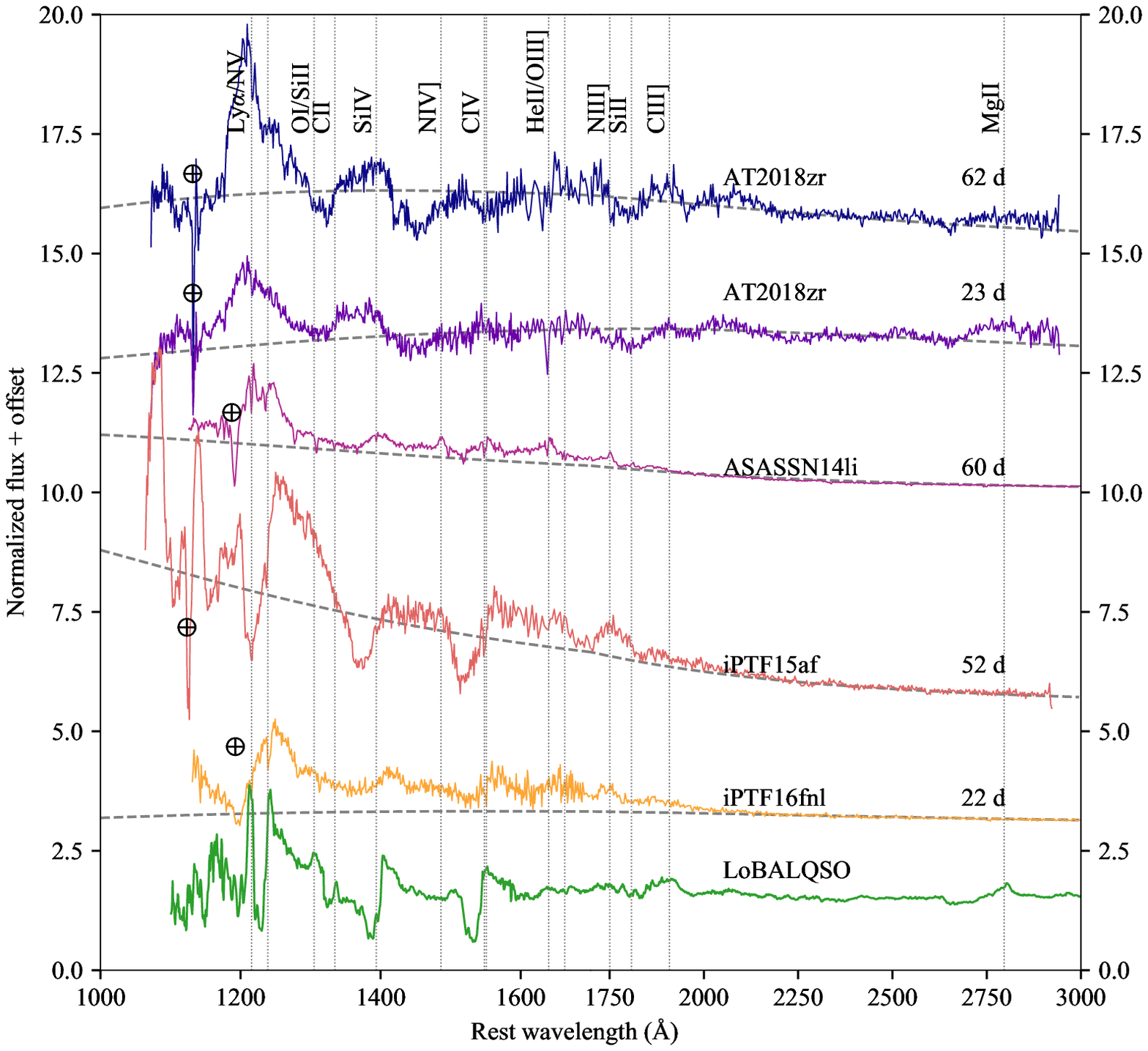}
\caption{Ultraviolet spectra of \optuv\ TDEs \citep{Hung2019}. Grey dashed lines mark the \bb\ spectra with temperatures measured from ultraviolet and optical photometry. Rest-frame days from peak luminosity are marked. TDE data from \cite{Cenko16, Brown18, Blagorodnova19, Hung2019}. A composite spectrum of low-ionization broad absorption line QSO (LoBALQSO; from \citealt{Brotherton2001}) is shown for comparison.}
\label{fig:uv_TDEs}       
\end{figure}

To date, ultraviolet spectroscopy has been obtained for five TDEs -- ASASSN-14li \cite[]{Cenko16}, iPTF16fnl \cite[]{Brown18}, PS16dtm \cite[]{Blanchard17}, iPTF15af \cite[]{Yang17, Blagorodnova19}, and AT\,2018zr \cite[]{Hung2019}, all of which have near-ultraviolet coverage (Fig. \ref{fig:uv_TDEs}). Far-ultraviolet spectra have been obtained for all but PS16dtm. While a single \bb\ spectrum extrapolated from optical and near-ultraviolet photometric observations captures the overall shape of the ultraviolet spectra, the continuum often appears to be more luminous than what the \bb\ predicts at the far-ultraviolet range. 

Ultraviolet spectra have revealed the presence of collisionally excited broad emission lines, such as Ly$\alpha$, C IV, N V and Si IV \citep{Cenko16,Brown18,Blagorodnova19}, similar to those seen in QSOs. On the other hand, the nitrogen-to-carbon ratios in these TDEs are anomalously high compared to typical QSOs, thus connections of TDEs with nitrogen-rich QSOs have been suggested \citep{Cenko16, Kochanek16}. 

All of these TDEs, except PS16dtm, also lack the Mg\,{\sc ii}\,$\lambda\lambda2796, 2803$ emission that is commonly seen in QSOs. 
Mg\,{\sc ii} has an ionization potential of 15 eV. The Mg\,{\sc ii}\,$\lambda\lambda2796, 2803$ emission arises from low ionization regions where the gas is at least partially neutral. If a hard continuum is present, the majority of the Mg atoms may be photoionized to Mg\,{\sc iii} or higher states and thus lead to the weakening or absence of Mg\,{\sc ii} features. The same hard spectrum may also be responsible for the production of Bowen fluorescence emission. However, as noted by \cite{leloudas19}, this hard ionizing source cannot be accounted for entirely by the near ultraviolet and optical \bb\ with a temperature on the order of $10^4$~K, as it provides too few photons at high energies.

Unlike in the optical spectra, {\it absorption} features have been identified in ultraviolet spectra. These are seen as broad ($\sim 10^3-10^4$\,km\,s$^{-1}$) features in iPTF16fnl, iPTF15af, PS16dtm, and AT\,2018zr and as narrow (few$\times$ 100\,km\,s$^{-1}$) ones in ASASSN-14li. The broad absorption features are seen in the high-ionization lines C IV, N V, and Si IV, similar to what is seen in Broad Absorption Line Quasars (BALQSOs), and tend to be blueshifted by $\sim10^3$--$10^4$\,km\,s$^{-1}$. The absorption troughs can either be attached (e.g. iPTF16fnl and iPTF15af) or detached (e.g. AT\,2018zr) from their emission peaks, which might be due to a viewing angle difference. The broad blueshifted absorption lines are interpreted as evidence for fast outflows of material accelerated by radiation. Although the driving mechanism for these outflows in the broad absorption-line TDEs is still unclear, it has been suggested that super-Eddington accretion \citep[e.g.][]{strubbe_quataert09,Strubbe11}, which might be common in TDEs around supermassive black holes with mass $<10^7~M_\odot$, could be capable of driving outflows to such high velocities ($\sim10^4$ km~s$^{-1}$). 

A sequence of ultraviolet spectra was obtained for two TDEs, iPTF16fnl and AT\,2018zr, revealing evolution in the absorption features. In iPTF16fnl, the ultraviolet emission features are found to be redshifted relative to the systematic velocity from the earlier spectra. This apparent redshift has been interpreted as a result of fast outflows \cite[]{Brown18}, where blueshifted broad absorption troughs eclipsed the bluer portion of the emission line profiles. In AT\,2018zr, \cite{Hung2019} found that the high-ionization broad absorption lines only appeared $\sim$2 months after maximum light (Fig. \ref{fig:uv_TDEs}). Expanding the sample of TDEs with multi-epoch ultraviolet spectroscopy will test whether this timescale is linked to the timescale of super-Eddington outflows, and whether outflows are ubiquitous among TDEs.

In AT\,2018zr, absorption features are also seen in the Balmer series and the metastable helium lines He\,{\sc i}\,$^*\lambda3889$ and  He\,{\sc i}\,$^*\lambda5876$ at optical wavelengths, with a FWHM of $\sim$3000\,km\,s$^{-1}$ and displaced by the same velocity ($0.05c$) as the ultraviolet absorption features \citep{Hung2019}. Such optical absorption features are rarely detected even in BALQSOs. Photoionization models indicate both high density ($\log ( n_H / {\rm cm}^{-3})\gtrsim 7$) and high column density ($\log ( N_H /{\rm cm}^{-2}) \gtrsim 23$) requirements must be met in order to produce the observed hydrogen and helium absorption lines \citep{Hamann2019}. Recent work on radiative transfer modelling of TDE outflows suggests that the column densities in TDE winds are as high as $>10^{25}$~cm$^{-2}$, depending on the inclination angle of the disk \citep{Parkinson2020}. This result is in agreement with the lower limit derived from the observed ultraviolet absorption lines.

\section{TDE Selection: Building a Sample of Events}
\label{sec:obsparams} 

We now distill the discussion of the previous sections on the photometric and spectroscopic properties into a set of selection criteria to define a sample of ``secure" \optuv\ TDEs (Table~\ref{tab:prop}). Due to the lack of a consistent theory that explains all \optuv\ TDE emission properties, our selection criteria are primarily motivated by observations. We realize that there is a somewhat circular logic here, identifying new events as \optuv\ TDEs based on those previously claimed to be such. There are likely additional TDEs that do not match these criteria, and therefore, in order to truly map the population of events, we must keep an open mind regarding newly discovered types of flares (yet, we must also be cautious of calling every new type of transient a TDE). Once a large enough sample is available we can revisit the criteria. That being said, after almost one decade of collecting \optuv\ data of transients found in galaxy centers, some clear patterns have emerged. Below we summarize four key properties that we use to compile the events in Table \ref{tab:prop}, which we consider to be a robust sample of \optuv\ TDEs. These criteria have been applied to events claimed to be TDEs in the literature.

\begin{itemize}
    \item {\bf Nuclear.} That is, a small offset ($d$) between the host centroid and the flare. This criterion is motivated by the assumptions that SMBHs in quiescent galaxies reside at the gravitational center of their host galaxy. For the purpose of populating Table~\ref{tab:prop} we use a cut of $d<0.3$", which is sufficient to establish a nuclear detection for the surveys summarised in section~\ref{sec:intro_hist}.  
    
    \item {\bf Spectroscopic.} As discussed in Section~\ref{sec:spectra}, nuclear flares that have blue colors, all have similar spectroscopic properties: broad helium and/or hydrogen on an otherwise predominantly featureless spectrum.
    In Table~\ref{tab:prop} we include also PS1-11af even though it showed no obvious  features in its optical spectrum.
    
    \item {\bf Photometric.} As discussed in Section~\ref{sec:lightcurves}, events that satisfy the above criteria share two key photometric properties: a high \bb\ temperature and no signs of cooling after peak luminosity. For the purpose of populating Table~\ref{tab:prop} we require $T>1.2 \times 10^4$\,K near peak. We note that for almost all TDEs in Table~\ref{tab:prop}, the temperature  measurements include ultraviolet photometry (from {\it Swift} or GALEX) and in most cases multiple ultraviolet observations allow a measurement of temperature evolution (see Table~\ref{tab:lc}). The three TDEs from PTF are an exception, their \bb\ temperatures are measured from the optical spectrum obtained near the peak of the flare (see \citealt{Arcavi14}). 
    
    \item {\bf Quiescent galaxy / not AGN.} This requirement is the most controversial because TDEs can also happen in AGN. Perhaps the TDE rate is even enhanced in AGN \citep{Stone18a}, and they might manifest themselves with distinct observational characteristics \citep{Chan2019,Trakhtenbrot2019_CLAGN}. Unfortunately, the origin of large amplitude AGN variability is poorly understood and such flares can masquerade as TDEs (see the \impostchap). For the purpose of populating Table~\ref{tab:prop} we define an event to be in an AGN if one of the following three properties is observed: (i) there is statistically-significant variability before the main flare, (ii) the signature of a dusty torus is detected using WISE colors \citep{Stern12}, or (iii) the host is classified as a Seyfert galaxy using the BPT diagram \citep{baldwin81}. These criteria select a relatively pure AGN sample. We do not reject events based on observational signs that indicate recent (but not on-going) activity of the supermassive black hole, such as a LINER spectrum \citep{Heckman80} or relic radio lobes.  
\end{itemize}
The above selection criteria yield 33 sources presented in Table \ref{tab:prop}. 

A few key photometric properties of our \optuv\ TDE sample are presented in Table~\ref{tab:lc}. For all sources, these features have been measured using the method presented in \citet{vanVelzen18_NedStark,vanvelzen2020ZTF}. Central to this approach is the use of a model that includes both the \bb\ light curve shape and temperature evolution. As such, the free parameters in this model can be measured simultaneously using all available photometric observations. The \bb\ temperature can be kept constant in the model in order to obtain the mean temperature ($\left< T \right>$).  Alternatively, a linear temperature change with time ($dT/dt$) can also be extracted. For the majority of the TDEs in our sample this approach is sufficient because the measured values of $dT/dt$ are consistent with zero (Table~\ref{tab:prop}). However to allow for more flexibility (e.g., a relatively rapid temperature increase as observed for AT\,2018zr), the light curve properties reported in Table~\ref{tab:lc} are obtained using a linear interpolation of the temperature on a grid with a spacing of 30 days (see \citealt{vanvelzen2020ZTF} for details and Fig.~\ref{fig:ZTF_LTR} for examples of the resulting temperature curves). In Table~\ref{tab:prop} we report both the exponential decay rate measured for the first 100 days of post-peak observations ($L_{\rm bb} \propto e^{-t/\tau}$) and the timescale measured from a power-law decay as applied to the first year of post-peak observations ($L_{\rm bb} \propto (t/t_0)^{p}$). For the power-law decay, we report both the results for a fixed power-law index (at the canonical value $p=-5/3$), as well as the results obtained when this index is left as a free parameter.

\begin{landscape}

\begin{table}[!ht]
\begin{center}
\begin{tabular}{llcclclrc}
\hline
\vspace{-0.2cm} \\
Discovery & IAU & R.A. / Decl. & Peak$^a$  & Redshift & Peak$^b$ & Spectroscopic &Host mass$^d$ & Discovery \\
Name & Name &   & Date &  & $M_g$ & Type (Subtype)$^c$ &  (log\,$M_\odot$) &  Ref.$^e$ \\
\hline
\hline
\vspace{-0.2cm} \\
SDSS-TDE2            &                 & 23:23:48.61\,$-$01:08:10.2  & 2007-06-15$^*$ &   0.2515  & $-21.3$ & H                                  &  $10.57_{0.10}^{0.17}$ & 1\\
PTF-09ge             &                 & 14:57:03.18\,$+$49:36:40.9  & 2009-06-10~    &    0.064  & $-19.8$ & He~(Fe?)                           &  $10.11_{0.12}^{0.13}$ & 2\\
PTF-09axc            &                 & 14:53:13.07\,$+$22:14:32.2  & 2009-07-04~    &   0.1146  & $-19.5$ & H                                  &  $10.12_{0.17}^{0.11}$ & 2\\
PTF-09djl            &                 & 16:33:55.97\,$+$30:14:16.6  & 2009-08-05~    &    0.184  & $-20.3$ & H                                  &  $9.95_{0.12}^{0.15}$ & 2\\
PS1-10jh             &                 & 16:09:28.27\,$+$53:40:23.9  & 2010-07-16~    &   0.1696  & $-19.6$ & He                                 &  $9.61_{0.13}^{0.10}$ & 3\\
PS1-11af             &                 & 09:57:26.81\,$+$03:14:00.9  & 2011-01-17~    &   0.4046  & $-20.2$ & Unknown                            &  $10.21_{0.19}^{0.21}$ & 4\\
ASASSN-14ae          &                 & 11:08:40.11\,$+$34:05:52.2  & 2014-01-27$^*$ &   0.0436  & $-19.6$ & H\,$\rightarrow$\,H+He             &  $10.02_{0.17}^{0.09}$ & 5\\
ASASSN-14li          &                 & 12:48:15.23\,$+$17:46:26.4  & 2014-11-28$^*$ &  0.02058  & $-18.1$ & H+He~(N)                           &  $9.69_{0.10}^{0.05}$ & 6\\
iPTF-15af            &                 & 08:48:28.13\,$+$22:03:33.4  & 2015-02-08~    &  0.07897  & $-18.0$ & H+He~(N+O)                         &  $10.33_{0.18}^{0.11}$ & 7\\
ASASSN-15oi          &                 & 20:39:09.14\,$-$30:45:20.6  & 2015-08-13$^*$ &   0.0484  & $-21.9$ & He~(Fe?+O?)                           &  $10.02_{0.04}^{0.04}$ & 8\\
OGLE16aaa            &                 & 01:07:20.76\,$-$64:16:20.4  & 2016-01-27~    &   0.1655  & $-20.7$ & H+He                              &  $10.47_{0.11}^{0.09}$ & 9\\
iPTF-16axa           &                 & 17:03:34.34\,$+$30:35:36.6  & 2016-05-15$^*$ &    0.108  & $-19.0$ & H+He~(N+O?)                           &  $10.18_{0.14}^{0.10}$ & 10\\
iPTF-16fnl           &                 & 00:29:57.05\,$+$32:53:37.2  & 2016-08-26~    &   0.0163  & $-17.5$ & H+He~(N)                           &  $9.35_{0.15}^{0.12}$ & 11\\
PS17dhz              & AT\,2017eqx     & 22:26:48.37\,$+$17:08:52.3  & 2017-06-06~    &   0.1089  & $-18.8$ & H+He\,$\rightarrow$\,He~(N+Fe?)    &  $9.44_{0.13}^{0.11}$ & 12\\
PS18kh               & AT\,2018zr      & 07:56:54.54\,$+$34:15:43.6  & 2018-03-03~    &    0.071  & $-19.9$ & H                                  &  $10.03_{0.18}^{0.09}$ & 13\\
ZTF18aahqkbt         & AT\,2018bsi     & 08:15:26.63\,$+$45:35:32.0  & 2018-04-09$^*$ &    0.051  & $-19.0$ & H+He~(N)                           &  $10.61_{0.06}^{0.05}$ & 14\\
ASASSN-18ul          & AT\,2018fyk     & 22:50:16.09\,$-$44:51:53.5  & 2018-07-18$^*$ &    0.059  & $-19.6$ & H+He~(O+Fe)                        &  $10.57_{0.21}^{0.12}$ & 15\\
ASASSN-18pg          & AT\,2018dyb     & 16:10:58.87\,$-$60:55:24.3  & 2018-08-16~    &    0.018  & $-19.8$ & H+He~(N+O)                         &  $9.83_{0.15}^{0.08}$ & 16\\
ATLAS18way           & AT\,2018hco     & 01:07:33.63\,$+$23:28:34.2  & 2018-10-10~    &    0.088  & $-19.4$ & H                                  &  $9.90_{0.18}^{0.09}$ & 14\\
ASASSN-18zj          & AT\,2018hyz     & 10:06:50.88\,$+$01:41:33.9  & 2018-11-05$^*$ &  0.04573  & $-20.3$ & H                                  &  $9.75_{0.26}^{0.12}$ & 14\\
ATLAS18yzs           & AT\,2018iih     & 17:28:03.92\,$+$30:41:31.5  & 2018-11-20~    &    0.212  & $-21.6$ & He                                 &  $10.76_{0.15}^{0.09}$ & 14\\
ZTF18actaqdw         & AT\,2018lni     & 04:09:37.65\,$+$73:53:41.6  & 2018-12-08~    &    0.138  & $-19.4$ & H+He~(N+O)                         &  $9.94_{0.15}^{0.10}$ & 14\\
ZTF19aabbnzo         & AT\,2018lna     & 07:03:18.64\,$+$23:01:44.6  & 2019-01-25~    &    0.091  & $-19.5$ & H+He~(N?)                               &  $9.47_{0.09}^{0.12}$ & 14\\
ZTF19aakswrb         & AT\,2019bhf     & 15:09:15.97\,$+$16:14:22.5  & 2019-02-25~    &   0.1206  & $-20.1$ & H                                  &  $10.21_{0.12}^{0.15}$ & 14\\
ASASSN-19bt          & AT\,2019ahk     & 07:00:11.39\,$-$66:02:24.7  & 2019-03-06~    &   0.0262  & $-20.7$ & H                                  &  $9.74_{0.10}^{0.09}$ & 17\\
ZTF19aakiwze         & AT\,2019cho     & 12:55:09.20\,$+$49:31:09.9  & 2019-02-21~    &    0.193  & $-20.7$ & H+He~(N)                           &  $10.10_{0.16}^{0.17}$ & 14\\
ASASSN-19dj          & AT\,2019azh     & 08:13:16.95\,$+$22:38:53.8  & 2019-03-16~    &   0.0222  & $-19.9$ & H+He                               &  $9.86_{0.14}^{0.15}$ & 18\\
ZTF19aapreis         & AT\,2019dsg     & 20:57:02.97\,$+$14:12:16.2  & 2019-04-27~    &   0.0512  & $-19.2$ & H+He~(N+O)                         &  $10.36_{0.12}^{0.17}$ & 14\\
Gaia19bpt            & AT\,2019ehz     & 14:09:41.90\,$+$55:29:27.7  & 2019-05-08~    &    0.074  & $-19.6$ & H                                  &  $9.67_{0.15}^{0.09}$ & 14\\
ZTF19abidbya         & AT\,2019lwu     & 23:11:12.30\,$-$01:00:10.8  & 2019-07-26~    &    0.117  & $-19.8$ & H                                  &  $9.87_{0.15}^{0.12}$ & 14\\
ZTF19abhhjcc         & AT\,2019meg     & 18:45:16.19\,$+$44:26:18.9  & 2019-08-01~    &    0.152  & $-20.2$ & H                                  &  $9.69_{0.05}^{0.05}$ & 14\\
ATLAS19qqu           & AT\,2019mha     & 16:16:27.78\,$+$56:25:56.3  & 2019-08-09~    &    0.148  & $-19.9$ & H                                  &  $10.05_{0.15}^{0.11}$ & 14\\
ZTF19abzrhgq         & AT\,2019qiz     & 04:46:37.88\,$-$10:13:34.8  & 2019-10-05~    &   0.0151  & $-18.6$ & H+He~(N)                           &  $10.03_{0.14}^{0.10}$ & 14\\

\hline
\end{tabular}
\end{center}
\caption{Discovery information of optically-selected TDEs with both spectroscopic and photometric observations.\label{tab:prop}}
\vspace{0.2cm} 
\footnotesize{$^a$The peak date is obtained using a light curve model to interpolate between gaps in the data (see \citealt{vanvelzen2020ZTF}). The sources that are only detected post-peak are indicated with an asterisk - for these we provide the date of the first detection}. \\ 
 \footnotesize{$^b$The maximum observed optical luminosity, k-corrected to the $g$-band in the rest-frame of the TDE using the best-fit blackbody temperature to make the spectral correction.}\\
 \footnotesize{$^c$Arrows denote a change that occurred in the spectral type with time}.\\
 \footnotesize{$^d$The host stellar mass as estimated from archival optical photometry, obtained before the TDE occurred (see \citealt{vanvelzen2020ZTF})}.\\
  \footnotesize{$^e$The discovery paper for this source (first journal article to present a classification and observed properties):
  1: \cite{vanVelzen10}, 2: \cite{Arcavi14}, 3: \cite{Gezari12}, 4: \cite{Chornock14}, 5: \cite{Holoien14}, 6: \cite{Miller15}, 7: \cite{Blagorodnova19}, 8: \cite{Holoien16b}, 9: \cite{Wyrzykowski17}, 10: \cite{Hung17}, 11: \cite{Blagorodnova17}, 12:\citep{Nicholl2019}, 13: \cite{Holoien18a}, 14: \cite{vanvelzen2020ZTF}, 15: \cite{wever19}, 16: \cite{leloudas19}, 17: \cite{Holoien19_bt}, 18: \cite{Liu2019}}\\
\end{table}

\end{landscape}
\clearpage

\clearpage
\begin{landscape}

\begin{table}

\begin{tabular}{l l c c c c c c c c c}
\hline
\vspace{-0.2cm} \\
Survey                  & IAU & $L_{\rm bb}$        & $\left<T\right>_{\rm 100d}$               & $dT/dt$ & $t_{\rm peak/max}$ & $\sigma$          & $\tau$            & $p$               & $t_0$           & $t_0|_{p=-5/3}$   \\
name & name& $\log$ erg/s & $\log$ K     & $10^{2}$ K/day & MJD            & $\log$ day   & $\log$ day   &                   & $\log$ day   & $\log$ day     \\
 \hline
 \hline 
 \vspace{-0.2cm} \\
SDSS-TDE2            &                 &  $44.54_{0.06}^{0.08}$ &  $4.35_{0.03}^{0.03}$ &  $0.39_{0.25}^{0.22}$ &  $54267.0             $ & --            &  $2.08_{0.08}^{0.09}$ &  $-0.7_{0.2}^{0.2}$ &  $1.51_{0.32}^{0.35}$ &  $2.32_{0.13}^{0.12}$\\
PTF-09ge             &                 &  $44.04_{0.01}^{0.01}$ &  $4.31_{0.00}^{0.00}$ &  --  &  $54992.9_{1.0}^{1.0}$ &  $1.3_{0.02}^{0.02}$ &  $1.78_{0.02}^{0.02}$ &  $-1.7_{0.2}^{0.2}$ &  $1.85_{0.10}^{0.08}$ &  $1.74_{0.03}^{0.04}$\\
PTF-09axc            &                 &  $43.46_{0.02}^{0.03}$ &  $4.08_{0.00}^{0.00}$ &  --  &  $55016.4_{6.0}^{9.3}$ &  $0.9_{0.37}^{0.23}$ &  $2.45_{0.43}^{0.39}$ &  $-1.5_{0.3}^{0.3}$ &  $2.21_{0.17}^{0.14}$ &  $2.22_{0.07}^{0.08}$\\
PTF-09djl            &                 &  $44.42_{0.04}^{0.04}$ &  $4.41_{0.00}^{0.00}$ &  --  &  $55048.3_{4.6}^{5.6}$ &  $1.1_{0.14}^{0.07}$ &  $2.12_{0.33}^{0.52}$ &  $-1.7_{0.3}^{0.2}$ &  $1.67_{0.14}^{0.11}$ &  $1.63_{0.05}^{0.06}$\\
PS1-10jh             &                 &  $44.47_{0.07}^{0.07}$ &  $4.49_{0.03}^{0.03}$ &  $0.43_{0.29}^{0.43}$ &  $55393.8_{1.4}^{1.4}$ &  $1.3_{0.02}^{0.02}$ &  $1.56_{0.03}^{0.03}$ &  $-1.5_{0.6}^{0.2}$ &  $1.44_{0.22}^{0.41}$ &  $1.55_{0.07}^{0.07}$\\
PS1-11af             &                 &  $44.16_{0.03}^{0.03}$ &  $4.31_{0.01}^{0.01}$ &  $-0.80_{0.28}^{0.39}$ &  $55578.9_{1.2}^{1.2}$ &  $0.9_{0.04}^{0.04}$ &  $1.69_{0.02}^{0.02}$ &  $-2.1_{0.5}^{0.5}$ &  $1.73_{0.17}^{0.16}$ &  $1.54_{0.09}^{0.11}$\\
ASASSN-14ae          &                 &  $43.87_{0.01}^{0.01}$ &  $4.27_{0.01}^{0.01}$ &  $0.24_{0.13}^{0.14}$ &  $56684.9             $ & --            &  $1.45_{0.01}^{0.01}$ &  $-2.4_{0.2}^{0.2}$ &  $1.52_{0.07}^{0.05}$ &  $1.21_{0.02}^{0.03}$\\
ASASSN-14li          &                 &  $43.66_{0.02}^{0.02}$ &  $4.51_{0.01}^{0.01}$ &  $-0.21_{0.07}^{0.07}$ &  $56989.4             $ & --            &  $1.77_{0.01}^{0.01}$ &  $-1.4_{0.1}^{0.1}$ &  $1.69_{0.09}^{0.09}$ &  $1.80_{0.04}^{0.04}$\\
iPTF-15af            &                 &  $44.10_{0.08}^{0.10}$ &  $4.70_{0.03}^{0.04}$ &  $1.42_{0.80}^{0.43}$ &  $57061.6_{1.8}^{1.6}$ &  $1.5_{0.03}^{0.02}$ &  $1.91_{0.02}^{0.03}$ &  $-1.6_{0.6}^{0.6}$ &  $1.91_{0.22}^{0.22}$ &  $1.91_{0.16}^{0.15}$\\
ASASSN-15oi          &                 &  $44.45_{0.01}^{0.01}$ &  $4.14_{0.01}^{0.01}$ &  $1.95_{0.09}^{0.04}$ &  $57247.9             $ & --            &  $1.39_{0.01}^{0.01}$ &  $-2.2_{0.2}^{0.2}$ &  $1.46_{0.09}^{0.07}$ &  $1.34_{0.06}^{0.06}$\\
OGLE16aaa            &                 &  $44.22_{0.01}^{0.01}$ &  $4.23_{0.00}^{0.00}$ &  $-0.11_{0.06}^{0.07}$ &  $57414.3_{1.7}^{1.5}$ &  $1.3_{0.03}^{0.03}$ &  $2.01_{0.05}^{0.06}$ &  $-2.2_{0.4}^{0.3}$ &  $2.19_{0.09}^{0.09}$ &  $2.03_{0.03}^{0.03}$\\
iPTF-16axa           &                 &  $43.82_{0.02}^{0.03}$ &  $4.37_{0.01}^{0.01}$ &  $-0.52_{0.24}^{0.29}$ &  $57523.4             $ & --            &  $1.73_{0.01}^{0.01}$ &  $-1.6_{0.2}^{0.3}$ &  $1.54_{0.16}^{0.13}$ &  $1.56_{0.06}^{0.08}$\\
iPTF-16fnl           &                 &  $43.18_{0.02}^{0.03}$ &  $4.40_{0.01}^{0.01}$ &  $-0.14_{0.13}^{0.16}$ &  $57626.5_{11.6}^{3.6}$ &  $1.3_{0.18}^{0.12}$ &  $1.36_{0.01}^{0.01}$ &  $-2.1_{0.2}^{0.2}$ &  $1.35_{0.09}^{0.08}$ &  $1.18_{0.05}^{0.05}$\\
PS17dhz              & AT\,2017eqx     &  $43.82_{0.05}^{0.03}$ &  $4.31_{0.01}^{0.01}$ &  $-0.39_{0.23}^{0.25}$ &  $57910.4_{7.1}^{3.9}$ &  $1.2_{0.20}^{0.19}$ &  $1.59_{0.02}^{0.02}$ &  $-2.0_{0.3}^{0.3}$ &  $1.47_{0.13}^{0.10}$ &  $1.31_{0.06}^{0.09}$\\
PS18kh               & AT\,2018zr      &  $43.78_{0.02}^{0.02}$ &  $4.14_{0.01}^{0.01}$ &  $0.46_{0.05}^{0.05}$ &  $58180.7_{1.0}^{1.1}$ &  $1.0_{0.03}^{0.04}$ &  $1.88_{0.03}^{0.03}$ &  $-0.8_{0.1}^{0.0}$ &  $1.23_{0.12}^{0.15}$ &  $2.14_{0.03}^{0.03}$\\
ZTF18aahqkbt         & AT\,2018bsi     &  $43.87_{0.08}^{0.08}$ &  $4.37_{0.03}^{0.03}$ &  $0.21_{0.30}^{0.63}$ &  $58217.2             $ & --            &  $1.94_{0.10}^{0.12}$ &  $-1.9_{0.6}^{0.4}$ &  $1.92_{0.21}^{0.25}$ &  $1.83_{0.07}^{0.09}$\\
ASASSN-18ul          & AT\,2018fyk     &  $44.48_{0.03}^{0.04}$ &  $4.56_{0.02}^{0.02}$ &  $-0.02_{0.12}^{0.12}$ &  $58317.5             $ & --            &  $1.99_{0.02}^{0.02}$ &  $-1.9_{0.1}^{0.1}$ &  $2.14_{0.06}^{0.06}$ &  $1.76_{0.02}^{0.03}$\\
ASASSN-18pg          & AT\,2018dyb     &  $44.16_{0.01}^{0.01}$ &  $4.37_{0.01}^{0.01}$ &  $-0.25_{0.03}^{0.03}$ &  $58346.7_{1.3}^{1.4}$ &  $1.5_{0.03}^{0.01}$ &  $1.59_{0.01}^{0.01}$ &  $-1.9_{0.1}^{0.1}$ &  $1.49_{0.05}^{0.05}$ &  $1.29_{0.03}^{0.03}$\\
ATLAS18way           & AT\,2018hco     &  $44.25_{0.04}^{0.04}$ &  $4.39_{0.01}^{0.01}$ &  $0.05_{0.09}^{0.10}$ &  $58401.8_{1.9}^{1.7}$ &  $0.9_{0.04}^{0.04}$ &  $2.03_{0.04}^{0.04}$ &  $-1.2_{0.2}^{0.2}$ &  $1.73_{0.15}^{0.16}$ &  $2.00_{0.07}^{0.06}$\\
ASASSN-18zj          & AT\,2018hyz     &  $44.11_{0.01}^{0.01}$ &  $4.25_{0.01}^{0.01}$ &  $0.18_{0.05}^{0.05}$ &  $58428.0             $ & --            &  $1.72_{0.01}^{0.01}$ &  $-1.1_{0.1}^{0.1}$ &  $1.29_{0.06}^{0.07}$ &  $1.71_{0.01}^{0.01}$\\
ATLAS18yzs           & AT\,2018iih     &  $44.62_{0.03}^{0.04}$ &  $4.23_{0.01}^{0.01}$ &  $0.19_{0.05}^{0.05}$ &  $58442.2_{0.8}^{1.6}$ &  $1.2_{0.01}^{0.01}$ &  $2.05_{0.04}^{0.04}$ &  $-0.9_{0.1}^{0.1}$ &  $1.61_{0.07}^{0.11}$ &  $2.06_{0.05}^{0.06}$\\
ZTF18actaqdw         & AT\,2018lni     &  $44.21_{0.17}^{0.29}$ &  $4.44_{0.07}^{0.09}$ &  $0.48_{0.47}^{0.46}$ &  $58460.3_{5.8}^{3.6}$ &  $0.8_{0.50}^{0.18}$ &  $2.16_{0.12}^{0.16}$ &  $-1.4_{1.6}^{0.8}$ &  $2.46_{0.45}^{0.37}$ &  $2.23_{0.19}^{0.22}$\\
ZTF19aabbnzo         & AT\,2018lna     &  $44.56_{0.06}^{0.06}$ &  $4.60_{0.02}^{0.03}$ &  $0.93_{0.92}^{0.72}$ &  $58508.3_{2.0}^{2.2}$ &  $1.2_{0.07}^{0.06}$ &  $1.65_{0.03}^{0.03}$ &  $-2.1_{0.7}^{0.7}$ &  $1.79_{0.21}^{0.24}$ &  $1.66_{0.15}^{0.16}$\\
ZTF19aakswrb         & AT\,2019bhf     &  $43.91_{0.05}^{0.04}$ &  $4.20_{0.02}^{0.02}$ &  $0.74_{0.14}^{0.15}$ &  $58539.3_{5.7}^{4.4}$ &  $0.6_{0.35}^{0.23}$ &  $1.70_{0.03}^{0.04}$ &  $-1.3_{0.4}^{0.4}$ &  $1.70_{0.20}^{0.23}$ &  $1.96_{0.09}^{0.07}$\\
ASASSN-19bt          & AT\,2019ahk     &  $44.08_{0.01}^{0.01}$ &  $4.21_{0.01}^{0.01}$ &  $0.10_{0.03}^{0.03}$ &  $58548.3_{0.9}^{0.8}$ &  $1.3_{0.01}^{0.01}$ &  $1.67_{0.01}^{0.01}$ &  $-1.6_{0.1}^{0.1}$ &  $1.66_{0.05}^{0.05}$ &  $1.72_{0.01}^{0.01}$\\
ZTF19aakiwze         & AT\,2019cho     &  $43.98_{0.01}^{0.01}$ &  $4.19_{0.01}^{0.01}$ &  $0.84_{0.23}^{0.21}$ &  $58535.3_{1.6}^{3.0}$ &  $0.8_{0.11}^{0.10}$ &  $1.94_{0.03}^{0.03}$ &  $-1.0_{0.5}^{0.4}$ &  $1.83_{0.13}^{0.25}$ &  $2.19_{0.16}^{0.13}$\\
ASASSN-19dj          & AT\,2019azh     &  $44.50_{0.02}^{0.02}$ &  $4.53_{0.01}^{0.01}$ &  $0.98_{0.19}^{0.21}$ &  $58558.6_{1.6}^{1.5}$ &  $1.3_{0.04}^{0.05}$ &  $1.83_{0.01}^{0.01}$ &  $-1.8_{0.1}^{0.2}$ &  $1.87_{0.08}^{0.08}$ &  $1.96_{0.04}^{0.05}$\\
ZTF19aapreis         & AT\,2019dsg     &  $44.26_{0.05}^{0.04}$ &  $4.49_{0.01}^{0.01}$ &  $-0.07_{0.12}^{0.14}$ &  $58600.2_{10.0}^{8.2}$ &  $1.2_{0.12}^{0.06}$ &  $1.80_{0.02}^{0.02}$ &  $-1.5_{0.1}^{0.1}$ &  $1.51_{0.10}^{0.11}$ &  $1.59_{0.06}^{0.06}$\\
Gaia19bpt            & AT\,2019ehz     &  $44.03_{0.02}^{0.01}$ &  $4.33_{0.01}^{0.01}$ &  $-0.40_{0.03}^{0.03}$ &  $58611.4_{0.6}^{0.6}$ &  $0.9_{0.02}^{0.03}$ &  $1.67_{0.01}^{0.01}$ &  $-1.8_{0.1}^{0.1}$ &  $1.73_{0.09}^{0.06}$ &  $1.61_{0.04}^{0.04}$\\
ZTF19abidbya         & AT\,2019lwu     &  $43.60_{0.04}^{0.03}$ &  $4.09_{0.01}^{0.01}$ &  $0.64_{0.19}^{0.14}$ &  $58690.2_{0.7}^{1.0}$ &  $0.4_{0.21}^{0.17}$ &  $1.68_{0.05}^{0.06}$ &  $-1.5_{0.5}^{0.4}$ &  $1.56_{0.18}^{0.20}$ &  $1.67_{0.08}^{0.09}$\\
ZTF19abhhjcc         & AT\,2019meg     &  $44.36_{0.04}^{0.03}$ &  $4.44_{0.01}^{0.01}$ &  $1.94_{0.09}^{0.05}$ &  $58696.4_{0.6}^{0.6}$ &  $0.9_{0.03}^{0.02}$ &  $1.72_{0.02}^{0.03}$ &  $-0.2_{0.5}^{0.1}$ &  $2.37_{0.66}^{0.44}$ &  $2.56_{0.18}^{0.15}$\\
ATLAS19qqu           & AT\,2019mha     &  $44.05_{0.05}^{0.06}$ &  $4.32_{0.03}^{0.03}$ &  $0.36_{1.07}^{0.88}$ &  $58704.0_{0.8}^{0.7}$ &  $1.1_{0.02}^{0.02}$ &  $1.23_{0.02}^{0.03}$ &  $-4.0_{0.5}^{0.7}$ &  $1.63_{0.12}^{0.11}$ &  $1.21_{0.09}^{0.10}$\\
ZTF19abzrhgq         & AT\,2019qiz     &  $43.44_{0.01}^{0.01}$ &  $4.28_{0.01}^{0.01}$ &  $-0.17_{0.09}^{0.10}$ &  $58761.4_{0.4}^{0.4}$ &  $0.8_{0.01}^{0.01}$ &  $1.48_{0.01}^{0.01}$ &  $-2.0_{0.1}^{0.1}$ &  $1.40_{0.06}^{0.06}$ &  $1.26_{0.03}^{0.02}$\\

\hline
\end{tabular}
\caption{Light curve properties of the TDEs listed in Table~\ref{tab:prop}.}\label{tab:lc}
\vspace{0.2cm} 
\footnotesize{Notes --- Light curve features are derived using the method presented in \citet{vanvelzen2020ZTF}. The column $L_{\rm bb}$ lists the peak \bb\ luminosity and $\left<T\right>_{\rm 100d}$ is the mean \bb\ temperature measured during the first 100~days post maximum light. A measurement of a linear temperature change during the first year of observations is listed as $dT/dt$ (only estimated for sources with sufficient light curve coverage). $t_{\rm peak/max}$ lists the best-fit time of peak, or, when no pre-peak detections are available, the time of the first detection. The parameters $\sigma$ and $\tau$ list the result of fitting a Gaussian rise and exponential decay, respectively, to the light curve, while $p$ and $t_0$ are the free parameters of a power-law decay ($L_{\rm bb}\propto (t/t_0)^p$). The last column lists the decay timescale obtained when the power-law index is fixed at $p=-5/3$. }
\end{table}
  
\end{landscape}
\clearpage

\section{Rate and Luminosity Function}
\label{sec:rates}

Measurements of the TDE rate have the potential to reveal how stars are fed to the loss cone (see the \ratechap\ in this book), which in turn can teach us about stellar dynamics in galaxy centers. This measurement, however, is not trivial, because it requires both a clear set of criteria for identifying TDEs and an understanding of the selection efficiency and biases of each search. 

The first comprehensive estimate of disruption rates was based on the TDE search in SDSS imaging data \citep{vanVelzen10} and is presented by \cite{vanVelzen14}. Since the SDSS search was based solely on photometry, the selection function was relatively straightforward. The SDSS-selected TDEs were not detected pre-peak. However, by using a light curve model based on the well-sampled TDE PS1-10jh \citep{Gezari12} this limitation can be addressed. The remaining systematic uncertainty due to the uncertainty of the TDE light curve was found to be a factor $\approx 2$. 

The approach of \citet{vanVelzen14} is to compute the detection efficiency of the two TDEs separately, and report the mean rate $\dot{N}$ based on the mean efficiency $\left<\epsilon\right>$:
\begin{equation}\label{eq:rate}
    \dot{N} = \frac{2}{\left<\epsilon\right> N_{\rm gal} \tau } 
\end{equation}
Here $N_{\rm gal}$ is the number of galaxies that were observed in the survey duration $\tau$. This approach yields a rate of $(1.5-2)_{-1.3}^{+2.7}\times 10^{-5}$\,galaxy$^{-1}$\,year$^{-1}$ (the sub/super scripts denote the statistical uncertainty while the range in parentheses approximates the systematic uncertainty due to the assumption of a light curve model). 

Based on two TDEs detected by the ASAS-SN survey (ASASSN-14ae and ASASSN-14l), \citet{Holoien16a} reported a rate of $(2.2-17.0) \times 10^{-5}$\,galaxy$^{-1}$\,year$^{-1}$ (here the range in parentheses denotes the 90\% confidence level statistical uncertainty). The authors used a similar approach to Eq.~\ref{eq:rate} and computed the efficiency under the assumption that the TDE absolute $V$-band peak magnitudes are uniformly distributed between $-19$ and $-20$. This assumption might be why \cite{Holoien16a} obtained a slightly higher rate compared to \cite{vanVelzen14}, who included SDSS-TDE2, an event that is brighter at peak ($M_g=-20.8$; Table~\ref{tab:prop}) than the range assumed by \cite{Holoien16a}, although we note that the two estimates are consistent within their Poisson uncertainty.

Samples based on luminous events produce a lower inferred rate according to Eq.~\ref{eq:rate}, since more luminous events have a higher efficiency $\epsilon$ (they can be detected to larger distance). Even when combining more luminous and less luminous events, the more luminous ones dominate the mean efficiency $\left<\epsilon\right>$, thus Eq.~\ref{eq:rate}, in general, always yields a rate that is weighted more strongly toward more luminous events. To solve this issue we have to consider the TDE luminosity function. 

The fact that the small flux-limited samples of TDEs span a wide range of luminosities (for e.g. the two TDEs found in the SDSS search had a luminosity difference of almost an order of magnitude), informs us that the luminosity function can not be flat: faint events must be more common than luminous ones. This effect is quantified in \citet{vanVelzen18}, who combine events from different surveys under the assumption that each survey samples the same TDE population. This approach yields a measure of the {\it shape} of the luminosity function: $dN /d \log L_g \propto L_g^{-1.3 \pm 0.3}$ (with $L_g$ the rest-frame g-band peak luminosity). Since the faint events dominate, \citet{vanVelzen18} find that extrapolating the observed absolute rate from SDSS to the faintest observed TDEs increases the inferred mean rate to about $1\times  10^{-4}$\,galaxy$^{-1}$\,year$^{-1}$. \citet{Hung18} find a similar result of $1.7_{-1.3}^{+2.7}\times 10^{-4}$\,galaxy$^{-1}$\,year$^{-1}$ by applying the \citet{vanVelzen18} luminosity function to compute the rate from the three TDEs discovered by the iPTF survey.

Since the $g$-band approaches the Rayleigh-Jeans tail of the SED, the steep $L_g$ luminosity function does not translate directly into a steep bolometric luminosity function. For example, \citet{vanvelzen2020ZTF} find that the TDE\,H+He and TDE\,H have a similar peak bolometric luminosity and occur at a similar rate in flux-limited samples, but since TDE\,H+He have a higher mean temperature, their optical luminosity is lower and their intrinsic rate is higher. Therefore, the temperature or radius of the \optuv\ photosphere plays an important role for determining the shape of the optical luminosity function. 

\begin{figure}
\includegraphics[trim={0 0 0 0}, clip, width=\textwidth]{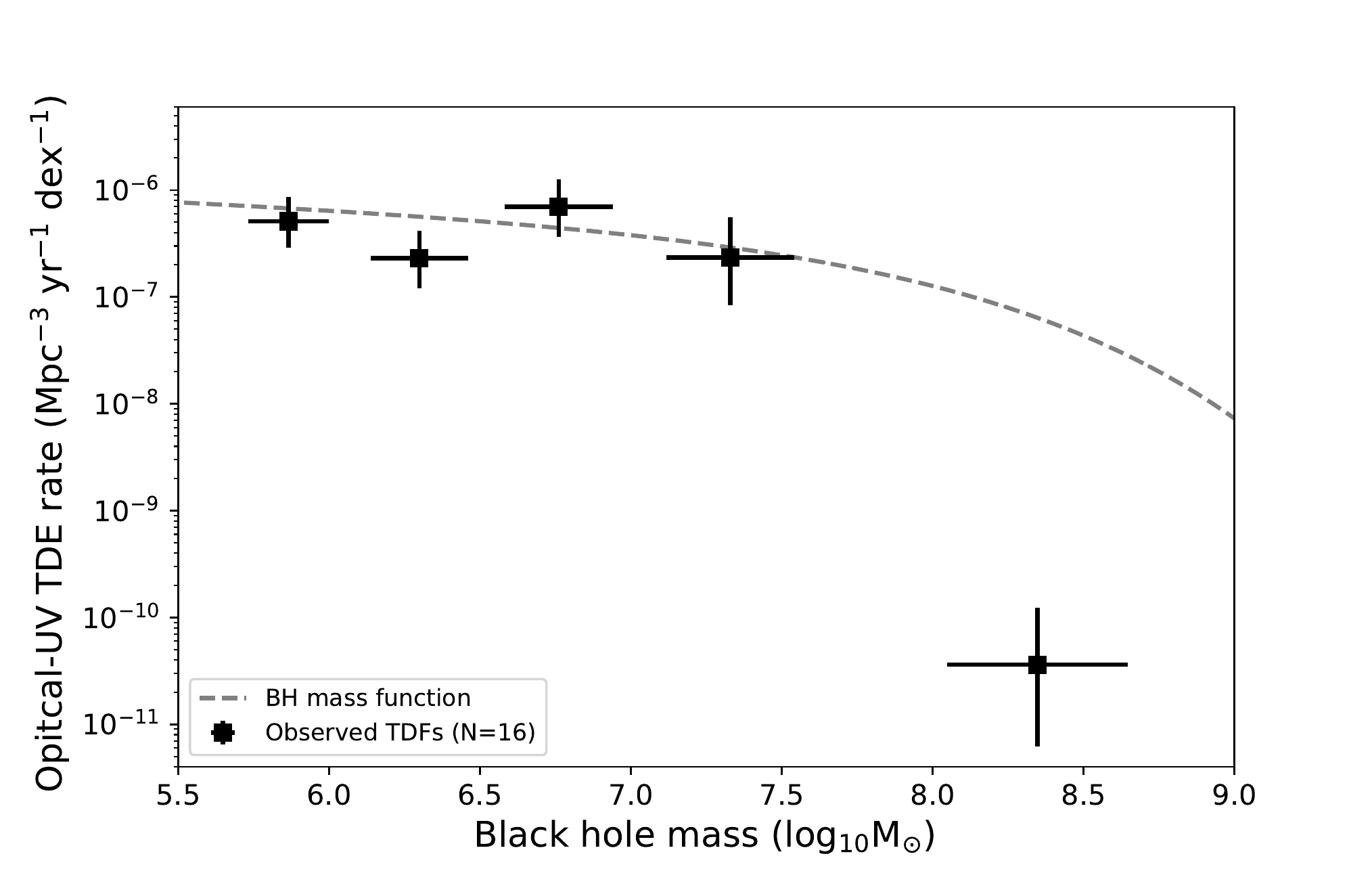}
\caption{The TDE black-hole mass function (updated version of Figure~3 from \citealt{vanVelzen18}). The dashed lines show the black-hole mass function \citep{Shankar04} multiplied by a constant factor of $6\times 10^{-5}$ yr$^{-1}$. The event in the highest mass bin is ASASSN-15lh (the origin of this transient, TDE or supernova, is controversial). Below the mass scale where stars can be swallowed by the black hole prior to disruption \citep{Hills75,Kesden12b}, the \optuv\ TDE mass function appears to trace the black-hole mass function.}
\label{fig:Mbhrate}       
\end{figure}

The work by \citet{vanVelzen18} also enables the first measurement of the TDE black-hole mass function. Figure~\ref{fig:Mbhrate} show the TDE rate as a function of black-hole mass, as estimated based on the $M$--$\sigma$ relation with velocity dispersion measurement from \cite{Wevers17,wever19}. The lack of TDEs from high mass black holes is consistent with the suppression of the TDE rate by the ``direct capture" of stars due to the black-hole event horizon \citep{Hills75}.  

\section{Concluding Remarks}
\label{sec:conclusions}

While the set of known \optuv\ TDEs is rapidly growing, many questions remain open regarding their nature, emission properties, and true diversity. Most prominent is the question of their emission source---is the emission coming entirely from outer shocks, entirely from reprocessed accretion emission, or some combination of both? 
Answering this question could help explain the observed emission properties and perhaps their expected diversity, thus helping to identify more events correctly. 

Observationally, the origin in the diversity of TDE spectral features, such as the species seen, the line profiles measured and their evolution, and the differences between the optical and ultraviolet spectra, will likely continue to be topics of study. In addition, the scale of the photometric diversity is tied to the question of how many rapidly declining events are being missed, affecting our ability to accurately estimate TDE rates and learn about stellar dynamics in galaxy centers. 

As more events are found, we will be able to continue to explore the classes of transients occurring in galaxy centers---increasing our sensitivity to new types of TDEs, mapping the diversity of the known types, and learning to distinguish between them and galaxy-central events that are not related to TDEs. Hopefully, this will help inform theoretical models on which disruption, accretion and emission mechanisms are realized in nature. In turn, such models could help find more events, and focus observational resources on the most critical phases and wavelengths. 

One of the main motivations for studying TDEs has been to reveal the population of otherwise quiescent SMBHs. Some of the questions above are still hindering the realization of the full potential of TDEs on that front. However, in the mean time, \optuv\ TDEs are teaching us a lot about accretion physics, emission mechanisms, galaxy dynamics, and more.
Given the increasing detection rate of (optical) transients, combined with added attention and followup resources invested in transients found in galaxy centers, we will learn more about all of these topics and likely make new and unexpected discoveries along the way.

\begin{acknowledgements}
We thank the International Space Science Institute for hosting us on several occasions to discuss the content and structure of this chapter, and its coordination with other chapters in this book. We are grateful to the two anonymous referees for their insightful comments and helpful suggestions. We thank Thomas Wevers for useful comments.
S. van Velzen is supported by the James Arthur Postdoctoral Fellowship. F. Onori acknowledges the support of the H2020 Hemera program, grant agreement No 730970, and the support from European Research Council Consolidator Grant 647208. I. Arcavi is a CIFAR Azrieli Global Scholar in the Gravity and the Extreme Universe Program and acknowledges support from that program, from the European Research Council (ERC) under the European Union’s Horizon 2020 research and innovation program (grant agreement number 852097), from the Israel Science Foundation (grant numbers 2108/18 and 2752/19), from the United States - Israel Binational Science Foundation (BSF), and from the Israeli Council for Higher Education Alon Fellowship.

\end{acknowledgements}

\bibliographystyle{aps-nameyear}     
\bibliography{example,general_desk}          

\begin{thebibliography}{130}
\ifx \bisbn   \undefined \def \bisbn  #1{ISBN #1}\fi
\ifx \binits  \undefined \def \binits#1{#1} \fi
\ifx \bauthor  \undefined \def \bauthor#1{#1} \fi
\ifx \bjtitle  \undefined \def \bjtitle#1{\textrm{#1}}\fi
\ifx \batitle  \undefined \def \batitle#1{#1} \fi
\ifx \bctitle  \undefined \def \bctitle#1{#1} \fi
\ifx \bvolume  \undefined \def \bvolume#1{\textbf{#1}}\fi
\ifx \byear  \undefined \def \byear#1{#1} \fi
\ifx \bissue  \undefined \def \bissue#1{#1} \fi
\ifx \bfpage  \undefined \def \bfpage#1{#1} \fi
\ifx \blpage  \undefined \def \blpage #1{#1} \fi
\ifx \burl  \undefined \def \burl#1{#1} \fi
\ifx \doiurl  \undefined \def \doiurl#1{#1} \fi
\ifx \betal  \undefined \def \betal{et al.} \fi
\ifx \binstitute  \undefined \def \binstitute#1{#1} \fi
\ifx \beditor  \undefined \def \beditor#1{#1} \fi
\ifx \bpublisher  \undefined \def \bpublisher#1{#1} \fi
\ifx \bbtitle  \undefined \def \bbtitle#1{\textit{#1}} \fi
\ifx \bedition  \undefined \def \bedition#1{#1} \fi
\ifx \bseriesno  \undefined \def \bseriesno#1{#1} \fi
\ifx \blocation  \undefined \def \blocation#1{#1} \fi
\ifx \bsertitle  \undefined \def \bsertitle#1{#1} \fi
\ifx \bsnm \undefined \def \bsnm#1{#1} \fi
\ifx \bsuffix \undefined \def \bsuffix#1{#1} \fi
\ifx \bparticle \undefined \def \bparticle#1{#1} \fi
\ifx \barticle \undefined \def \barticle#1{#1} \fi
\ifx \botherref \undefined \def \botherref #1{#1} \fi
\ifx \url \undefined \def \url#1{#1} \fi
\ifx \bchapter \undefined \def \bchapter#1{#1} \fi
\ifx \bbook \undefined \def \bbook#1{#1} \fi
\ifx \bcomment \undefined \def \bcomment#1{#1} \fi
\ifx \oauthor \undefined \def \oauthor#1{#1} \fi
\ifx \citeauthoryear \undefined \def \citeauthoryear#1{#1} \fi
\ifx \texttildelow  \undefined \def \texttildelow{\symbol{126}} \fi
\def \endbibitem {}
\ifx \bconflocation  \undefined \def \bconflocation#1{#1} \fi

\bibitem[\protect\citeauthoryear{{Alexander} et~al.}{2016}]{Alexander16}
\begin{barticle}
\bauthor{\binits{K.D.} \bsnm{{Alexander}}},
\bauthor{\binits{E.} \bsnm{{Berger}}},
\bauthor{\binits{J.} \bsnm{{Guillochon}}},
\bauthor{\binits{B.A.} \bsnm{{Zauderer}}},
\bauthor{\binits{P.K.G.} \bsnm{{Williams}}},
\batitle{{Discovery of an Outflow from Radio Observations of the Tidal
  Disruption Event ASASSN-14li}}.
\bjtitle{\apjl}
\bvolume{819},
\bfpage{25}
(\byear{2016}).
doi:\doiurl{10.3847/2041-8205/819/2/L25}
\end{barticle}
\endbibitem

\bibitem[\protect\citeauthoryear{{Arcavi} et~al.}{2014}]{Arcavi14}
\begin{barticle}
\bauthor{\binits{I.} \bsnm{{Arcavi}}},
\bauthor{\binits{A.} \bsnm{{Gal-Yam}}},
\bauthor{\binits{M.} \bsnm{{Sullivan}}},
\bauthor{\binits{Y.-C.} \bsnm{{Pan}}},
\bauthor{\binits{S.B.} \bsnm{{Cenko}}},
\bauthor{\binits{A.} \bsnm{{Horesh}}},
\bauthor{\binits{E.O.} \bsnm{{Ofek}}},
\bauthor{\binits{A.} \bsnm{{De Cia}}},
\bauthor{\binits{L.} \bsnm{{Yan}}},
\bauthor{\binits{C.-W.} \bsnm{{Yang}}},
\bauthor{\binits{D.A.} \bsnm{{Howell}}},
\bauthor{\binits{D.} \bsnm{{Tal}}},
\bauthor{\binits{S.R.} \bsnm{{Kulkarni}}},
\bauthor{\binits{S.P.} \bsnm{{Tendulkar}}},
\bauthor{\binits{S.} \bsnm{{Tang}}},
\bauthor{\binits{D.} \bsnm{{Xu}}},
\bauthor{\binits{A.} \bsnm{{Sternberg}}},
\bauthor{\binits{J.G.} \bsnm{{Cohen}}},
\bauthor{\binits{J.S.} \bsnm{{Bloom}}},
\bauthor{\binits{P.E.} \bsnm{{Nugent}}},
\bauthor{\binits{M.M.} \bsnm{{Kasliwal}}},
\bauthor{\binits{D.A.} \bsnm{{Perley}}},
\bauthor{\binits{R.M.} \bsnm{{Quimby}}},
\bauthor{\binits{A.A.} \bsnm{{Miller}}},
\bauthor{\binits{C.A.} \bsnm{{Theissen}}},
\bauthor{\binits{R.R.} \bsnm{{Laher}}},
\batitle{{A Continuum of H- to He-rich Tidal Disruption Candidates With a
  Preference for E+A Galaxies}}.
\bjtitle{\apj}
\bvolume{793},
\bfpage{38}
(\byear{2014}).
doi:\doiurl{10.1088/0004-637X/793/1/38}
\end{barticle}
\endbibitem

\bibitem[\protect\citeauthoryear{{Auchettl} et~al.}{2017}]{Auchettl16}
\begin{barticle}
\bauthor{\binits{K.} \bsnm{{Auchettl}}},
\bauthor{\binits{J.} \bsnm{{Guillochon}}},
\bauthor{\binits{E.} \bsnm{{Ramirez-Ruiz}}},
\batitle{{New Physical Insights about Tidal Disruption Events from a
  Comprehensive Observational Inventory at X-Ray Wavelengths}}.
\bjtitle{\apj}
\bvolume{838},
\bfpage{149}
(\byear{2017}).
doi:\doiurl{10.3847/1538-4357/aa633b}
\end{barticle}
\endbibitem

\bibitem[\protect\citeauthoryear{{Bade} et~al.}{1996}]{Bade96}
\begin{barticle}
\bauthor{\binits{N.} \bsnm{{Bade}}},
\bauthor{\binits{S.} \bsnm{{Komossa}}},
\bauthor{\binits{M.} \bsnm{{Dahlem}}},
\batitle{{Detection of an extremely soft X-ray outburst in the HII-like nucleus
  of NGC 5905.}}
\bjtitle{\aap}
\bvolume{309},
\bfpage{35}--\blpage{38}
(\byear{1996})
\end{barticle}
\endbibitem

\bibitem[\protect\citeauthoryear{{Baldwin} et~al.}{1981}]{baldwin81}
\begin{barticle}
\bauthor{\binits{J.A.} \bsnm{{Baldwin}}},
\bauthor{\binits{M.M.} \bsnm{{Phillips}}},
\bauthor{\binits{R.} \bsnm{{Terlevich}}},
\batitle{{Classification parameters for the emission-line spectra of
  extragalactic objects}}.
\bjtitle{\pasp}
\bvolume{93},
\bfpage{5}--\blpage{19}
(\byear{1981}).
doi:\doiurl{10.1086/130766}
\end{barticle}
\endbibitem

\bibitem[\protect\citeauthoryear{{Bellm} et~al.}{2019}]{Bellm19}
\begin{barticle}
\bauthor{\binits{E.C.} \bsnm{{Bellm}}},
\bauthor{\binits{S.R.} \bsnm{{Kulkarni}}},
\bauthor{\binits{M.J.} \bsnm{{Graham}}},
\bauthor{\binits{R.} \bsnm{{Dekany}}},
\bauthor{\binits{R.M.} \bsnm{{Smith}}},
\bauthor{\binits{R.} \bsnm{{Riddle}}},
\bauthor{\binits{F.J.} \bsnm{{Masci}}},
\bauthor{\binits{G.} \bsnm{{Helou}}},
\bauthor{\binits{T.A.} \bsnm{{Prince}}},
\bauthor{\binits{S.M.} \bsnm{{Adams}}},
\bauthor{\binits{C.} \bsnm{{Barbarino}}},
\bauthor{\binits{T.} \bsnm{{Barlow}}},
\bauthor{\binits{J.} \bsnm{{Bauer}}},
\bauthor{\binits{R.} \bsnm{{Beck}}},
\bauthor{\binits{J.} \bsnm{{Belicki}}},
\bauthor{\binits{R.} \bsnm{{Biswas}}},
\bauthor{\binits{N.} \bsnm{{Blagorodnova}}},
\bauthor{\binits{D.} \bsnm{{Bodewits}}},
\bauthor{\binits{B.} \bsnm{{Bolin}}},
\bauthor{\binits{V.} \bsnm{{Brinnel}}},
\bauthor{\binits{T.} \bsnm{{Brooke}}},
\bauthor{\binits{B.} \bsnm{{Bue}}},
\bauthor{\binits{M.} \bsnm{{Bulla}}},
\bauthor{\binits{R.} \bsnm{{Burruss}}},
\bauthor{\binits{S.B.} \bsnm{{Cenko}}},
\bauthor{\binits{C.-K.} \bsnm{{Chang}}},
\bauthor{\binits{A.} \bsnm{{Connolly}}},
\bauthor{\binits{M.} \bsnm{{Coughlin}}},
\bauthor{\binits{J.} \bsnm{{Cromer}}},
\bauthor{\binits{V.} \bsnm{{Cunningham}}},
\bauthor{\binits{K.} \bsnm{{De}}},
\bauthor{\binits{A.} \bsnm{{Delacroix}}},
\bauthor{\binits{V.} \bsnm{{Desai}}},
\bauthor{\binits{D.A.} \bsnm{{Duev}}},
\bauthor{\binits{G.} \bsnm{{Eadie}}},
\bauthor{\binits{T.L.} \bsnm{{Farnham}}},
\bauthor{\binits{M.} \bsnm{{Feeney}}},
\bauthor{\binits{U.} \bsnm{{Feindt}}},
\bauthor{\binits{D.} \bsnm{{Flynn}}},
\bauthor{\binits{A.} \bsnm{{Franckowiak}}},
\bauthor{\binits{S.} \bsnm{{Frederick}}},
\bauthor{\binits{C.} \bsnm{{Fremling}}},
\bauthor{\binits{A.} \bsnm{{Gal-Yam}}},
\bauthor{\binits{S.} \bsnm{{Gezari}}},
\bauthor{\binits{M.} \bsnm{{Giomi}}},
\bauthor{\binits{D.A.} \bsnm{{Goldstein}}},
\bauthor{\binits{V.Z.} \bsnm{{Golkhou}}},
\bauthor{\binits{A.} \bsnm{{Goobar}}},
\bauthor{\binits{S.} \bsnm{{Groom}}},
\bauthor{\binits{E.} \bsnm{{Hacopians}}},
\bauthor{\binits{D.} \bsnm{{Hale}}},
\bauthor{\binits{J.} \bsnm{{Henning}}},
\bauthor{\binits{A.Y.Q.} \bsnm{{Ho}}},
\bauthor{\binits{D.} \bsnm{{Hover}}},
\bauthor{\binits{J.} \bsnm{{Howell}}},
\bauthor{\binits{T.} \bsnm{{Hung}}},
\bauthor{\binits{D.} \bsnm{{Huppenkothen}}},
\bauthor{\binits{D.} \bsnm{{Imel}}},
\bauthor{\binits{W.-H.} \bsnm{{Ip}}},
\bauthor{\binits{{\v Z}.} \bsnm{{Ivezi{\'c}}}},
\bauthor{\binits{E.} \bsnm{{Jackson}}},
\bauthor{\binits{L.} \bsnm{{Jones}}},
\bauthor{\binits{M.} \bsnm{{Juric}}},
\bauthor{\binits{M.M.} \bsnm{{Kasliwal}}},
\bauthor{\binits{S.} \bsnm{{Kaspi}}},
\bauthor{\binits{S.} \bsnm{{Kaye}}},
\bauthor{\binits{M.S.P.} \bsnm{{Kelley}}},
\bauthor{\binits{M.} \bsnm{{Kowalski}}},
\bauthor{\binits{E.} \bsnm{{Kramer}}},
\bauthor{\binits{T.} \bsnm{{Kupfer}}},
\bauthor{\binits{W.} \bsnm{{Landry}}},
\bauthor{\binits{R.R.} \bsnm{{Laher}}},
\bauthor{\binits{C.-D.} \bsnm{{Lee}}},
\bauthor{\binits{H.W.} \bsnm{{Lin}}},
\bauthor{\binits{Z.-Y.} \bsnm{{Lin}}},
\bauthor{\binits{R.} \bsnm{{Lunnan}}},
\bauthor{\binits{M.} \bsnm{{Giomi}}},
\bauthor{\binits{A.} \bsnm{{Mahabal}}},
\bauthor{\binits{P.} \bsnm{{Mao}}},
\bauthor{\binits{A.A.} \bsnm{{Miller}}},
\bauthor{\binits{S.} \bsnm{{Monkewitz}}},
\bauthor{\binits{P.} \bsnm{{Murphy}}},
\bauthor{\binits{C.-C.} \bsnm{{Ngeow}}},
\bauthor{\binits{J.} \bsnm{{Nordin}}},
\bauthor{\binits{P.} \bsnm{{Nugent}}},
\bauthor{\binits{E.} \bsnm{{Ofek}}},
\bauthor{\binits{M.T.} \bsnm{{Patterson}}},
\bauthor{\binits{B.} \bsnm{{Penprase}}},
\bauthor{\binits{M.} \bsnm{{Porter}}},
\bauthor{\binits{L.} \bsnm{{Rauch}}},
\bauthor{\binits{U.} \bsnm{{Rebbapragada}}},
\bauthor{\binits{D.} \bsnm{{Reiley}}},
\bauthor{\binits{M.} \bsnm{{Rigault}}},
\bauthor{\binits{H.} \bsnm{{Rodriguez}}},
\bauthor{\binits{J.} \bsnm{{van Roestel}}},
\bauthor{\binits{B.} \bsnm{{Rusholme}}},
\bauthor{\binits{J.} \bsnm{{van Santen}}},
\bauthor{\binits{S.} \bsnm{{Schulze}}},
\bauthor{\binits{D.L.} \bsnm{{Shupe}}},
\bauthor{\binits{L.P.} \bsnm{{Singer}}},
\bauthor{\binits{M.T.} \bsnm{{Soumagnac}}},
\bauthor{\binits{R.} \bsnm{{Stein}}},
\bauthor{\binits{J.} \bsnm{{Surace}}},
\bauthor{\binits{J.} \bsnm{{Sollerman}}},
\bauthor{\binits{P.} \bsnm{{Szkody}}},
\bauthor{\binits{F.} \bsnm{{Taddia}}},
\bauthor{\binits{S.} \bsnm{{Terek}}},
\bauthor{\binits{A.} \bsnm{{Van Sistine}}},
\bauthor{\binits{S.} \bsnm{{van Velzen}}},
\bauthor{\binits{W.T.} \bsnm{{Vestrand}}},
\bauthor{\binits{R.} \bsnm{{Walters}}},
\bauthor{\binits{C.} \bsnm{{Ward}}},
\bauthor{\binits{Q.-Z.} \bsnm{{Ye}}},
\bauthor{\binits{P.-C.} \bsnm{{Yu}}},
\bauthor{\binits{L.} \bsnm{{Yan}}},
\bauthor{\binits{J.} \bsnm{{Zolkower}}},
\batitle{{The Zwicky Transient Facility: System Overview, Performance, and
  First Results}}.
\bjtitle{\pasp}
\bvolume{131}(\bissue{1}),
\bfpage{018002}
(\byear{2019}).
doi:\doiurl{10.1088/1538-3873/aaecbe}
\end{barticle}
\endbibitem

\bibitem[\protect\citeauthoryear{{Blagorodnova} et~al.}{2017}]{Blagorodnova17}
\begin{barticle}
\bauthor{\binits{N.} \bsnm{{Blagorodnova}}},
\bauthor{\binits{S.} \bsnm{{Gezari}}},
\bauthor{\binits{T.} \bsnm{{Hung}}},
\bauthor{\binits{S.R.} \bsnm{{Kulkarni}}},
\bauthor{\binits{S.B.} \bsnm{{Cenko}}},
\bauthor{\binits{D.R.} \bsnm{{Pasham}}},
\bauthor{\binits{L.} \bsnm{{Yan}}},
\bauthor{\binits{I.} \bsnm{{Arcavi}}},
\bauthor{\binits{S.} \bsnm{{Ben-Ami}}},
\bauthor{\binits{B.D.} \bsnm{{Bue}}},
\bauthor{\binits{T.} \bsnm{{Cantwell}}},
\bauthor{\binits{Y.} \bsnm{{Cao}}},
\bauthor{\binits{A.J.} \bsnm{{Castro-Tirado}}},
\bauthor{\binits{R.} \bsnm{{Fender}}},
\bauthor{\binits{C.} \bsnm{{Fremling}}},
\bauthor{\binits{A.} \bsnm{{Gal-Yam}}},
\bauthor{\binits{A.Y.Q.} \bsnm{{Ho}}},
\bauthor{\binits{A.} \bsnm{{Horesh}}},
\bauthor{\binits{G.} \bsnm{{Hosseinzadeh}}},
\bauthor{\binits{M.M.} \bsnm{{Kasliwal}}},
\bauthor{\binits{A.K.H.} \bsnm{{Kong}}},
\bauthor{\binits{R.R.} \bsnm{{Laher}}},
\bauthor{\binits{G.} \bsnm{{Leloudas}}},
\bauthor{\binits{R.} \bsnm{{Lunnan}}},
\bauthor{\binits{F.J.} \bsnm{{Masci}}},
\bauthor{\binits{K.} \bsnm{{Mooley}}},
\bauthor{\binits{J.D.} \bsnm{{Neill}}},
\bauthor{\binits{P.} \bsnm{{Nugent}}},
\bauthor{\binits{M.} \bsnm{{Powell}}},
\bauthor{\binits{A.F.} \bsnm{{Valeev}}},
\bauthor{\binits{P.M.} \bsnm{{Vreeswijk}}},
\bauthor{\binits{R.} \bsnm{{Walters}}},
\bauthor{\binits{P.} \bsnm{{Wozniak}}},
\batitle{{iPTF16fnl: A Faint and Fast Tidal Disruption Event in an E+A
  Galaxy}}.
\bjtitle{\apj}
\bvolume{844},
\bfpage{46}
(\byear{2017}).
doi:\doiurl{10.3847/1538-4357/aa7579}
\end{barticle}
\endbibitem

\bibitem[\protect\citeauthoryear{{Blagorodnova} et~al.}{2019}]{Blagorodnova19}
\begin{barticle}
\bauthor{\binits{N.} \bsnm{{Blagorodnova}}},
\bauthor{\binits{S.B.} \bsnm{{Cenko}}},
\bauthor{\binits{S.R.} \bsnm{{Kulkarni}}},
\bauthor{\binits{I.} \bsnm{{Arcavi}}},
\bauthor{\binits{J.S.} \bsnm{{Bloom}}},
\bauthor{\binits{G.} \bsnm{{Duggan}}},
\bauthor{\binits{A.V.} \bsnm{{Filippenko}}},
\bauthor{\binits{C.} \bsnm{{Fremling}}},
\bauthor{\binits{A.} \bsnm{{Horesh}}},
\bauthor{\binits{G.} \bsnm{{Hosseinzadeh}}},
\bauthor{\binits{E.} \bsnm{{Karamehmetoglu}}},
\bauthor{\binits{A.} \bsnm{{Levan}}},
\bauthor{\binits{F.J.} \bsnm{{Masci}}},
\bauthor{\binits{P.E.} \bsnm{{Nugent}}},
\bauthor{\binits{D.R.} \bsnm{{Pasham}}},
\bauthor{\binits{S.} \bsnm{{Veilleux}}},
\bauthor{\binits{R.} \bsnm{{Walters}}},
\bauthor{\binits{L.} \bsnm{{Yan}}},
\bauthor{\binits{W.} \bsnm{{Zheng}}},
\batitle{{The Broad Absorption Line Tidal Disruption Event iPTF15af: Optical
  and Ultraviolet Evolution}}.
\bjtitle{\apj}
\bvolume{873}(\bissue{1}),
\bfpage{92}
(\byear{2019}).
doi:\doiurl{10.3847/1538-4357/ab04b0}
\end{barticle}
\endbibitem

\bibitem[\protect\citeauthoryear{{Blanchard} et~al.}{2017}]{Blanchard17}
\begin{barticle}
\bauthor{\binits{P.K.} \bsnm{{Blanchard}}},
\bauthor{\binits{M.} \bsnm{{Nicholl}}},
\bauthor{\binits{E.} \bsnm{{Berger}}},
\bauthor{\binits{J.} \bsnm{{Guillochon}}},
\bauthor{\binits{R.} \bsnm{{Margutti}}},
\bauthor{\binits{R.} \bsnm{{Chornock}}},
\bauthor{\binits{K.D.} \bsnm{{Alexander}}},
\bauthor{\binits{J.} \bsnm{{Leja}}},
\bauthor{\binits{M.R.} \bsnm{{Drout}}},
\batitle{{PS16dtm: A Tidal Disruption Event in a Narrow-line Seyfert 1
  Galaxy}}.
\bjtitle{\apj}
\bvolume{843},
\bfpage{106}
(\byear{2017}).
doi:\doiurl{10.3847/1538-4357/aa77f7}
\end{barticle}
\endbibitem

\bibitem[\protect\citeauthoryear{{Bogdanovi{\'c}} et~al.}{2004}]{bogdanovic04}
\begin{barticle}
\bauthor{\binits{T.} \bsnm{{Bogdanovi{\'c}}}},
\bauthor{\binits{M.} \bsnm{{Eracleous}}},
\bauthor{\binits{S.} \bsnm{{Mahadevan}}},
\bauthor{\binits{S.} \bsnm{{Sigurdsson}}},
\bauthor{\binits{P.} \bsnm{{Laguna}}},
\batitle{{Tidal Disruption of a Star by a Black Hole: Observational
  Signature}}.
\bjtitle{\apj}
\bvolume{610},
\bfpage{707}--\blpage{721}
(\byear{2004}).
doi:\doiurl{10.1086/421758}
\end{barticle}
\endbibitem

\bibitem[\protect\citeauthoryear{{Bowen}}{1928}]{Bowen1928}
\begin{barticle}
\bauthor{\binits{I.S.} \bsnm{{Bowen}}},
\batitle{{The origin of the nebular lines and the structure of the planetary
  nebulae.}}
\bjtitle{\apj}
\bvolume{67},
\bfpage{1}--\blpage{15}
(\byear{1928}).
doi:\doiurl{10.1086/143091}
\end{barticle}
\endbibitem

\bibitem[\protect\citeauthoryear{{Brotherton} et~al.}{2001}]{Brotherton2001}
\begin{barticle}
\bauthor{\binits{M.S.} \bsnm{{Brotherton}}},
\bauthor{\binits{H.D.} \bsnm{{Tran}}},
\bauthor{\binits{R.H.} \bsnm{{Becker}}},
\bauthor{\binits{M.D.} \bsnm{{Gregg}}},
\bauthor{\binits{S.A.} \bsnm{{Laurent-Muehleisen}}},
\bauthor{\binits{R.L.} \bsnm{{White}}},
\batitle{{Composite Spectra from the FIRST Bright Quasar Survey}}.
\bjtitle{\apj}
\bvolume{546},
\bfpage{775}--\blpage{781}
(\byear{2001}).
doi:\doiurl{10.1086/318309}
\end{barticle}
\endbibitem

\bibitem[\protect\citeauthoryear{{Brown} et~al.}{2016}]{Brown16}
\begin{barticle}
\bauthor{\binits{J.S.} \bsnm{{Brown}}},
\bauthor{\binits{B.J.} \bsnm{{Shappee}}},
\bauthor{\binits{T.} \bsnm{{W.-S Holoien}}},
\bauthor{\binits{K.Z.} \bsnm{{Stanek}}},
\bauthor{\binits{C.S.} \bsnm{{Kochanek}}},
\bauthor{\binits{J.L.} \bsnm{{Prieto}}},
\batitle{{Hello Darkness My Old Friend: The Fading of the Nearby TDE
  ASASSN-14ae}}.
\bjtitle{\mnras}
\bvolume{462},
\bfpage{3993}--\blpage{4000}
(\byear{2016}).
doi:\doiurl{10.1093/mnras/stw1928}
\end{barticle}
\endbibitem

\bibitem[\protect\citeauthoryear{{Brown} et~al.}{2017a}]{Brown17a}
\begin{barticle}
\bauthor{\binits{J.S.} \bsnm{{Brown}}},
\bauthor{\binits{T.W.-S.} \bsnm{{Holoien}}},
\bauthor{\binits{K.} \bsnm{{Auchettl}}},
\bauthor{\binits{K.Z.} \bsnm{{Stanek}}},
\bauthor{\binits{C.S.} \bsnm{{Kochanek}}},
\bauthor{\binits{B.J.} \bsnm{{Shappee}}},
\bauthor{\binits{J.L.} \bsnm{{Prieto}}},
\bauthor{\binits{D.} \bsnm{{Grupe}}},
\batitle{{The Long Term Evolution of ASASSN-14li}}.
\bjtitle{\mnras}
\bvolume{466}(\bissue{4}),
\bfpage{4904}--\blpage{4916}
(\byear{2017}a).
doi:\doiurl{10.1093/mnras/stx033}
\end{barticle}
\endbibitem

\bibitem[\protect\citeauthoryear{{Brown} et~al.}{2017b}]{Brown16b}
\begin{barticle}
\bauthor{\binits{J.S.} \bsnm{{Brown}}},
\bauthor{\binits{T.W.-S.} \bsnm{{Holoien}}},
\bauthor{\binits{K.} \bsnm{{Auchettl}}},
\bauthor{\binits{K.Z.} \bsnm{{Stanek}}},
\bauthor{\binits{C.S.} \bsnm{{Kochanek}}},
\bauthor{\binits{B.J.} \bsnm{{Shappee}}},
\bauthor{\binits{J.L.} \bsnm{{Prieto}}},
\bauthor{\binits{D.} \bsnm{{Grupe}}},
\batitle{{The Long Term Evolution of ASASSN-14li}}.
\bjtitle{\mnras}
\bvolume{466},
\bfpage{4904}--\blpage{4916}
(\byear{2017}b).
doi:\doiurl{10.1093/mnras/stx033}
\end{barticle}
\endbibitem

\bibitem[\protect\citeauthoryear{{Brown} et~al.}{2018}]{Brown18}
\begin{barticle}
\bauthor{\binits{J.S.} \bsnm{{Brown}}},
\bauthor{\binits{C.S.} \bsnm{{Kochanek}}},
\bauthor{\binits{T.W.-S.} \bsnm{{Holoien}}},
\bauthor{\binits{K.Z.} \bsnm{{Stanek}}},
\bauthor{\binits{K.} \bsnm{{Auchettl}}},
\bauthor{\binits{B.J.} \bsnm{{Shappee}}},
\bauthor{\binits{J.L.} \bsnm{{Prieto}}},
\bauthor{\binits{N.} \bsnm{{Morrell}}},
\bauthor{\binits{E.} \bsnm{{Falco}}},
\bauthor{\binits{J.} \bsnm{{Strader}}},
\bauthor{\binits{L.} \bsnm{{Chomiuk}}},
\bauthor{\binits{R.} \bsnm{{Post}}},
\bauthor{\binits{S.} \bsnm{{Villanueva}} \bsuffix{Jr.}},
\bauthor{\binits{S.} \bsnm{{Mathur}}},
\bauthor{\binits{S.} \bsnm{{Dong}}},
\bauthor{\binits{P.} \bsnm{{Chen}}},
\bauthor{\binits{S.} \bsnm{{Bose}}},
\batitle{{The ultraviolet spectroscopic evolution of the low-luminosity tidal
  disruption event iPTF16fnl}}.
\bjtitle{\mnras}
\bvolume{473},
\bfpage{1130}--\blpage{1144}
(\byear{2018}).
doi:\doiurl{10.1093/mnras/stx2372}
\end{barticle}
\endbibitem

\bibitem[\protect\citeauthoryear{{Cannizzo} et~al.}{1990}]{Cannizzo90}
\begin{barticle}
\bauthor{\binits{J.K.} \bsnm{{Cannizzo}}},
\bauthor{\binits{H.M.} \bsnm{{Lee}}},
\bauthor{\binits{J.} \bsnm{{Goodman}}},
\batitle{{The disk accretion of a tidally disrupted star onto a massive black
  hole}}.
\bjtitle{\apj}
\bvolume{351},
\bfpage{38}--\blpage{46}
(\byear{1990}).
doi:\doiurl{10.1086/168442}
\end{barticle}
\endbibitem

\bibitem[\protect\citeauthoryear{{Cenko} et~al.}{2016}]{Cenko16}
\begin{barticle}
\bauthor{\binits{S.B.} \bsnm{{Cenko}}},
\bauthor{\binits{A.} \bsnm{{Cucchiara}}},
\bauthor{\binits{N.} \bsnm{{Roth}}},
\bauthor{\binits{S.} \bsnm{{Veilleux}}},
\bauthor{\binits{J.X.} \bsnm{{Prochaska}}},
\bauthor{\binits{L.} \bsnm{{Yan}}},
\bauthor{\binits{J.} \bsnm{{Guillochon}}},
\bauthor{\binits{W.P.} \bsnm{{Maksym}}},
\bauthor{\binits{I.} \bsnm{{Arcavi}}},
\bauthor{\binits{N.R.} \bsnm{{Butler}}},
\bauthor{\binits{A.V.} \bsnm{{Filippenko}}},
\bauthor{\binits{A.S.} \bsnm{{Fruchter}}},
\bauthor{\binits{S.} \bsnm{{Gezari}}},
\bauthor{\binits{D.} \bsnm{{Kasen}}},
\bauthor{\binits{A.J.} \bsnm{{Levan}}},
\bauthor{\binits{J.M.} \bsnm{{Miller}}},
\bauthor{\binits{D.R.} \bsnm{{Pasham}}},
\bauthor{\binits{E.} \bsnm{{Ramirez-Ruiz}}},
\bauthor{\binits{L.E.} \bsnm{{Strubbe}}},
\bauthor{\binits{N.R.} \bsnm{{Tanvir}}},
\bauthor{\binits{F.} \bsnm{{Tombesi}}},
\batitle{{An Ultraviolet Spectrum of the Tidal Disruption Flare ASASSN-14li}}.
\bjtitle{\apjl}
\bvolume{818},
\bfpage{32}
(\byear{2016}).
doi:\doiurl{10.3847/2041-8205/818/2/L32}
\end{barticle}
\endbibitem

\bibitem[\protect\citeauthoryear{{Chambers}}{2007}]{Chambers07}
\begin{bchapter}
\bauthor{\binits{K.C.} \bsnm{{Chambers}}},
\bctitle{{The PS1 System and Science Mission}},
in \bbtitle{Bulletin of the AAS},
vol. \bseriesno{38},
\byear{2007},
p. \bfpage{995}
\end{bchapter}
\endbibitem

\bibitem[\protect\citeauthoryear{{Chambers} et~al.}{2016}]{Chambers16}
\begin{botherref}
\oauthor{\binits{K.C.} \bsnm{{Chambers}}},
\oauthor{\binits{E.A.} \bsnm{{Magnier}}},
\oauthor{\binits{N.} \bsnm{{Metcalfe}}},
\oauthor{\binits{H.A.} \bsnm{{Flewelling}}},
\oauthor{\binits{M.E.} \bsnm{{Huber}}},
\oauthor{\binits{C.Z.} \bsnm{{Waters}}},
\oauthor{\binits{L.} \bsnm{{Denneau}}},
\oauthor{\binits{P.W.} \bsnm{{Draper}}},
\oauthor{\binits{D.} \bsnm{{Farrow}}},
\oauthor{\binits{D.P.} \bsnm{{Finkbeiner}}},
\oauthor{\binits{C.} \bsnm{{Holmberg}}},
\oauthor{\binits{J.} \bsnm{{Koppenhoefer}}},
\oauthor{\binits{P.A.} \bsnm{{Price}}},
\oauthor{\binits{R.P.} \bsnm{{Saglia}}},
\oauthor{\binits{E.F.} \bsnm{{Schlafly}}},
\oauthor{\binits{S.J.} \bsnm{{Smartt}}},
\oauthor{\binits{W.} \bsnm{{Sweeney}}},
\oauthor{\binits{R.J.} \bsnm{{Wainscoat}}},
\oauthor{\binits{W.S.} \bsnm{{Burgett}}},
\oauthor{\binits{T.} \bsnm{{Grav}}},
\oauthor{\binits{J.N.} \bsnm{{Heasley}}},
\oauthor{\binits{K.W.} \bsnm{{Hodapp}}},
\oauthor{\binits{R.} \bsnm{{Jedicke}}},
\oauthor{\binits{N.} \bsnm{{Kaiser}}},
\oauthor{\binits{R.-P.} \bsnm{{Kudritzki}}},
\oauthor{\binits{G.A.} \bsnm{{Luppino}}},
\oauthor{\binits{R.H.} \bsnm{{Lupton}}},
\oauthor{\binits{D.G.} \bsnm{{Monet}}},
\oauthor{\binits{J.S.} \bsnm{{Morgan}}},
\oauthor{\binits{P.M.} \bsnm{{Onaka}}},
\oauthor{\binits{C.W.} \bsnm{{Stubbs}}},
\oauthor{\binits{J.L.} \bsnm{{Tonry}}},
\oauthor{\binits{E.} \bsnm{{Banados}}},
\oauthor{\binits{E.F.} \bsnm{{Bell}}},
\oauthor{\binits{R.} \bsnm{{Bender}}},
\oauthor{\binits{E.J.} \bsnm{{Bernard}}},
\oauthor{\binits{M.T.} \bsnm{{Botticella}}},
\oauthor{\binits{S.} \bsnm{{Casertano}}},
\oauthor{\binits{S.} \bsnm{{Chastel}}},
\oauthor{\binits{W.-P.} \bsnm{{Chen}}},
\oauthor{\binits{X.} \bsnm{{Chen}}},
\oauthor{\binits{S.} \bsnm{{Cole}}},
\oauthor{\binits{N.} \bsnm{{Deacon}}},
\oauthor{\binits{C.} \bsnm{{Frenk}}},
\oauthor{\binits{A.} \bsnm{{Fitzsimmons}}},
\oauthor{\binits{S.} \bsnm{{Gezari}}},
\oauthor{\binits{C.} \bsnm{{Goessl}}},
\oauthor{\binits{T.} \bsnm{{Goggia}}},
\oauthor{\binits{B.} \bsnm{{Goldman}}},
\oauthor{\binits{E.K.} \bsnm{{Grebel}}},
\oauthor{\binits{N.C.} \bsnm{{Hambly}}},
\oauthor{\binits{G.} \bsnm{{Hasinger}}},
\oauthor{\binits{A.F.} \bsnm{{Heavens}}},
\oauthor{\binits{T.M.} \bsnm{{Heckman}}},
\oauthor{\binits{R.} \bsnm{{Henderson}}},
\oauthor{\binits{T.} \bsnm{{Henning}}},
\oauthor{\binits{M.} \bsnm{{Holman}}},
\oauthor{\binits{U.} \bsnm{{Hopp}}},
\oauthor{\binits{W.-H.} \bsnm{{Ip}}},
\oauthor{\binits{S.} \bsnm{{Isani}}},
\oauthor{\binits{C.D.} \bsnm{{Keyes}}},
\oauthor{\binits{A.} \bsnm{{Koekemoer}}},
\oauthor{\binits{R.} \bsnm{{Kotak}}},
\oauthor{\binits{K.S.} \bsnm{{Long}}},
\oauthor{\binits{J.R.} \bsnm{{Lucey}}},
\oauthor{\binits{M.} \bsnm{{Liu}}},
\oauthor{\binits{N.F.} \bsnm{{Martin}}},
\oauthor{\binits{B.} \bsnm{{McLean}}},
\oauthor{\binits{E.} \bsnm{{Morganson}}},
\oauthor{\binits{D.N.A.} \bsnm{{Murphy}}},
\oauthor{\binits{M.A.} \bsnm{{Nieto-Santisteban}}},
\oauthor{\binits{P.} \bsnm{{Norberg}}},
\oauthor{\binits{J.A.} \bsnm{{Peacock}}},
\oauthor{\binits{E.A.} \bsnm{{Pier}}},
\oauthor{\binits{M.} \bsnm{{Postman}}},
\oauthor{\binits{N.} \bsnm{{Primak}}},
\oauthor{\binits{C.} \bsnm{{Rae}}},
\oauthor{\binits{A.} \bsnm{{Rest}}},
\oauthor{\binits{A.} \bsnm{{Riess}}},
\oauthor{\binits{A.} \bsnm{{Riffeser}}},
\oauthor{\binits{H.W.} \bsnm{{Rix}}},
\oauthor{\binits{S.} \bsnm{{Roser}}},
\oauthor{\binits{E.} \bsnm{{Schilbach}}},
\oauthor{\binits{A.S.B.} \bsnm{{Schultz}}},
\oauthor{\binits{D.} \bsnm{{Scolnic}}},
\oauthor{\binits{A.} \bsnm{{Szalay}}},
\oauthor{\binits{S.} \bsnm{{Seitz}}},
\oauthor{\binits{B.} \bsnm{{Shiao}}},
\oauthor{\binits{E.} \bsnm{{Small}}},
\oauthor{\binits{K.W.} \bsnm{{Smith}}},
\oauthor{\binits{D.} \bsnm{{Soderblom}}},
\oauthor{\binits{A.N.} \bsnm{{Taylor}}},
\oauthor{\binits{A.R.} \bsnm{{Thakar}}},
\oauthor{\binits{J.} \bsnm{{Thiel}}},
\oauthor{\binits{D.} \bsnm{{Thilker}}},
\oauthor{\binits{Y.} \bsnm{{Urata}}},
\oauthor{\binits{J.} \bsnm{{Valenti}}},
\oauthor{\binits{F.} \bsnm{{Walter}}},
\oauthor{\binits{S.P.} \bsnm{{Watters}}},
\oauthor{\binits{S.} \bsnm{{Werner}}},
\oauthor{\binits{R.} \bsnm{{White}}},
\oauthor{\binits{W.M.} \bsnm{{Wood-Vasey}}},
\oauthor{\binits{R.} \bsnm{{Wyse}}},
{The Pan-STARRS1 Surveys}.
ArXiv e-prints
(2016)
\end{botherref}
\endbibitem

\bibitem[\protect\citeauthoryear{{Chan} et~al.}{2019}]{Chan2019}
\begin{barticle}
\bauthor{\binits{C.-H.} \bsnm{{Chan}}},
\bauthor{\binits{T.} \bsnm{{Piran}}},
\bauthor{\binits{J.H.} \bsnm{{Krolik}}},
\bauthor{\binits{D.} \bsnm{{Saban}}},
\batitle{{Tidal Disruption Events in Active Galactic Nuclei}}.
\bjtitle{\apj}
\bvolume{881}(\bissue{2}),
\bfpage{113}
(\byear{2019}).
doi:\doiurl{10.3847/1538-4357/ab2b40}
\end{barticle}
\endbibitem

\bibitem[\protect\citeauthoryear{{Chornock} et~al.}{2014}]{Chornock14}
\begin{barticle}
\bauthor{\binits{R.} \bsnm{{Chornock}}},
\bauthor{\binits{E.} \bsnm{{Berger}}},
\bauthor{\binits{S.} \bsnm{{Gezari}}},
\bauthor{\binits{B.A.} \bsnm{{Zauderer}}},
\bauthor{\binits{A.} \bsnm{{Rest}}},
\bauthor{\binits{L.} \bsnm{{Chomiuk}}},
\bauthor{\binits{A.} \bsnm{{Kamble}}},
\bauthor{\binits{A.M.} \bsnm{{Soderberg}}},
\bauthor{\binits{I.} \bsnm{{Czekala}}},
\bauthor{\binits{J.} \bsnm{{Dittmann}}},
\bauthor{\binits{M.} \bsnm{{Drout}}},
\bauthor{\binits{R.J.} \bsnm{{Foley}}},
\bauthor{\binits{W.} \bsnm{{Fong}}},
\bauthor{\binits{M.E.} \bsnm{{Huber}}},
\bauthor{\binits{R.P.} \bsnm{{Kirshner}}},
\bauthor{\binits{A.} \bsnm{{Lawrence}}},
\bauthor{\binits{R.} \bsnm{{Lunnan}}},
\bauthor{\binits{G.H.} \bsnm{{Marion}}},
\bauthor{\binits{G.} \bsnm{{Narayan}}},
\bauthor{\binits{A.G.} \bsnm{{Riess}}},
\bauthor{\binits{K.C.} \bsnm{{Roth}}},
\bauthor{\binits{N.E.} \bsnm{{Sanders}}},
\bauthor{\binits{D.} \bsnm{{Scolnic}}},
\bauthor{\binits{S.J.} \bsnm{{Smartt}}},
\bauthor{\binits{K.} \bsnm{{Smith}}},
\bauthor{\binits{C.W.} \bsnm{{Stubbs}}},
\bauthor{\binits{J.L.} \bsnm{{Tonry}}},
\bauthor{\binits{W.S.} \bsnm{{Burgett}}},
\bauthor{\binits{K.C.} \bsnm{{Chambers}}},
\bauthor{\binits{H.} \bsnm{{Flewelling}}},
\bauthor{\binits{K.W.} \bsnm{{Hodapp}}},
\bauthor{\binits{N.} \bsnm{{Kaiser}}},
\bauthor{\binits{E.A.} \bsnm{{Magnier}}},
\bauthor{\binits{D.C.} \bsnm{{Martin}}},
\bauthor{\binits{J.D.} \bsnm{{Neill}}},
\bauthor{\binits{P.A.} \bsnm{{Price}}},
\bauthor{\binits{R.} \bsnm{{Wainscoat}}},
\batitle{{The Ultraviolet-bright, Slowly Declining Transient PS1-11af as a
  Partial Tidal Disruption Event}}.
\bjtitle{\apj}
\bvolume{780},
\bfpage{44}
(\byear{2014}).
doi:\doiurl{10.1088/0004-637X/780/1/44}
\end{barticle}
\endbibitem

\bibitem[\protect\citeauthoryear{{Dai} et~al.}{2018}]{Dai18}
\begin{barticle}
\bauthor{\binits{L.} \bsnm{{Dai}}},
\bauthor{\binits{J.C.} \bsnm{{McKinney}}},
\bauthor{\binits{N.} \bsnm{{Roth}}},
\bauthor{\binits{E.} \bsnm{{Ramirez-Ruiz}}},
\bauthor{\binits{M.C.} \bsnm{{Miller}}},
\batitle{{A Unified Model for Tidal Disruption Events}}.
\bjtitle{\apj}
\bvolume{859},
\bfpage{20}
(\byear{2018}).
doi:\doiurl{10.3847/2041-8213/aab429}
\end{barticle}
\endbibitem

\bibitem[\protect\citeauthoryear{{Dong} et~al.}{2016}]{Dong16}
\begin{barticle}
\bauthor{\binits{S.} \bsnm{{Dong}}},
\bauthor{\binits{B.J.} \bsnm{{Shappee}}},
\bauthor{\binits{J.L.} \bsnm{{Prieto}}},
\bauthor{\binits{S.W.} \bsnm{{Jha}}},
\bauthor{\binits{K.Z.} \bsnm{{Stanek}}},
\bauthor{\binits{T.W.-S.} \bsnm{{Holoien}}},
\bauthor{\binits{C.S.} \bsnm{{Kochanek}}},
\bauthor{\binits{T.A.} \bsnm{{Thompson}}},
\bauthor{\binits{N.} \bsnm{{Morrell}}},
\bauthor{\binits{I.B.} \bsnm{{Thompson}}},
\bauthor{\binits{U.} \bsnm{{Basu}}},
\bauthor{\binits{J.F.} \bsnm{{Beacom}}},
\bauthor{\binits{D.} \bsnm{{Bersier}}},
\bauthor{\binits{J.} \bsnm{{Brimacombe}}},
\bauthor{\binits{J.S.} \bsnm{{Brown}}},
\bauthor{\binits{F.} \bsnm{{Bufano}}},
\bauthor{\binits{P.} \bsnm{{Chen}}},
\bauthor{\binits{E.} \bsnm{{Conseil}}},
\bauthor{\binits{A.B.} \bsnm{{Danilet}}},
\bauthor{\binits{E.} \bsnm{{Falco}}},
\bauthor{\binits{D.} \bsnm{{Grupe}}},
\bauthor{\binits{S.} \bsnm{{Kiyota}}},
\bauthor{\binits{G.} \bsnm{{Masi}}},
\bauthor{\binits{B.} \bsnm{{Nicholls}}},
\bauthor{\binits{F.} \bsnm{{Olivares E.}}},
\bauthor{\binits{G.} \bsnm{{Pignata}}},
\bauthor{\binits{G.} \bsnm{{Pojmanski}}},
\bauthor{\binits{G.V.} \bsnm{{Simonian}}},
\bauthor{\binits{D.M.} \bsnm{{Szczygiel}}},
\bauthor{\binits{P.R.} \bsnm{{Wo{\'z}niak}}},
\batitle{{ASASSN-15lh: A highly super-luminous supernova}}.
\bjtitle{Science}
\bvolume{351},
\bfpage{257}--\blpage{260}
(\byear{2016}).
doi:\doiurl{10.1126/science.aac9613}
\end{barticle}
\endbibitem

\bibitem[\protect\citeauthoryear{{Esquej} et~al.}{2007}]{Esquej07}
\begin{barticle}
\bauthor{\binits{P.} \bsnm{{Esquej}}},
\bauthor{\binits{R.D.} \bsnm{{Saxton}}},
\bauthor{\binits{M.J.} \bsnm{{Freyberg}}},
\bauthor{\binits{A.M.} \bsnm{{Read}}},
\bauthor{\binits{B.} \bsnm{{Altieri}}},
\bauthor{\binits{M.} \bsnm{{Sanchez-Portal}}},
\bauthor{\binits{G.} \bsnm{{Hasinger}}},
\batitle{{Candidate tidal disruption events from the XMM-Newton slew survey}}.
\bjtitle{\aap}
\bvolume{462},
\bfpage{49}--\blpage{52}
(\byear{2007}).
doi:\doiurl{10.1051/0004-6361:20066072}
\end{barticle}
\endbibitem

\bibitem[\protect\citeauthoryear{{Esquej} et~al.}{2008}]{Esquej08}
\begin{barticle}
\bauthor{\binits{P.} \bsnm{{Esquej}}},
\bauthor{\binits{R.D.} \bsnm{{Saxton}}},
\bauthor{\binits{S.} \bsnm{{Komossa}}},
\bauthor{\binits{A.M.} \bsnm{{Read}}},
\bauthor{\binits{M.J.} \bsnm{{Freyberg}}},
\bauthor{\binits{G.} \bsnm{{Hasinger}}},
\bauthor{\binits{D.A.} \bsnm{{Garc{\'{\i}}a-Hern{\'a}ndez}}},
\bauthor{\binits{H.} \bsnm{{Lu}}},
\bauthor{\binits{J.R.} \bsnm{{Zaur{\'{\i}}n}}},
\bauthor{\binits{M.} \bsnm{{S{\'a}nchez-Portal}}},
\bauthor{\binits{H.} \bsnm{{Zhou}}},
\batitle{{Evolution of tidal disruption candidates discovered by XMM-Newton}}.
\bjtitle{\aap}
\bvolume{489},
\bfpage{543}--\blpage{554}
(\byear{2008}).
doi:\doiurl{10.1051/0004-6361:200810110}
\end{barticle}
\endbibitem

\bibitem[\protect\citeauthoryear{{Frederick} et~al.}{2019}]{Frederick19}
\begin{barticle}
\bauthor{\binits{S.} \bsnm{{Frederick}}},
\bauthor{\binits{S.} \bsnm{{Gezari}}},
\bauthor{\binits{M.J.} \bsnm{{Graham}}},
\bauthor{\binits{S.B.} \bsnm{{Cenko}}},
\bauthor{\binits{S.} \bsnm{{van Velzen}}},
\bauthor{\binits{D.} \bsnm{{Stern}}},
\bauthor{\binits{N.} \bsnm{{Blagorodnova}}},
\bauthor{\binits{S.R.} \bsnm{{Kulkarni}}},
\bauthor{\binits{L.} \bsnm{{Yan}}},
\bauthor{\binits{K.} \bsnm{{De}}},
\bauthor{\binits{U.C.} \bsnm{{Fremling}}},
\bauthor{\binits{T.} \bsnm{{Hung}}},
\bauthor{\binits{E.} \bsnm{{Kara}}},
\bauthor{\binits{D.L.} \bsnm{{Shupe}}},
\bauthor{\binits{C.} \bsnm{{Ward}}},
\bauthor{\binits{E.C.} \bsnm{{Bellm}}},
\bauthor{\binits{R.} \bsnm{{Dekany}}},
\bauthor{\binits{D.A.} \bsnm{{Duev}}},
\bauthor{\binits{U.} \bsnm{{Feindt}}},
\bauthor{\binits{M.} \bsnm{{Giomi}}},
\bauthor{\binits{T.} \bsnm{{Kupfer}}},
\bauthor{\binits{R.R.} \bsnm{{Laher}}},
\bauthor{\binits{F.J.} \bsnm{{Masci}}},
\bauthor{\binits{A.A.} \bsnm{{Miller}}},
\bauthor{\binits{J.D.} \bsnm{{Neill}}},
\bauthor{\binits{C.-C.} \bsnm{{Ngeow}}},
\bauthor{\binits{M.T.} \bsnm{{Patterson}}},
\bauthor{\binits{M.} \bsnm{{Porter}}},
\bauthor{\binits{B.} \bsnm{{Rusholme}}},
\bauthor{\binits{J.} \bsnm{{Sollerman}}},
\bauthor{\binits{R.} \bsnm{{Walters}}},
\batitle{{A New Class of Changing-look LINERs}}.
\bjtitle{\apj}
\bvolume{883}(\bissue{1}),
\bfpage{31}
(\byear{2019}).
doi:\doiurl{10.3847/1538-4357/ab3a38}
\end{barticle}
\endbibitem

\bibitem[\protect\citeauthoryear{{Frieman} et~al.}{2008}]{frieman08}
\begin{barticle}
\bauthor{\binits{J.A.} \bsnm{{Frieman}}},
\bauthor{\binits{B.} \bsnm{{Bassett}}},
\bauthor{\binits{A.} \bsnm{{Becker}}},
\bauthor{\binits{C.} \bsnm{{Choi}}},
\bauthor{\binits{D.} \bsnm{{Cinabro}}},
\bauthor{\binits{F.} \bsnm{{DeJongh}}},
\bauthor{\binits{D.L.} \bsnm{{Depoy}}},
\bauthor{\binits{B.} \bsnm{{Dilday}}},
\bauthor{\binits{M.} \bsnm{{Doi}}},
\bauthor{\binits{P.M.} \bsnm{{Garnavich}}},
\bauthor{\binits{C.J.} \bsnm{{Hogan}}},
\bauthor{\binits{J.} \bsnm{{Holtzman}}},
\bauthor{\binits{M.} \bsnm{{Im}}},
\bauthor{\binits{S.} \bsnm{{Jha}}},
\bauthor{\binits{R.} \bsnm{{Kessler}}},
\bauthor{\binits{K.} \bsnm{{Konishi}}},
\bauthor{\binits{H.} \bsnm{{Lampeitl}}},
\bauthor{\binits{J.} \bsnm{{Marriner}}},
\bauthor{\binits{J.L.} \bsnm{{Marshall}}},
\bauthor{\binits{D.} \bsnm{{McGinnis}}},
\bauthor{\binits{G.} \bsnm{{Miknaitis}}},
\bauthor{\binits{R.C.} \bsnm{{Nichol}}},
\bauthor{\binits{J.L.} \bsnm{{Prieto}}},
\bauthor{\binits{A.G.} \bsnm{{Riess}}},
\bauthor{\binits{M.W.} \bsnm{{Richmond}}},
\bauthor{\binits{R.} \bsnm{{Romani}}},
\bauthor{\binits{M.} \bsnm{{Sako}}},
\bauthor{\binits{D.P.} \bsnm{{Schneider}}},
\bauthor{\binits{M.} \bsnm{{Smith}}},
\bauthor{\binits{N.} \bsnm{{Takanashi}}},
\bauthor{\binits{K.} \bsnm{{Tokita}}},
\bauthor{\binits{K.} \bsnm{{van der Heyden}}},
\bauthor{\binits{N.} \bsnm{{Yasuda}}},
\bauthor{\binits{C.} \bsnm{{Zheng}}},
\bauthor{\binits{J.} \bsnm{{Adelman-McCarthy}}},
\bauthor{\binits{J.} \bsnm{{Annis}}},
\bauthor{\binits{R.J.} \bsnm{{Assef}}},
\bauthor{\binits{J.} \bsnm{{Barentine}}},
\bauthor{\binits{R.} \bsnm{{Bender}}},
\bauthor{\binits{R.D.} \bsnm{{Blandford}}},
\bauthor{\binits{W.N.} \bsnm{{Boroski}}},
\bauthor{\binits{M.} \bsnm{{Bremer}}},
\bauthor{\binits{H.} \bsnm{{Brewington}}},
\bauthor{\binits{C.A.} \bsnm{{Collins}}},
\bauthor{\binits{A.} \bsnm{{Crotts}}},
\bauthor{\binits{J.} \bsnm{{Dembicky}}},
\bauthor{\binits{J.} \bsnm{{Eastman}}},
\bauthor{\binits{A.} \bsnm{{Edge}}},
\bauthor{\binits{E.} \bsnm{{Edmondson}}},
\bauthor{\binits{E.} \bsnm{{Elson}}},
\bauthor{\binits{M.E.} \bsnm{{Eyler}}},
\bauthor{\binits{A.V.} \bsnm{{Filippenko}}},
\bauthor{\binits{R.J.} \bsnm{{Foley}}},
\bauthor{\binits{S.} \bsnm{{Frank}}},
\bauthor{\binits{A.} \bsnm{{Goobar}}},
\bauthor{\binits{T.} \bsnm{{Gueth}}},
\bauthor{\binits{J.E.} \bsnm{{Gunn}}},
\bauthor{\binits{M.} \bsnm{{Harvanek}}},
\bauthor{\binits{U.} \bsnm{{Hopp}}},
\bauthor{\binits{Y.} \bsnm{{Ihara}}},
\bauthor{\binits{{\v Z}.} \bsnm{{Ivezi{\'c}}}},
\bauthor{\binits{S.} \bsnm{{Kahn}}},
\bauthor{\binits{J.} \bsnm{{Kaplan}}},
\bauthor{\binits{S.} \bsnm{{Kent}}},
\bauthor{\binits{W.} \bsnm{{Ketzeback}}},
\bauthor{\binits{S.J.} \bsnm{{Kleinman}}},
\bauthor{\binits{W.} \bsnm{{Kollatschny}}},
\bauthor{\binits{R.G.} \bsnm{{Kron}}},
\bauthor{\binits{J.} \bsnm{{Krzesi{\'n}ski}}},
\bauthor{\binits{D.} \bsnm{{Lamenti}}},
\bauthor{\binits{G.} \bsnm{{Leloudas}}},
\bauthor{\binits{H.} \bsnm{{Lin}}},
\bauthor{\binits{D.C.} \bsnm{{Long}}},
\bauthor{\binits{J.} \bsnm{{Lucey}}},
\bauthor{\binits{R.H.} \bsnm{{Lupton}}},
\bauthor{\binits{E.} \bsnm{{Malanushenko}}},
\bauthor{\binits{V.} \bsnm{{Malanushenko}}},
\bauthor{\binits{R.J.} \bsnm{{McMillan}}},
\bauthor{\binits{J.} \bsnm{{Mendez}}},
\bauthor{\binits{C.W.} \bsnm{{Morgan}}},
\bauthor{\binits{T.} \bsnm{{Morokuma}}},
\bauthor{\binits{A.} \bsnm{{Nitta}}},
\bauthor{\binits{L.} \bsnm{{Ostman}}},
\bauthor{\binits{K.} \bsnm{{Pan}}},
\bauthor{\binits{C.M.} \bsnm{{Rockosi}}},
\bauthor{\binits{A.K.} \bsnm{{Romer}}},
\bauthor{\binits{P.} \bsnm{{Ruiz-Lapuente}}},
\bauthor{\binits{G.} \bsnm{{Saurage}}},
\bauthor{\binits{K.} \bsnm{{Schlesinger}}},
\bauthor{\binits{S.A.} \bsnm{{Snedden}}},
\bauthor{\binits{J.} \bsnm{{Sollerman}}},
\bauthor{\binits{C.} \bsnm{{Stoughton}}},
\bauthor{\binits{M.} \bsnm{{Stritzinger}}},
\bauthor{\binits{M.} \bsnm{{Subba Rao}}},
\bauthor{\binits{D.} \bsnm{{Tucker}}},
\bauthor{\binits{P.} \bsnm{{Vaisanen}}},
\bauthor{\binits{L.C.} \bsnm{{Watson}}},
\bauthor{\binits{S.} \bsnm{{Watters}}},
\bauthor{\binits{J.C.} \bsnm{{Wheeler}}},
\bauthor{\binits{B.} \bsnm{{Yanny}}},
\bauthor{\binits{D.} \bsnm{{York}}},
\batitle{{The Sloan Digital Sky Survey-II Supernova Survey: Technical
  Summary}}.
\bjtitle{\aj}
\bvolume{135},
\bfpage{338}--\blpage{347}
(\byear{2008}).
doi:\doiurl{10.1088/0004-6256/135/1/338}
\end{barticle}
\endbibitem

\bibitem[\protect\citeauthoryear{{Gaskell} and {Rojas Lobos}}{2014}]{Gaskell14}
\begin{barticle}
\bauthor{\binits{C.M.} \bsnm{{Gaskell}}},
\bauthor{\binits{P.A.} \bsnm{{Rojas Lobos}}},
\batitle{{The production of strong, broad He II emission after the tidal
  disruption of a main-sequence star by a supermassive black hole}}.
\bjtitle{\mnras}
\bvolume{438},
\bfpage{36}--\blpage{40}
(\byear{2014}).
doi:\doiurl{10.1093/mnrasl/slt154}
\end{barticle}
\endbibitem

\bibitem[\protect\citeauthoryear{{Gezari} et~al.}{2006}]{Gezari06}
\begin{barticle}
\bauthor{\binits{S.} \bsnm{{Gezari}}},
\bauthor{\binits{D.C.} \bsnm{{Martin}}},
\bauthor{\binits{B.} \bsnm{{Milliard}}},
\bauthor{\binits{S.} \bsnm{{Basa}}},
\bauthor{\binits{J.P.} \bsnm{{Halpern}}},
\bauthor{\binits{K.} \bsnm{{Forster}}},
\bauthor{\binits{P.G.} \bsnm{{Friedman}}},
\bauthor{\binits{P.} \bsnm{{Morrissey}}},
\bauthor{\binits{S.G.} \bsnm{{Neff}}},
\bauthor{\binits{D.} \bsnm{{Schiminovich}}},
\bauthor{\binits{M.} \bsnm{{Seibert}}},
\bauthor{\binits{T.} \bsnm{{Small}}},
\bauthor{\binits{T.K.} \bsnm{{Wyder}}},
\batitle{{Ultraviolet Detection of the Tidal Disruption of a Star by a
  Supermassive Black Hole}}.
\bjtitle{\apjl}
\bvolume{653},
\bfpage{25}--\blpage{28}
(\byear{2006}).
doi:\doiurl{10.1086/509918}
\end{barticle}
\endbibitem

\bibitem[\protect\citeauthoryear{{Gezari} et~al.}{2008}]{Gezari08}
\begin{barticle}
\bauthor{\binits{S.} \bsnm{{Gezari}}},
\bauthor{\binits{S.} \bsnm{{Basa}}},
\bauthor{\binits{D.C.} \bsnm{{Martin}}},
\bauthor{\binits{G.} \bsnm{{Bazin}}},
\bauthor{\binits{K.} \bsnm{{Forster}}},
\bauthor{\binits{B.} \bsnm{{Milliard}}},
\bauthor{\binits{J.P.} \bsnm{{Halpern}}},
\bauthor{\binits{P.G.} \bsnm{{Friedman}}},
\bauthor{\binits{P.} \bsnm{{Morrissey}}},
\bauthor{\binits{S.G.} \bsnm{{Neff}}},
\bauthor{\binits{D.} \bsnm{{Schiminovich}}},
\bauthor{\binits{M.} \bsnm{{Seibert}}},
\bauthor{\binits{T.} \bsnm{{Small}}},
\bauthor{\binits{T.K.} \bsnm{{Wyder}}},
\batitle{{UV/Optical Detections of Candidate Tidal Disruption Events by GALEX
  and CFHTLS}}.
\bjtitle{\apj}
\bvolume{676},
\bfpage{944}--\blpage{969}
(\byear{2008}).
doi:\doiurl{10.1086/529008}
\end{barticle}
\endbibitem

\bibitem[\protect\citeauthoryear{{Gezari} et~al.}{2009}]{Gezari09}
\begin{barticle}
\bauthor{\binits{S.} \bsnm{{Gezari}}},
\bauthor{\binits{T.} \bsnm{{Heckman}}},
\bauthor{\binits{S.B.} \bsnm{{Cenko}}},
\bauthor{\binits{M.} \bsnm{{Eracleous}}},
\bauthor{\binits{K.} \bsnm{{Forster}}},
\bauthor{\binits{T.S.} \bsnm{{Gon{\c c}alves}}},
\bauthor{\binits{D.C.} \bsnm{{Martin}}},
\bauthor{\binits{P.} \bsnm{{Morrissey}}},
\bauthor{\binits{S.G.} \bsnm{{Neff}}},
\bauthor{\binits{M.} \bsnm{{Seibert}}},
\bauthor{\binits{D.} \bsnm{{Schiminovich}}},
\bauthor{\binits{T.K.} \bsnm{{Wyder}}},
\batitle{{Luminous Thermal Flares from Quiescent Supermassive Black Holes}}.
\bjtitle{\apj}
\bvolume{698},
\bfpage{1367}--\blpage{1379}
(\byear{2009}).
doi:\doiurl{10.1088/0004-637X/698/2/1367}
\end{barticle}
\endbibitem

\bibitem[\protect\citeauthoryear{{Gezari} et~al.}{2012}]{Gezari12}
\begin{barticle}
\bauthor{\binits{S.} \bsnm{{Gezari}}},
\bauthor{\binits{R.} \bsnm{{Chornock}}},
\bauthor{\binits{A.} \bsnm{{Rest}}},
\bauthor{\binits{M.E.} \bsnm{{Huber}}},
\bauthor{\binits{K.} \bsnm{{Forster}}},
\bauthor{\binits{E.} \bsnm{{Berger}}},
\bauthor{\binits{P.J.} \bsnm{{Challis}}},
\bauthor{\binits{J.D.} \bsnm{{Neill}}},
\bauthor{\binits{D.C.} \bsnm{{Martin}}},
\bauthor{\binits{T.} \bsnm{{Heckman}}},
\bauthor{\binits{A.} \bsnm{{Lawrence}}},
\bauthor{\binits{C.} \bsnm{{Norman}}},
\bauthor{\binits{G.} \bsnm{{Narayan}}},
\bauthor{\binits{R.J.} \bsnm{{Foley}}},
\bauthor{\binits{G.H.} \bsnm{{Marion}}},
\bauthor{\binits{D.} \bsnm{{Scolnic}}},
\bauthor{\binits{L.} \bsnm{{Chomiuk}}},
\bauthor{\binits{A.} \bsnm{{Soderberg}}},
\bauthor{\binits{K.} \bsnm{{Smith}}},
\bauthor{\binits{R.P.} \bsnm{{Kirshner}}},
\bauthor{\binits{A.G.} \bsnm{{Riess}}},
\bauthor{\binits{S.J.} \bsnm{{Smartt}}},
\bauthor{\binits{C.W.} \bsnm{{Stubbs}}},
\bauthor{\binits{J.L.} \bsnm{{Tonry}}},
\bauthor{\binits{W.M.} \bsnm{{Wood-Vasey}}},
\bauthor{\binits{W.S.} \bsnm{{Burgett}}},
\bauthor{\binits{K.C.} \bsnm{{Chambers}}},
\bauthor{\binits{T.} \bsnm{{Grav}}},
\bauthor{\binits{J.N.} \bsnm{{Heasley}}},
\bauthor{\binits{N.} \bsnm{{Kaiser}}},
\bauthor{\binits{R.-P.} \bsnm{{Kudritzki}}},
\bauthor{\binits{E.A.} \bsnm{{Magnier}}},
\bauthor{\binits{J.S.} \bsnm{{Morgan}}},
\bauthor{\binits{P.A.} \bsnm{{Price}}},
\batitle{{An ultraviolet-optical flare from the tidal disruption of a
  helium-rich stellar core}}.
\bjtitle{\nat}
\bvolume{485},
\bfpage{217}--\blpage{220}
(\byear{2012}).
doi:\doiurl{10.1038/nature10990}
\end{barticle}
\endbibitem

\bibitem[\protect\citeauthoryear{{Gezari} et~al.}{2013}]{Gezari13}
\begin{barticle}
\bauthor{\binits{S.} \bsnm{{Gezari}}},
\bauthor{\binits{D.C.} \bsnm{{Martin}}},
\bauthor{\binits{K.} \bsnm{{Forster}}},
\bauthor{\binits{J.D.} \bsnm{{Neill}}},
\bauthor{\binits{M.} \bsnm{{Huber}}},
\bauthor{\binits{T.} \bsnm{{Heckman}}},
\bauthor{\binits{L.} \bsnm{{Bianchi}}},
\bauthor{\binits{P.} \bsnm{{Morrissey}}},
\bauthor{\binits{S.G.} \bsnm{{Neff}}},
\bauthor{\binits{M.} \bsnm{{Seibert}}},
\bauthor{\binits{D.} \bsnm{{Schiminovich}}},
\bauthor{\binits{T.K.} \bsnm{{Wyder}}},
\bauthor{\binits{W.S.} \bsnm{{Burgett}}},
\bauthor{\binits{K.C.} \bsnm{{Chambers}}},
\bauthor{\binits{N.} \bsnm{{Kaiser}}},
\bauthor{\binits{E.A.} \bsnm{{Magnier}}},
\bauthor{\binits{P.A.} \bsnm{{Price}}},
\bauthor{\binits{J.L.} \bsnm{{Tonry}}},
\batitle{{The GALEX Time Domain Survey. I. Selection and Classification of Over
  a Thousand Ultraviolet Variable Sources}}.
\bjtitle{\apj}
\bvolume{766},
\bfpage{60}
(\byear{2013}).
doi:\doiurl{10.1088/0004-637X/766/1/60}
\end{barticle}
\endbibitem

\bibitem[\protect\citeauthoryear{{Gezari} et~al.}{2015}]{Gezari15}
\begin{barticle}
\bauthor{\binits{S.} \bsnm{{Gezari}}},
\bauthor{\binits{R.} \bsnm{{Chornock}}},
\bauthor{\binits{A.} \bsnm{{Lawrence}}},
\bauthor{\binits{A.} \bsnm{{Rest}}},
\bauthor{\binits{D.O.} \bsnm{{Jones}}},
\bauthor{\binits{E.} \bsnm{{Berger}}},
\bauthor{\binits{P.M.} \bsnm{{Challis}}},
\bauthor{\binits{G.} \bsnm{{Narayan}}},
\batitle{{PS1-10jh Continues to Follow the Fallback Accretion Rate of a Tidally
  Disrupted Star}}.
\bjtitle{\apjl}
\bvolume{815},
\bfpage{5}
(\byear{2015}).
doi:\doiurl{10.1088/2041-8205/815/1/L5}
\end{barticle}
\endbibitem

\bibitem[\protect\citeauthoryear{{Godoy-Rivera} et~al.}{2017a}]{Godoy-Rivera17}
\begin{barticle}
\bauthor{\binits{D.} \bsnm{{Godoy-Rivera}}},
\bauthor{\binits{K.Z.} \bsnm{{Stanek}}},
\bauthor{\binits{C.S.} \bsnm{{Kochanek}}},
\bauthor{\binits{P.} \bsnm{{Chen}}},
\bauthor{\binits{S.} \bsnm{{Dong}}},
\bauthor{\binits{J.L.} \bsnm{{Prieto}}},
\bauthor{\binits{B.J.} \bsnm{{Shappee}}},
\bauthor{\binits{S.W.} \bsnm{{Jha}}},
\bauthor{\binits{R.J.} \bsnm{{Foley}}},
\bauthor{\binits{Y.-C.} \bsnm{{Pan}}},
\bauthor{\binits{T.W.-S.} \bsnm{{Holoien}}},
\bauthor{\binits{T.A.} \bsnm{{Thompson}}},
\bauthor{\binits{D.} \bsnm{{Grupe}}},
\bauthor{\binits{J.F.} \bsnm{{Beacom}}},
\batitle{{The unexpected, long-lasting, UV rebrightening of the superluminous
  supernova ASASSN-15lh}}.
\bjtitle{\mnras}
\bvolume{466},
\bfpage{1428}--\blpage{1443}
(\byear{2017}a).
doi:\doiurl{10.1093/mnras/stw3237}
\end{barticle}
\endbibitem

\bibitem[\protect\citeauthoryear{{Godoy-Rivera}
  et~al.}{2017b}]{Godoy-Rivera2017}
\begin{barticle}
\bauthor{\binits{D.} \bsnm{{Godoy-Rivera}}},
\bauthor{\binits{K.Z.} \bsnm{{Stanek}}},
\bauthor{\binits{C.S.} \bsnm{{Kochanek}}},
\bauthor{\binits{P.} \bsnm{{Chen}}},
\bauthor{\binits{S.} \bsnm{{Dong}}},
\bauthor{\binits{J.L.} \bsnm{{Prieto}}},
\bauthor{\binits{B.J.} \bsnm{{Shappee}}},
\bauthor{\binits{S.W.} \bsnm{{Jha}}},
\bauthor{\binits{R.J.} \bsnm{{Foley}}},
\bauthor{\binits{Y.-C.} \bsnm{{Pan}}},
\bauthor{\binits{T.W.-S.} \bsnm{{Holoien}}},
\bauthor{\binits{T.A.} \bsnm{{Thompson}}},
\bauthor{\binits{D.} \bsnm{{Grupe}}},
\bauthor{\binits{J.F.} \bsnm{{Beacom}}},
\batitle{{The unexpected, long-lasting, UV rebrightening of the superluminous
  supernova ASASSN-15lh}}.
\bjtitle{\mnras}
\bvolume{466}(\bissue{2}),
\bfpage{1428}--\blpage{1443}
(\byear{2017}b).
doi:\doiurl{10.1093/mnras/stw3237}
\end{barticle}
\endbibitem

\bibitem[\protect\citeauthoryear{{Gomez} et~al.}{2020}]{gomez20}
\begin{botherref}
\oauthor{\binits{S.} \bsnm{{Gomez}}},
\oauthor{\binits{M.} \bsnm{{Nicholl}}},
\oauthor{\binits{P.} \bsnm{{Short}}},
\oauthor{\binits{R.} \bsnm{{Margutti}}},
\oauthor{\binits{K.D.} \bsnm{{Alexander}}},
\oauthor{\binits{P.K.} \bsnm{{Blanchard}}},
\oauthor{\binits{E.} \bsnm{{Berger}}},
\oauthor{\binits{T.} \bsnm{{Eftekhari}}},
\oauthor{\binits{S.} \bsnm{{Schulze}}},
\oauthor{\binits{J.} \bsnm{{Anderson}}},
\oauthor{\binits{I.} \bsnm{{Arcavi}}},
\oauthor{\binits{R.} \bsnm{{Chornock}}},
\oauthor{\binits{P.S.} \bsnm{{Cowperthwaite}}},
\oauthor{\binits{L.} \bsnm{{Galbany}}},
\oauthor{\binits{L.J.} \bsnm{{Herzog}}},
\oauthor{\binits{D.} \bsnm{{Hiramatsu}}},
\oauthor{\binits{G.} \bsnm{{Hosseinzadeh}}},
\oauthor{\binits{T.} \bsnm{{Laskar}}},
\oauthor{\binits{T.E.} \bsnm{{M{\"u}ller Bravo}}},
\oauthor{\binits{L.} \bsnm{{Patton}}},
\oauthor{\binits{G.} \bsnm{{Terreran}}},
{The Tidal Disruption Event AT 2018hyz II: Light Curve Modeling of a Partially
  Disrupted Star}.
arXiv e-prints,
2003--05469
(2020)
\end{botherref}
\endbibitem

\bibitem[\protect\citeauthoryear{{Greiner} et~al.}{2000}]{Greiner00}
\begin{barticle}
\bauthor{\binits{J.} \bsnm{{Greiner}}},
\bauthor{\binits{R.} \bsnm{{Schwarz}}},
\bauthor{\binits{S.} \bsnm{{Zharikov}}},
\bauthor{\binits{M.} \bsnm{{Orio}}},
\batitle{{RX J1420.4+5334 - another tidal disruption event?}}
\bjtitle{\aap}
\bvolume{362},
\bfpage{25}--\blpage{28}
(\byear{2000})
\end{barticle}
\endbibitem

\bibitem[\protect\citeauthoryear{{Grupe} et~al.}{1999}]{Grupe99}
\begin{barticle}
\bauthor{\binits{D.} \bsnm{{Grupe}}},
\bauthor{\binits{H.-C.} \bsnm{{Thomas}}},
\bauthor{\binits{K.M.} \bsnm{{Leighly}}},
\batitle{{RX J1624.9+7554: a new X-ray transient AGN}}.
\bjtitle{\aap}
\bvolume{350},
\bfpage{31}--\blpage{34}
(\byear{1999})
\end{barticle}
\endbibitem

\bibitem[\protect\citeauthoryear{{Guillochon} and
  {Ramirez-Ruiz}}{2013}]{Guillochon13}
\begin{barticle}
\bauthor{\binits{J.} \bsnm{{Guillochon}}},
\bauthor{\binits{E.} \bsnm{{Ramirez-Ruiz}}},
\batitle{{Hydrodynamical Simulations to Determine the Feeding Rate of Black
  Holes by the Tidal Disruption of Stars: The Importance of the Impact
  Parameter and Stellar Structure}}.
\bjtitle{\apj}
\bvolume{767},
\bfpage{25}
(\byear{2013}).
doi:\doiurl{10.1088/0004-637X/767/1/25}
\end{barticle}
\endbibitem

\bibitem[\protect\citeauthoryear{{Guillochon} et~al.}{2014}]{guillochon14}
\begin{barticle}
\bauthor{\binits{J.} \bsnm{{Guillochon}}},
\bauthor{\binits{H.} \bsnm{{Manukian}}},
\bauthor{\binits{E.} \bsnm{{Ramirez-Ruiz}}},
\batitle{{PS1-10jh: The Disruption of a Main-sequence Star of Near-solar
  Composition}}.
\bjtitle{\apj}
\bvolume{783},
\bfpage{23}
(\byear{2014}).
doi:\doiurl{10.1088/0004-637X/783/1/23}
\end{barticle}
\endbibitem

\bibitem[\protect\citeauthoryear{{G{\"u}ltekin} et~al.}{2009}]{Gultekin09}
\begin{barticle}
\bauthor{\binits{K.} \bsnm{{G{\"u}ltekin}}},
\bauthor{\binits{D.O.} \bsnm{{Richstone}}},
\bauthor{\binits{K.} \bsnm{{Gebhardt}}},
\bauthor{\binits{T.R.} \bsnm{{Lauer}}},
\bauthor{\binits{S.} \bsnm{{Tremaine}}},
\bauthor{\binits{M.C.} \bsnm{{Aller}}},
\bauthor{\binits{R.} \bsnm{{Bender}}},
\bauthor{\binits{A.} \bsnm{{Dressler}}},
\bauthor{\binits{S.M.} \bsnm{{Faber}}},
\bauthor{\binits{A.V.} \bsnm{{Filippenko}}},
\bauthor{\binits{R.} \bsnm{{Green}}},
\bauthor{\binits{L.C.} \bsnm{{Ho}}},
\bauthor{\binits{J.} \bsnm{{Kormendy}}},
\bauthor{\binits{J.} \bsnm{{Magorrian}}},
\bauthor{\binits{J.} \bsnm{{Pinkney}}},
\bauthor{\binits{C.} \bsnm{{Siopis}}},
\batitle{{The M-{$\sigma$} and M-L Relations in Galactic Bulges, and
  Determinations of Their Intrinsic Scatter}}.
\bjtitle{\apj}
\bvolume{698},
\bfpage{198}--\blpage{221}
(\byear{2009}).
doi:\doiurl{10.1088/0004-637X/698/1/198}
\end{barticle}
\endbibitem

\bibitem[\protect\citeauthoryear{{Hamann} et~al.}{2019}]{Hamann2019}
\begin{barticle}
\bauthor{\binits{F.} \bsnm{{Hamann}}},
\bauthor{\binits{T.M.} \bsnm{{Tripp}}},
\bauthor{\binits{D.} \bsnm{{Rupke}}},
\bauthor{\binits{S.} \bsnm{{Veilleux}}},
\batitle{{On the emergence of THOUSANDS of absorption lines in the quasar PG
  1411+442: a clumpy high-column density outflow from the broad emission-line
  region?}}
\bjtitle{\mnras}
\bvolume{487}(\bissue{4}),
\bfpage{5041}--\blpage{5061}
(\byear{2019}).
doi:\doiurl{10.1093/mnras/stz1408}
\end{barticle}
\endbibitem

\bibitem[\protect\citeauthoryear{{Heckman}}{1980}]{Heckman80}
\begin{barticle}
\bauthor{\binits{T.M.} \bsnm{{Heckman}}},
\batitle{{An optical and radio survey of the nuclei of bright galaxies -
  Activity in normal galactic nuclei}}.
\bjtitle{\aap}
\bvolume{87},
\bfpage{152}--\blpage{164}
(\byear{1980})
\end{barticle}
\endbibitem

\bibitem[\protect\citeauthoryear{{Hills}}{1975}]{Hills75}
\begin{barticle}
\bauthor{\binits{J.G.} \bsnm{{Hills}}},
\batitle{{Possible power source of Seyfert galaxies and QSOs}}.
\bjtitle{\nat}
\bvolume{254},
\bfpage{295}--\blpage{298}
(\byear{1975}).
doi:\doiurl{10.1038/254295a0}
\end{barticle}
\endbibitem

\bibitem[\protect\citeauthoryear{{Holoien} et~al.}{2014}]{Holoien14}
\begin{barticle}
\bauthor{\binits{T.W.-S.} \bsnm{{Holoien}}},
\bauthor{\binits{J.L.} \bsnm{{Prieto}}},
\bauthor{\binits{D.} \bsnm{{Bersier}}},
\bauthor{\binits{C.S.} \bsnm{{Kochanek}}},
\bauthor{\binits{K.Z.} \bsnm{{Stanek}}},
\bauthor{\binits{B.J.} \bsnm{{Shappee}}},
\bauthor{\binits{D.} \bsnm{{Grupe}}},
\bauthor{\binits{U.} \bsnm{{Basu}}},
\bauthor{\binits{J.F.} \bsnm{{Beacom}}},
\bauthor{\binits{J.} \bsnm{{Brimacombe}}},
\bauthor{\binits{J.S.} \bsnm{{Brown}}},
\bauthor{\binits{A.B.} \bsnm{{Davis}}},
\bauthor{\binits{J.} \bsnm{{Jencson}}},
\bauthor{\binits{G.} \bsnm{{Pojmanski}}},
\bauthor{\binits{D.M.} \bsnm{{Szczygie{\l}}}},
\batitle{{ASASSN-14ae: a tidal disruption event at 200 Mpc}}.
\bjtitle{\mnras}
\bvolume{445},
\bfpage{3263}--\blpage{3277}
(\byear{2014}).
doi:\doiurl{10.1093/mnras/stu1922}
\end{barticle}
\endbibitem

\bibitem[\protect\citeauthoryear{{Holoien} et~al.}{2016a}]{Holoien16b}
\begin{barticle}
\bauthor{\binits{T.W.-S.} \bsnm{{Holoien}}},
\bauthor{\binits{C.S.} \bsnm{{Kochanek}}},
\bauthor{\binits{J.L.} \bsnm{{Prieto}}},
\bauthor{\binits{D.} \bsnm{{Grupe}}},
\bauthor{\binits{P.} \bsnm{{Chen}}},
\bauthor{\binits{D.} \bsnm{{Godoy-Rivera}}},
\bauthor{\binits{K.Z.} \bsnm{{Stanek}}},
\bauthor{\binits{B.J.} \bsnm{{Shappee}}},
\bauthor{\binits{S.} \bsnm{{Dong}}},
\bauthor{\binits{J.S.} \bsnm{{Brown}}},
\bauthor{\binits{U.} \bsnm{{Basu}}},
\bauthor{\binits{J.F.} \bsnm{{Beacom}}},
\bauthor{\binits{D.} \bsnm{{Bersier}}},
\bauthor{\binits{J.} \bsnm{{Brimacombe}}},
\bauthor{\binits{E.K.} \bsnm{{Carlson}}},
\bauthor{\binits{E.} \bsnm{{Falco}}},
\bauthor{\binits{E.} \bsnm{{Johnston}}},
\bauthor{\binits{B.F.} \bsnm{{Madore}}},
\bauthor{\binits{G.} \bsnm{{Pojmanski}}},
\bauthor{\binits{M.} \bsnm{{Seibert}}},
\batitle{{ASASSN-15oi: a rapidly evolving, luminous tidal disruption event at
  216 Mpc}}.
\bjtitle{\mnras}
\bvolume{463},
\bfpage{3813}--\blpage{3828}
(\byear{2016}a).
doi:\doiurl{10.1093/mnras/stw2272}
\end{barticle}
\endbibitem

\bibitem[\protect\citeauthoryear{{Holoien} et~al.}{2016b}]{Holoien16a}
\begin{barticle}
\bauthor{\binits{T.W.-S.} \bsnm{{Holoien}}},
\bauthor{\binits{C.S.} \bsnm{{Kochanek}}},
\bauthor{\binits{J.L.} \bsnm{{Prieto}}},
\bauthor{\binits{K.Z.} \bsnm{{Stanek}}},
\bauthor{\binits{S.} \bsnm{{Dong}}},
\bauthor{\binits{B.J.} \bsnm{{Shappee}}},
\bauthor{\binits{D.} \bsnm{{Grupe}}},
\bauthor{\binits{J.S.} \bsnm{{Brown}}},
\bauthor{\binits{U.} \bsnm{{Basu}}},
\bauthor{\binits{J.F.} \bsnm{{Beacom}}},
\bauthor{\binits{D.} \bsnm{{Bersier}}},
\bauthor{\binits{J.} \bsnm{{Brimacombe}}},
\bauthor{\binits{A.B.} \bsnm{{Danilet}}},
\bauthor{\binits{E.} \bsnm{{Falco}}},
\bauthor{\binits{Z.} \bsnm{{Guo}}},
\bauthor{\binits{J.} \bsnm{{Jose}}},
\bauthor{\binits{G.J.} \bsnm{{Herczeg}}},
\bauthor{\binits{F.} \bsnm{{Long}}},
\bauthor{\binits{G.} \bsnm{{Pojmanski}}},
\bauthor{\binits{G.V.} \bsnm{{Simonian}}},
\bauthor{\binits{D.M.} \bsnm{{Szczygie{\l}}}},
\bauthor{\binits{T.A.} \bsnm{{Thompson}}},
\bauthor{\binits{J.R.} \bsnm{{Thorstensen}}},
\bauthor{\binits{R.M.} \bsnm{{Wagner}}},
\bauthor{\binits{P.R.} \bsnm{{Wo{\'z}niak}}},
\batitle{{Six months of multiwavelength follow-up of the tidal disruption
  candidate ASASSN-14li and implied TDE rates from ASAS-SN}}.
\bjtitle{\mnras}
\bvolume{455},
\bfpage{2918}--\blpage{2935}
(\byear{2016}b).
doi:\doiurl{10.1093/mnras/stv2486}
\end{barticle}
\endbibitem

\bibitem[\protect\citeauthoryear{{Holoien} et~al.}{2016c}]{Holoien16}
\begin{barticle}
\bauthor{\binits{T.W.-S.} \bsnm{{Holoien}}},
\bauthor{\binits{C.S.} \bsnm{{Kochanek}}},
\bauthor{\binits{J.L.} \bsnm{{Prieto}}},
\bauthor{\binits{K.Z.} \bsnm{{Stanek}}},
\bauthor{\binits{S.} \bsnm{{Dong}}},
\bauthor{\binits{B.J.} \bsnm{{Shappee}}},
\bauthor{\binits{D.} \bsnm{{Grupe}}},
\bauthor{\binits{J.S.} \bsnm{{Brown}}},
\bauthor{\binits{U.} \bsnm{{Basu}}},
\bauthor{\binits{J.F.} \bsnm{{Beacom}}},
\bauthor{\binits{D.} \bsnm{{Bersier}}},
\bauthor{\binits{J.} \bsnm{{Brimacombe}}},
\bauthor{\binits{A.B.} \bsnm{{Danilet}}},
\bauthor{\binits{E.} \bsnm{{Falco}}},
\bauthor{\binits{Z.} \bsnm{{Guo}}},
\bauthor{\binits{J.} \bsnm{{Jose}}},
\bauthor{\binits{G.J.} \bsnm{{Herczeg}}},
\bauthor{\binits{F.} \bsnm{{Long}}},
\bauthor{\binits{G.} \bsnm{{Pojmanski}}},
\bauthor{\binits{G.V.} \bsnm{{Simonian}}},
\bauthor{\binits{D.M.} \bsnm{{Szczygie{\l}}}},
\bauthor{\binits{T.A.} \bsnm{{Thompson}}},
\bauthor{\binits{J.R.} \bsnm{{Thorstensen}}},
\bauthor{\binits{R.M.} \bsnm{{Wagner}}},
\bauthor{\binits{P.R.} \bsnm{{Wo{\'z}niak}}},
\batitle{{Six months of multiwavelength follow-up of the tidal disruption
  candidate ASASSN-14li and implied TDE rates from ASAS-SN}}.
\bjtitle{\mnras}
\bvolume{455},
\bfpage{2918}--\blpage{2935}
(\byear{2016}c).
doi:\doiurl{10.1093/mnras/stv2486}
\end{barticle}
\endbibitem

\bibitem[\protect\citeauthoryear{{Holoien} et~al.}{2019a}]{Holoien18a}
\begin{barticle}
\bauthor{\binits{T.W.-S.} \bsnm{{Holoien}}},
\bauthor{\binits{M.E.} \bsnm{{Huber}}},
\bauthor{\binits{B.J.} \bsnm{{Shappee}}},
\bauthor{\binits{M.} \bsnm{{Eracleous}}},
\bauthor{\binits{K.} \bsnm{{Auchettl}}},
\bauthor{\binits{J.S.} \bsnm{{Brown}}},
\bauthor{\binits{M.A.} \bsnm{{Tucker}}},
\bauthor{\binits{K.C.} \bsnm{{Chambers}}},
\bauthor{\binits{C.S.} \bsnm{{Kochanek}}},
\bauthor{\binits{K.Z.} \bsnm{{Stanek}}},
\bauthor{\binits{A.} \bsnm{{Rest}}},
\bauthor{\binits{D.} \bsnm{{Bersier}}},
\bauthor{\binits{R.S.} \bsnm{{Post}}},
\bauthor{\binits{G.} \bsnm{{Aldering}}},
\bauthor{\binits{K.A.} \bsnm{{Ponder}}},
\bauthor{\binits{J.D.} \bsnm{{Simon}}},
\bauthor{\binits{E.} \bsnm{{Kankare}}},
\bauthor{\binits{D.} \bsnm{{Dong}}},
\bauthor{\binits{G.} \bsnm{{Hallinan}}},
\bauthor{\binits{N.A.} \bsnm{{Reddy}}},
\bauthor{\binits{R.L.} \bsnm{{Sanders}}},
\bauthor{\binits{M.W.} \bsnm{{Topping}}},
\bauthor{\bsnm{{Pan-STARRS}}},
\bauthor{\binits{J.} \bsnm{{Bulger}}},
\bauthor{\binits{T.B.} \bsnm{{Lowe}}},
\bauthor{\binits{E.A.} \bsnm{{Magnier}}},
\bauthor{\binits{A.S.B.} \bsnm{{Schultz}}},
\bauthor{\binits{C.Z.} \bsnm{{Waters}}},
\bauthor{\binits{M.} \bsnm{{Willman}}},
\bauthor{\binits{D.} \bsnm{{Wright}}},
\bauthor{\binits{D.R.} \bsnm{{Young}}},
\bauthor{\bsnm{{ASAS-SN}}},
\bauthor{\binits{S.} \bsnm{{Dong}}},
\bauthor{\binits{J.L.} \bsnm{{Prieto}}},
\bauthor{\binits{T.A.} \bsnm{{Thompson}}},
\bauthor{\bsnm{{ATLAS}}},
\bauthor{\binits{L.} \bsnm{{Denneau}}},
\bauthor{\binits{H.} \bsnm{{Flewelling}}},
\bauthor{\binits{A.N.} \bsnm{{Heinze}}},
\bauthor{\binits{S.J.} \bsnm{{Smartt}}},
\bauthor{\binits{K.W.} \bsnm{{Smith}}},
\bauthor{\binits{B.} \bsnm{{Stalder}}},
\bauthor{\binits{J.L.} \bsnm{{Tonry}}},
\bauthor{\binits{H.} \bsnm{{Weiland}}},
\batitle{{PS18kh: A New Tidal Disruption Event with a Non-axisymmetric
  Accretion Disk}}.
\bjtitle{\apj}
\bvolume{880}(\bissue{2}),
\bfpage{120}
(\byear{2019}a).
doi:\doiurl{10.3847/1538-4357/ab2ae1}
\end{barticle}
\endbibitem

\bibitem[\protect\citeauthoryear{{Holoien} et~al.}{2019b}]{Holoien19_kh}
\begin{barticle}
\bauthor{\binits{T.W.-S.} \bsnm{{Holoien}}},
\bauthor{\binits{M.E.} \bsnm{{Huber}}},
\bauthor{\binits{B.J.} \bsnm{{Shappee}}},
\bauthor{\binits{M.} \bsnm{{Eracleous}}},
\bauthor{\binits{K.} \bsnm{{Auchettl}}},
\bauthor{\binits{J.S.} \bsnm{{Brown}}},
\bauthor{\binits{M.A.} \bsnm{{Tucker}}},
\bauthor{\binits{K.C.} \bsnm{{Chambers}}},
\bauthor{\binits{C.S.} \bsnm{{Kochanek}}},
\bauthor{\binits{K.Z.} \bsnm{{Stanek}}},
\bauthor{\binits{A.} \bsnm{{Rest}}},
\bauthor{\binits{D.} \bsnm{{Bersier}}},
\bauthor{\binits{R.S.} \bsnm{{Post}}},
\bauthor{\binits{G.} \bsnm{{Aldering}}},
\bauthor{\binits{K.A.} \bsnm{{Ponder}}},
\bauthor{\binits{J.D.} \bsnm{{Simon}}},
\bauthor{\binits{E.} \bsnm{{Kankare}}},
\bauthor{\binits{D.} \bsnm{{Dong}}},
\bauthor{\binits{G.} \bsnm{{Hallinan}}},
\bauthor{\binits{N.A.} \bsnm{{Reddy}}},
\bauthor{\binits{R.L.} \bsnm{{Sanders}}},
\bauthor{\binits{M.W.} \bsnm{{Topping}}},
\bauthor{\bsnm{{Pan-STARRS}}},
\bauthor{\binits{J.} \bsnm{{Bulger}}},
\bauthor{\binits{T.B.} \bsnm{{Lowe}}},
\bauthor{\binits{E.A.} \bsnm{{Magnier}}},
\bauthor{\binits{A.S.B.} \bsnm{{Schultz}}},
\bauthor{\binits{C.Z.} \bsnm{{Waters}}},
\bauthor{\binits{M.} \bsnm{{Willman}}},
\bauthor{\binits{D.} \bsnm{{Wright}}},
\bauthor{\binits{D.R.} \bsnm{{Young}}},
\bauthor{\bsnm{{ASAS-SN}}},
\bauthor{\binits{S.} \bsnm{{Dong}}},
\bauthor{\binits{J.L.} \bsnm{{Prieto}}},
\bauthor{\binits{T.A.} \bsnm{{Thompson}}},
\bauthor{\bsnm{{ATLAS}}},
\bauthor{\binits{L.} \bsnm{{Denneau}}},
\bauthor{\binits{H.} \bsnm{{Flewelling}}},
\bauthor{\binits{A.N.} \bsnm{{Heinze}}},
\bauthor{\binits{S.J.} \bsnm{{Smartt}}},
\bauthor{\binits{K.W.} \bsnm{{Smith}}},
\bauthor{\binits{B.} \bsnm{{Stalder}}},
\bauthor{\binits{J.L.} \bsnm{{Tonry}}},
\bauthor{\binits{H.} \bsnm{{Weiland}}},
\batitle{{PS18kh: A New Tidal Disruption Event with a Non-axisymmetric
  Accretion Disk}}.
\bjtitle{\apj}
\bvolume{880}(\bissue{2}),
\bfpage{120}
(\byear{2019}b).
doi:\doiurl{10.3847/1538-4357/ab2ae1}
\end{barticle}
\endbibitem

\bibitem[\protect\citeauthoryear{{Holoien} et~al.}{2019c}]{Holoien19_bt}
\begin{barticle}
\bauthor{\binits{T.W.-S.} \bsnm{{Holoien}}},
\bauthor{\binits{P.J.} \bsnm{{Vallely}}},
\bauthor{\binits{K.} \bsnm{{Auchettl}}},
\bauthor{\binits{K.Z.} \bsnm{{Stanek}}},
\bauthor{\binits{C.S.} \bsnm{{Kochanek}}},
\bauthor{\binits{K.D.} \bsnm{{French}}},
\bauthor{\binits{J.L.} \bsnm{{Prieto}}},
\bauthor{\binits{B.J.} \bsnm{{Shappee}}},
\bauthor{\binits{J.S.} \bsnm{{Brown}}},
\bauthor{\binits{M.M.} \bsnm{{Fausnaugh}}},
\bauthor{\binits{S.} \bsnm{{Dong}}},
\bauthor{\binits{T.A.} \bsnm{{Thompson}}},
\bauthor{\binits{S.} \bsnm{{Bose}}},
\bauthor{\binits{J.M.M.} \bsnm{{Neustadt}}},
\bauthor{\binits{P.} \bsnm{{Cacella}}},
\bauthor{\binits{J.} \bsnm{{Brimacombe}}},
\bauthor{\binits{M.R.} \bsnm{{Kendurkar}}},
\bauthor{\binits{R.L.} \bsnm{{Beaton}}},
\bauthor{\binits{K.} \bsnm{{Boutsia}}},
\bauthor{\binits{L.} \bsnm{{Chomiuk}}},
\bauthor{\binits{T.} \bsnm{{Connor}}},
\bauthor{\binits{N.} \bsnm{{Morrell}}},
\bauthor{\binits{A.B.} \bsnm{{Newman}}},
\bauthor{\binits{G.C.} \bsnm{{Rudie}}},
\bauthor{\binits{L.} \bsnm{{Shishkovksy}}},
\bauthor{\binits{J.} \bsnm{{Strader}}},
\batitle{{Discovery and Early Evolution of ASASSN-19bt, the First TDE Detected
  by TESS}}.
\bjtitle{\apj}
\bvolume{883}(\bissue{2}),
\bfpage{111}
(\byear{2019}c).
doi:\doiurl{10.3847/1538-4357/ab3c66}
\end{barticle}
\endbibitem

\bibitem[\protect\citeauthoryear{{Holoien} et~al.}{2020}]{Holoien2020_18pg}
\begin{botherref}
\oauthor{\binits{T.W.-S.} \bsnm{{Holoien}}},
\oauthor{\binits{K.} \bsnm{{Auchettl}}},
\oauthor{\binits{M.A.} \bsnm{{Tucker}}},
\oauthor{\binits{B.J.} \bsnm{{Shappee}}},
\oauthor{\binits{S.G.} \bsnm{{Patel}}},
\oauthor{\binits{J.C.A.} \bsnm{{Miller-Jones}}},
\oauthor{\binits{B.} \bsnm{{Mockler}}},
\oauthor{\binits{D.N.} \bsnm{{Groenewald}}},
\oauthor{\binits{J.S.} \bsnm{{Brown}}},
\oauthor{\binits{C.S.} \bsnm{{Kochanek}}},
\oauthor{\binits{K.Z.} \bsnm{{Stanek}}},
\oauthor{\binits{P.} \bsnm{{Chen}}},
\oauthor{\binits{S.} \bsnm{{Dong}}},
\oauthor{\binits{J.L.} \bsnm{{Prieto}}},
\oauthor{\binits{T.A.} \bsnm{{Thompson}}},
\oauthor{\binits{R.L.} \bsnm{{Beaton}}},
\oauthor{\binits{T.} \bsnm{{Connor}}},
\oauthor{\binits{P.S.} \bsnm{{Cowperthwaite}}},
\oauthor{\binits{L.} \bsnm{{Dahmen}}},
\oauthor{\binits{K.D.} \bsnm{{French}}},
\oauthor{\binits{N.} \bsnm{{Morrell}}},
\oauthor{\binits{D.A.H.} \bsnm{{Buckley}}},
\oauthor{\binits{M.} \bsnm{{Gromadzki}}},
\oauthor{\binits{R.} \bsnm{{Roy}}},
\oauthor{\binits{D.A.} \bsnm{{Coulter}}},
\oauthor{\binits{G.} \bsnm{{Dimitriadis}}},
\oauthor{\binits{R.J.} \bsnm{{Foley}}},
\oauthor{\binits{C.D.} \bsnm{{Kilpatrick}}},
\oauthor{\binits{A.L.} \bsnm{{Piro}}},
\oauthor{\binits{C.} \bsnm{{Rojas-Bravo}}},
\oauthor{\binits{M.R.} \bsnm{{Siebert}}},
\oauthor{\binits{S.} \bsnm{{van Velzen}}},
{The Rise and Fall of ASASSN-18pg: Following a TDE from Early To Late Times}.
arXiv e-prints,
2003--13693
(2020)
\end{botherref}
\endbibitem

\bibitem[\protect\citeauthoryear{{Huber} et~al.}{2015}]{Huber15}
\begin{barticle}
\bauthor{\binits{M.} \bsnm{{Huber}}},
\bauthor{\binits{K.C.} \bsnm{{Chambers}}},
\bauthor{\binits{H.} \bsnm{{Flewelling}}},
\bauthor{\binits{M.} \bsnm{{Willman}}},
\bauthor{\binits{N.} \bsnm{{Primak}}},
\bauthor{\binits{A.} \bsnm{{Schultz}}},
\bauthor{\binits{B.} \bsnm{{Gibson}}},
\bauthor{\binits{E.} \bsnm{{Magnier}}},
\bauthor{\binits{C.} \bsnm{{Waters}}},
\bauthor{\binits{J.} \bsnm{{Tonry}}},
\bauthor{\binits{R.J.} \bsnm{{Wainscoat}}},
\bauthor{\binits{K.W.} \bsnm{{Smith}}},
\bauthor{\binits{D.} \bsnm{{Wright}}},
\bauthor{\binits{S.J.} \bsnm{{Smartt}}},
\bauthor{\binits{R.J.} \bsnm{{Foley}}},
\bauthor{\binits{S.W.} \bsnm{{Jha}}},
\bauthor{\binits{A.} \bsnm{{Rest}}},
\bauthor{\binits{D.} \bsnm{{Scolnic}}},
\batitle{{The Pan-STARRS Survey for Transients (PSST) - first announcement and
  public release}}.
\bjtitle{The Astronomer's Telegram}
\bvolume{7153},
\bfpage{1}
(\byear{2015})
\end{barticle}
\endbibitem

\bibitem[\protect\citeauthoryear{{Hung} et~al.}{2017}]{Hung17}
\begin{barticle}
\bauthor{\binits{T.} \bsnm{{Hung}}},
\bauthor{\binits{S.} \bsnm{{Gezari}}},
\bauthor{\binits{N.} \bsnm{{Blagorodnova}}},
\bauthor{\binits{N.} \bsnm{{Roth}}},
\bauthor{\binits{S.B.} \bsnm{{Cenko}}},
\bauthor{\binits{S.R.} \bsnm{{Kulkarni}}},
\bauthor{\binits{A.} \bsnm{{Horesh}}},
\bauthor{\binits{I.} \bsnm{{Arcavi}}},
\bauthor{\binits{C.} \bsnm{{McCully}}},
\bauthor{\binits{L.} \bsnm{{Yan}}},
\bauthor{\binits{R.} \bsnm{{Lunnan}}},
\bauthor{\binits{C.} \bsnm{{Fremling}}},
\bauthor{\binits{Y.} \bsnm{{Cao}}},
\bauthor{\binits{P.E.} \bsnm{{Nugent}}},
\bauthor{\binits{P.} \bsnm{{Wozniak}}},
\batitle{{Revisiting Optical Tidal Disruption Events with iPTF16axa}}.
\bjtitle{\apj}
\bvolume{842},
\bfpage{29}
(\byear{2017}).
doi:\doiurl{10.3847/1538-4357/aa7337}
\end{barticle}
\endbibitem

\bibitem[\protect\citeauthoryear{{Hung} et~al.}{2018}]{Hung18}
\begin{barticle}
\bauthor{\binits{T.} \bsnm{{Hung}}},
\bauthor{\binits{S.} \bsnm{{Gezari}}},
\bauthor{\binits{S.B.} \bsnm{{Cenko}}},
\bauthor{\binits{S.} \bsnm{{van Velzen}}},
\bauthor{\binits{N.} \bsnm{{Blagorodnova}}},
\bauthor{\binits{L.} \bsnm{{Yan}}},
\bauthor{\binits{S.R.} \bsnm{{Kulkarni}}},
\bauthor{\binits{R.} \bsnm{{Lunnan}}},
\bauthor{\binits{T.} \bsnm{{Kupfer}}},
\bauthor{\binits{G.} \bsnm{{Leloudas}}},
\bauthor{\binits{A.K.H.} \bsnm{{Kong}}},
\bauthor{\binits{P.E.} \bsnm{{Nugent}}},
\bauthor{\binits{C.} \bsnm{{Fremling}}},
\bauthor{\binits{R.R.} \bsnm{{Laher}}},
\bauthor{\binits{F.J.} \bsnm{{Masci}}},
\bauthor{\binits{Y.} \bsnm{{Cao}}},
\bauthor{\binits{R.} \bsnm{{Roy}}},
\bauthor{\binits{T.} \bsnm{{Petrushevska}}},
\batitle{{Sifting for Sapphires: Systematic Selection of Tidal Disruption
  Events in iPTF}}.
\bjtitle{The Astrophysical Journal Supplement Series}
\bvolume{238},
\bfpage{15}
(\byear{2018}).
doi:\doiurl{10.3847/1538-4365/aad8b1}
\end{barticle}
\endbibitem

\bibitem[\protect\citeauthoryear{{Hung} et~al.}{2019}]{Hung2019}
\begin{barticle}
\bauthor{\binits{T.} \bsnm{{Hung}}},
\bauthor{\binits{S.B.} \bsnm{{Cenko}}},
\bauthor{\binits{N.} \bsnm{{Roth}}},
\bauthor{\binits{S.} \bsnm{{Gezari}}},
\bauthor{\binits{S.} \bsnm{{Veilleux}}},
\bauthor{\binits{S.} \bsnm{{van Velzen}}},
\bauthor{\binits{C.M.} \bsnm{{Gaskell}}},
\bauthor{\binits{R.J.} \bsnm{{Foley}}},
\bauthor{\binits{N.} \bsnm{{Blagorodnova}}},
\bauthor{\binits{L.} \bsnm{{Yan}}},
\bauthor{\binits{M.J.} \bsnm{{Graham}}},
\bauthor{\binits{J.S.} \bsnm{{Brown}}},
\bauthor{\binits{M.R.} \bsnm{{Siebert}}},
\bauthor{\binits{S.} \bsnm{{Frederick}}},
\bauthor{\binits{C.} \bsnm{{Ward}}},
\bauthor{\binits{P.} \bsnm{{Gatkine}}},
\bauthor{\binits{A.} \bsnm{{Gal-Yam}}},
\bauthor{\binits{Y.} \bsnm{{Yang}}},
\bauthor{\binits{S.} \bsnm{{Schulze}}},
\bauthor{\binits{G.} \bsnm{{Dimitriadis}}},
\bauthor{\binits{T.} \bsnm{{Kupfer}}},
\bauthor{\binits{D.L.} \bsnm{{Shupe}}},
\bauthor{\binits{B.} \bsnm{{Rusholme}}},
\bauthor{\binits{F.J.} \bsnm{{Masci}}},
\bauthor{\binits{R.} \bsnm{{Riddle}}},
\bauthor{\binits{M.T.} \bsnm{{Soumagnac}}},
\bauthor{\binits{J.} \bsnm{{van Roestel}}},
\bauthor{\binits{R.} \bsnm{{Dekany}}},
\batitle{{Discovery of Highly Blueshifted Broad Balmer and Metastable Helium
  Absorption Lines in a Tidal Disruption Event}}.
\bjtitle{\apj}
\bvolume{879}(\bissue{2}),
\bfpage{119}
(\byear{2019}).
doi:\doiurl{10.3847/1538-4357/ab24de}
\end{barticle}
\endbibitem

\bibitem[\protect\citeauthoryear{{Hung} et~al.}{2020}]{Hung2020}
\begin{botherref}
\oauthor{\binits{T.} \bsnm{{Hung}}},
\oauthor{\binits{R.J.} \bsnm{{Foley}}},
\oauthor{\binits{E.} \bsnm{{Ramirez-Ruiz}}},
\oauthor{\binits{J.L.} \bsnm{{Dai}}},
\oauthor{\binits{K.} \bsnm{{Auchettl}}},
\oauthor{\binits{C.D.} \bsnm{{Kilpatrick}}},
\oauthor{\binits{B.} \bsnm{{Mockler}}},
\oauthor{\binits{J.} \bsnm{{Brown}}},
\oauthor{\binits{D.A.} \bsnm{{Coulter}}},
\oauthor{\binits{G.} \bsnm{{Dimitriadis}}},
\oauthor{\binits{T.} \bsnm{{Holoien}}},
\oauthor{\binits{J.} \bsnm{{Law-Smith}}},
\oauthor{\binits{A.L.} \bsnm{{Piro}}},
\oauthor{\binits{A.} \bsnm{{Rest}}},
\oauthor{\binits{C.} \bsnm{{Rojas-Bravo}}},
\oauthor{\binits{M.R.} \bsnm{{Siebert}}},
{Prompt Accretion Disk Formation in an X-Ray Faint Tidal Disruption Event}.
arXiv e-prints,
2003--09427
(2020)
\end{botherref}
\endbibitem

\bibitem[\protect\citeauthoryear{{Jiang} et~al.}{2016}]{Jiang16}
\begin{barticle}
\bauthor{\binits{N.} \bsnm{{Jiang}}},
\bauthor{\binits{L.} \bsnm{{Dou}}},
\bauthor{\binits{T.} \bsnm{{Wang}}},
\bauthor{\binits{C.} \bsnm{{Yang}}},
\bauthor{\binits{J.} \bsnm{{Lyu}}},
\bauthor{\binits{H.} \bsnm{{Zhou}}},
\batitle{{The WISE Detection of an Infrared Echo in Tidal Disruption Event
  ASASSN-14li}}.
\bjtitle{\apjl}
\bvolume{828},
\bfpage{14}
(\byear{2016}).
doi:\doiurl{10.3847/2041-8205/828/1/L14}
\end{barticle}
\endbibitem

\bibitem[\protect\citeauthoryear{{Kajava} et~al.}{2020}]{Kajava20}
\begin{botherref}
\oauthor{\binits{J.J.E.} \bsnm{{Kajava}}},
\oauthor{\binits{M.} \bsnm{{Giustini}}},
\oauthor{\binits{R.D.} \bsnm{{Saxton}}},
\oauthor{\binits{G.} \bsnm{{Miniutti}}},
{Rapid late-time X-ray brightening of the tidal disruption event OGLE16aaa}.
arXiv e-prints,
2006--11179
(2020)
\end{botherref}
\endbibitem

\bibitem[\protect\citeauthoryear{{Kastner} and {Bhatia}}{1996}]{Kastner1996}
\begin{barticle}
\bauthor{\binits{S.O.} \bsnm{{Kastner}}},
\bauthor{\binits{A.K.} \bsnm{{Bhatia}}},
\batitle{{The Bowen fluorescence lines: overview and re-analysis of the
  observations.}}
\bjtitle{\mnras}
\bvolume{279}(\bissue{4}),
\bfpage{1137}--\blpage{1156}
(\byear{1996}).
doi:\doiurl{10.1093/mnras/279.4.1137}
\end{barticle}
\endbibitem

\bibitem[\protect\citeauthoryear{{Kesden}}{2012a}]{Kesden12}
\begin{barticle}
\bauthor{\binits{M.} \bsnm{{Kesden}}},
\batitle{{Black-hole spin dependence in the light curves of tidal disruption
  events}}.
\bjtitle{\prd}
\bvolume{86}(\bissue{6}),
\bfpage{064026}
(\byear{2012}a).
doi:\doiurl{10.1103/PhysRevD.86.064026}
\end{barticle}
\endbibitem

\bibitem[\protect\citeauthoryear{{Kesden}}{2012b}]{Kesden12b}
\begin{barticle}
\bauthor{\binits{M.} \bsnm{{Kesden}}},
\batitle{{Tidal-disruption rate of stars by spinning supermassive black
  holes}}.
\bjtitle{\prd}
\bvolume{85}(\bissue{2}),
\bfpage{024037}
(\byear{2012}b).
doi:\doiurl{10.1103/PhysRevD.85.024037}
\end{barticle}
\endbibitem

\bibitem[\protect\citeauthoryear{{Kochanek}}{2016}]{Kochanek16}
\begin{barticle}
\bauthor{\binits{C.S.} \bsnm{{Kochanek}}},
\batitle{{Tidal disruption event demographics}}.
\bjtitle{\mnras}
\bvolume{461},
\bfpage{371}--\blpage{384}
(\byear{2016}).
doi:\doiurl{10.1093/mnras/stw1290}
\end{barticle}
\endbibitem

\bibitem[\protect\citeauthoryear{{Komossa} and
  {Greiner}}{1999}]{KomossaGreiner99}
\begin{barticle}
\bauthor{\binits{S.} \bsnm{{Komossa}}},
\bauthor{\binits{J.} \bsnm{{Greiner}}},
\batitle{{Discovery of a giant and luminous X-ray outburst from the optically
  inactive galaxy pair RX J1242.6-1119}}.
\bjtitle{\aap}
\bvolume{349},
\bfpage{45}--\blpage{48}
(\byear{1999})
\end{barticle}
\endbibitem

\bibitem[\protect\citeauthoryear{{Komossa} et~al.}{2008}]{Komossa2008}
\begin{barticle}
\bauthor{\binits{S.} \bsnm{{Komossa}}},
\bauthor{\binits{H.} \bsnm{{Zhou}}},
\bauthor{\binits{T.} \bsnm{{Wang}}},
\bauthor{\binits{M.} \bsnm{{Ajello}}},
\bauthor{\binits{J.} \bsnm{{Ge}}},
\bauthor{\binits{J.} \bsnm{{Greiner}}},
\bauthor{\binits{H.} \bsnm{{Lu}}},
\bauthor{\binits{M.} \bsnm{{Salvato}}},
\bauthor{\binits{R.} \bsnm{{Saxton}}},
\bauthor{\binits{H.} \bsnm{{Shan}}},
\bauthor{\binits{D.} \bsnm{{Xu}}},
\bauthor{\binits{W.} \bsnm{{Yuan}}},
\batitle{{Discovery of Superstrong, Fading, Iron Line Emission and
  Double-peaked Balmer Lines of the Galaxy SDSS J095209.56+214313.3: The Light
  Echo of a Huge Flare}}.
\bjtitle{\apjl}
\bvolume{678}(\bissue{1}),
\bfpage{13}
(\byear{2008}).
doi:\doiurl{10.1086/588281}
\end{barticle}
\endbibitem

\bibitem[\protect\citeauthoryear{{Kr{\"u}hler} et~al.}{2018}]{Kruhler18}
\begin{barticle}
\bauthor{\binits{T.} \bsnm{{Kr{\"u}hler}}},
\bauthor{\binits{M.} \bsnm{{Fraser}}},
\bauthor{\binits{G.} \bsnm{{Leloudas}}},
\bauthor{\binits{S.} \bsnm{{Schulze}}},
\bauthor{\binits{N.C.} \bsnm{{Stone}}},
\bauthor{\binits{S.} \bsnm{{van Velzen}}},
\bauthor{\binits{R.} \bsnm{{Amorin}}},
\bauthor{\binits{J.} \bsnm{{Hjorth}}},
\bauthor{\binits{P.G.} \bsnm{{Jonker}}},
\bauthor{\binits{D.A.} \bsnm{{Kann}}},
\bauthor{\binits{S.} \bsnm{{Kim}}},
\bauthor{\binits{H.} \bsnm{{Kuncarayakti}}},
\bauthor{\binits{A.} \bsnm{{Mehner}}},
\bauthor{\binits{A.} \bsnm{{Nicuesa Guelbenzu}}},
\batitle{{The supermassive black hole coincident with the luminous transient
  ASASSN-15lh}}.
\bjtitle{\aap}
\bvolume{610},
\bfpage{14}
(\byear{2018}).
doi:\doiurl{10.1051/0004-6361/201731773}
\end{barticle}
\endbibitem

\bibitem[\protect\citeauthoryear{{Law} et~al.}{2009}]{Law09}
\begin{barticle}
\bauthor{\binits{N.M.} \bsnm{{Law}}},
\bauthor{\binits{S.R.} \bsnm{{Kulkarni}}},
\bauthor{\binits{R.G.} \bsnm{{Dekany}}},
\bauthor{\binits{E.O.} \bsnm{{Ofek}}},
\bauthor{\binits{R.M.} \bsnm{{Quimby}}},
\bauthor{\binits{P.E.} \bsnm{{Nugent}}},
\bauthor{\binits{J.} \bsnm{{Surace}}},
\bauthor{\binits{C.C.} \bsnm{{Grillmair}}},
\bauthor{\binits{J.S.} \bsnm{{Bloom}}},
\bauthor{\binits{M.M.} \bsnm{{Kasliwal}}},
\bauthor{\binits{L.} \bsnm{{Bildsten}}},
\bauthor{\binits{T.} \bsnm{{Brown}}},
\bauthor{\binits{S.B.} \bsnm{{Cenko}}},
\bauthor{\binits{D.} \bsnm{{Ciardi}}},
\bauthor{\binits{E.} \bsnm{{Croner}}},
\bauthor{\binits{S.G.} \bsnm{{Djorgovski}}},
\bauthor{\binits{J.} \bsnm{{van Eyken}}},
\bauthor{\binits{A.V.} \bsnm{{Filippenko}}},
\bauthor{\binits{D.B.} \bsnm{{Fox}}},
\bauthor{\binits{A.} \bsnm{{Gal-Yam}}},
\bauthor{\binits{D.} \bsnm{{Hale}}},
\bauthor{\binits{N.} \bsnm{{Hamam}}},
\bauthor{\binits{G.} \bsnm{{Helou}}},
\bauthor{\binits{J.} \bsnm{{Henning}}},
\bauthor{\binits{D.A.} \bsnm{{Howell}}},
\bauthor{\binits{J.} \bsnm{{Jacobsen}}},
\bauthor{\binits{R.} \bsnm{{Laher}}},
\bauthor{\binits{S.} \bsnm{{Mattingly}}},
\bauthor{\binits{D.} \bsnm{{McKenna}}},
\bauthor{\binits{A.} \bsnm{{Pickles}}},
\bauthor{\binits{D.} \bsnm{{Poznanski}}},
\bauthor{\binits{G.} \bsnm{{Rahmer}}},
\bauthor{\binits{A.} \bsnm{{Rau}}},
\bauthor{\binits{W.} \bsnm{{Rosing}}},
\bauthor{\binits{M.} \bsnm{{Shara}}},
\bauthor{\binits{R.} \bsnm{{Smith}}},
\bauthor{\binits{D.} \bsnm{{Starr}}},
\bauthor{\binits{M.} \bsnm{{Sullivan}}},
\bauthor{\binits{V.} \bsnm{{Velur}}},
\bauthor{\binits{R.} \bsnm{{Walters}}},
\bauthor{\binits{J.} \bsnm{{Zolkower}}},
\batitle{{The Palomar Transient Factory: System Overview, Performance, and
  First Results}}.
\bjtitle{\pasp}
\bvolume{121},
\bfpage{1395}--\blpage{1408}
(\byear{2009}).
doi:\doiurl{10.1086/648598}
\end{barticle}
\endbibitem

\bibitem[\protect\citeauthoryear{{Leloudas} et~al.}{2016}]{Leloudas16}
\begin{barticle}
\bauthor{\binits{G.} \bsnm{{Leloudas}}},
\bauthor{\binits{M.} \bsnm{{Fraser}}},
\bauthor{\binits{N.C.} \bsnm{{Stone}}},
\bauthor{\binits{S.} \bsnm{{van Velzen}}},
\bauthor{\binits{P.G.} \bsnm{{Jonker}}},
\bauthor{\binits{I.} \bsnm{{Arcavi}}},
\bauthor{\binits{C.} \bsnm{{Fremling}}},
\bauthor{\binits{J.R.} \bsnm{{Maund}}},
\bauthor{\binits{S.J.} \bsnm{{Smartt}}},
\bauthor{\binits{T.} \bsnm{{Kr{\`\i}hler}}},
\bauthor{\binits{J.C.A.} \bsnm{{Miller-Jones}}},
\bauthor{\binits{P.M.} \bsnm{{Vreeswijk}}},
\bauthor{\binits{A.} \bsnm{{Gal-Yam}}},
\bauthor{\binits{P.A.} \bsnm{{Mazzali}}},
\bauthor{\binits{A.} \bsnm{{De Cia}}},
\bauthor{\binits{D.A.} \bsnm{{Howell}}},
\bauthor{\binits{C.} \bsnm{{Inserra}}},
\bauthor{\binits{F.} \bsnm{{Patat}}},
\bauthor{\binits{A.} \bsnm{{de Ugarte Postigo}}},
\bauthor{\binits{O.} \bsnm{{Yaron}}},
\bauthor{\binits{C.} \bsnm{{Ashall}}},
\bauthor{\binits{I.} \bsnm{{Bar}}},
\bauthor{\binits{H.} \bsnm{{Campbell}}},
\bauthor{\binits{T.-W.} \bsnm{{Chen}}},
\bauthor{\binits{M.} \bsnm{{Childress}}},
\bauthor{\binits{N.} \bsnm{{Elias-Rosa}}},
\bauthor{\binits{J.} \bsnm{{Harmanen}}},
\bauthor{\binits{G.} \bsnm{{Hosseinzadeh}}},
\bauthor{\binits{J.} \bsnm{{Johansson}}},
\bauthor{\binits{T.} \bsnm{{Kangas}}},
\bauthor{\binits{E.} \bsnm{{Kankare}}},
\bauthor{\binits{S.} \bsnm{{Kim}}},
\bauthor{\binits{H.} \bsnm{{Kuncarayakti}}},
\bauthor{\binits{J.} \bsnm{{Lyman}}},
\bauthor{\binits{M.R.} \bsnm{{Magee}}},
\bauthor{\binits{K.} \bsnm{{Maguire}}},
\bauthor{\binits{D.} \bsnm{{Malesani}}},
\bauthor{\binits{S.} \bsnm{{Mattila}}},
\bauthor{\binits{C.V.} \bsnm{{McCully}}},
\bauthor{\binits{M.} \bsnm{{Nicholl}}},
\bauthor{\binits{S.} \bsnm{{Prentice}}},
\bauthor{\binits{C.} \bsnm{{Romero-Ca{\~n}izales}}},
\bauthor{\binits{S.} \bsnm{{Schulze}}},
\bauthor{\binits{K.W.} \bsnm{{Smith}}},
\bauthor{\binits{J.} \bsnm{{Sollerman}}},
\bauthor{\binits{M.} \bsnm{{Sullivan}}},
\bauthor{\binits{B.E.} \bsnm{{Tucker}}},
\bauthor{\binits{S.} \bsnm{{Valenti}}},
\bauthor{\binits{J.C.} \bsnm{{Wheeler}}},
\bauthor{\binits{D.R.} \bsnm{{Young}}},
\batitle{{The superluminous transient ASASSN-15lh as a tidal disruption event
  from a Kerr black hole}}.
\bjtitle{Nature Astronomy}
\bvolume{1},
\bfpage{0002}
(\byear{2016}).
doi:\doiurl{10.1038/s41550-016-0002}
\end{barticle}
\endbibitem

\bibitem[\protect\citeauthoryear{{Leloudas} et~al.}{2019}]{leloudas19}
\begin{barticle}
\bauthor{\binits{G.} \bsnm{{Leloudas}}},
\bauthor{\binits{L.} \bsnm{{Dai}}},
\bauthor{\binits{I.} \bsnm{{Arcavi}}},
\bauthor{\binits{P.M.} \bsnm{{Vreeswijk}}},
\bauthor{\binits{B.} \bsnm{{Mockler}}},
\bauthor{\binits{R.} \bsnm{{Roy}}},
\bauthor{\binits{D.B.} \bsnm{{Malesani}}},
\bauthor{\binits{S.} \bsnm{{Schulze}}},
\bauthor{\binits{T.} \bsnm{{Wevers}}},
\bauthor{\binits{M.} \bsnm{{Fraser}}},
\bauthor{\binits{E.} \bsnm{{Ramirez-Ruiz}}},
\bauthor{\binits{K.} \bsnm{{Auchettl}}},
\bauthor{\binits{J.} \bsnm{{Burke}}},
\bauthor{\binits{G.} \bsnm{{Cannizzaro}}},
\bauthor{\binits{P.} \bsnm{{Charalampopoulos}}},
\bauthor{\binits{T.-W.} \bsnm{{Chen}}},
\bauthor{\binits{A.a.} \bsnm{{Cikota}}},
\bauthor{\binits{M.} \bsnm{{Della Valle}}},
\bauthor{\binits{L.} \bsnm{{Galbany}}},
\bauthor{\binits{M.} \bsnm{{Gromadzki}}},
\bauthor{\binits{K.E.} \bsnm{{Heintz}}},
\bauthor{\binits{D.} \bsnm{{Hiramatsu}}},
\bauthor{\binits{P.G.} \bsnm{{Jonker}}},
\bauthor{\binits{Z.} \bsnm{{Kostrzewa-Rutkowska}}},
\bauthor{\binits{K.} \bsnm{{Maguire}}},
\bauthor{\binits{I.} \bsnm{{Mandel}}},
\bauthor{\binits{M.} \bsnm{{Nicholl}}},
\bauthor{\binits{F.} \bsnm{{Onori}}},
\bauthor{\binits{N.} \bsnm{{Roth}}},
\bauthor{\binits{S.J.} \bsnm{{Smartt}}},
\bauthor{\binits{L.} \bsnm{{Wyrzykowski}}},
\bauthor{\binits{D.R.} \bsnm{{Young}}},
\batitle{{The Spectral Evolution of AT 2018dyb and the Presence of Metal Lines
  in Tidal Disruption Events}}.
\bjtitle{\apj}
\bvolume{887}(\bissue{2}),
\bfpage{218}
(\byear{2019}).
doi:\doiurl{10.3847/1538-4357/ab5792}
\end{barticle}
\endbibitem

\bibitem[\protect\citeauthoryear{{Lin} et~al.}{2011}]{Lin11}
\begin{barticle}
\bauthor{\binits{D.} \bsnm{{Lin}}},
\bauthor{\binits{E.R.} \bsnm{{Carrasco}}},
\bauthor{\binits{D.} \bsnm{{Grupe}}},
\bauthor{\binits{N.A.} \bsnm{{Webb}}},
\bauthor{\binits{D.} \bsnm{{Barret}}},
\bauthor{\binits{S.A.} \bsnm{{Farrell}}},
\batitle{{Discovery of an Ultrasoft X-Ray Transient Source in the 2XMM Catalog:
  A Tidal Disruption Event Candidate}}.
\bjtitle{\apj}
\bvolume{738},
\bfpage{52}
(\byear{2011}).
doi:\doiurl{10.1088/0004-637X/738/1/52}
\end{barticle}
\endbibitem

\bibitem[\protect\citeauthoryear{{Liu} et~al.}{2017}]{Liu17}
\begin{barticle}
\bauthor{\binits{F.K.} \bsnm{{Liu}}},
\bauthor{\binits{Z.Q.} \bsnm{{Zhou}}},
\bauthor{\binits{R.} \bsnm{{Cao}}},
\bauthor{\binits{L.C.} \bsnm{{Ho}}},
\bauthor{\binits{S.} \bsnm{{Komossa}}},
\batitle{{Disc origin of broad optical emission lines of the TDE candidate
  PTF09djl}}.
\bjtitle{\mnras}
\bvolume{472}(\bissue{1}),
\bfpage{99}--\blpage{103}
(\byear{2017}).
doi:\doiurl{10.1093/mnrasl/slx147}
\end{barticle}
\endbibitem

\bibitem[\protect\citeauthoryear{{Liu} et~al.}{2019}]{Liu2019}
\begin{botherref}
\oauthor{\binits{X.-L.} \bsnm{{Liu}}},
\oauthor{\binits{L.-M.} \bsnm{{Dou}}},
\oauthor{\binits{R.-F.} \bsnm{{Shen}}},
\oauthor{\binits{J.-H.} \bsnm{{Chen}}},
{The UV/optical peak and X-ray brightening in TDE candidate AT2019azh: A case
  of stream-stream collision and delayed accretion}.
arXiv e-prints,
1912--06081
(2019)
\end{botherref}
\endbibitem

\bibitem[\protect\citeauthoryear{{Lodato} et~al.}{2009}]{Lodato09}
\begin{barticle}
\bauthor{\binits{G.} \bsnm{{Lodato}}},
\bauthor{\binits{A.R.} \bsnm{{King}}},
\bauthor{\binits{J.E.} \bsnm{{Pringle}}},
\batitle{{Stellar disruption by a supermassive black hole: is the light curve
  really proportional to $t^{-5/3}$?}}
\bjtitle{\mnras}
\bvolume{392},
\bfpage{332}--\blpage{340}
(\byear{2009}).
doi:\doiurl{10.1111/j.1365-2966.2008.14049.x}
\end{barticle}
\endbibitem

\bibitem[\protect\citeauthoryear{{Loeb} and {Ulmer}}{1997}]{LoebUlmer97}
\begin{barticle}
\bauthor{\binits{A.} \bsnm{{Loeb}}},
\bauthor{\binits{A.} \bsnm{{Ulmer}}},
\batitle{{Optical Appearance of the Debris of a Star Disrupted by a Massive
  Black Hole}}.
\bjtitle{\apj}
\bvolume{489},
\bfpage{573}--\blpage{578}
(\byear{1997}).
doi:\doiurl{10.1086/304814}
\end{barticle}
\endbibitem

\bibitem[\protect\citeauthoryear{{Lu} and {Bonnerot}}{2020}]{Lu20}
\begin{barticle}
\bauthor{\binits{W.} \bsnm{{Lu}}},
\bauthor{\binits{C.} \bsnm{{Bonnerot}}},
\batitle{{Self-intersection of the fallback stream in tidal disruption
  events}}.
\bjtitle{\mnras}
\bvolume{492}(\bissue{1}),
\bfpage{686}--\blpage{707}
(\byear{2020}).
doi:\doiurl{10.1093/mnras/stz3405}
\end{barticle}
\endbibitem

\bibitem[\protect\citeauthoryear{{Margutti} et~al.}{2017}]{Margutti17}
\begin{barticle}
\bauthor{\binits{R.} \bsnm{{Margutti}}},
\bauthor{\binits{B.D.} \bsnm{{Metzger}}},
\bauthor{\binits{R.} \bsnm{{Chornock}}},
\bauthor{\binits{D.} \bsnm{{Milisavljevic}}},
\bauthor{\binits{E.} \bsnm{{Berger}}},
\bauthor{\binits{P.K.} \bsnm{{Blanchard}}},
\bauthor{\binits{C.} \bsnm{{Guidorzi}}},
\bauthor{\binits{G.} \bsnm{{Migliori}}},
\bauthor{\binits{A.} \bsnm{{Kamble}}},
\bauthor{\binits{R.} \bsnm{{Lunnan}}},
\bauthor{\binits{M.} \bsnm{{Nicholl}}},
\bauthor{\binits{D.L.} \bsnm{{Coppejans}}},
\bauthor{\binits{S.} \bsnm{{Dall'Osso}}},
\bauthor{\binits{M.R.} \bsnm{{Drout}}},
\bauthor{\binits{R.} \bsnm{{Perna}}},
\bauthor{\binits{B.} \bsnm{{Sbarufatti}}},
\batitle{{X-Rays from the Location of the Double-humped Transient
  ASASSN-15lh}}.
\bjtitle{\apj}
\bvolume{836},
\bfpage{25}
(\byear{2017}).
doi:\doiurl{10.3847/1538-4357/836/1/25}
\end{barticle}
\endbibitem

\bibitem[\protect\citeauthoryear{{Martin} et~al.}{2005}]{Martin05}
\begin{barticle}
\bauthor{\binits{D.C.} \bsnm{{Martin}}},
\bauthor{\binits{J.} \bsnm{{Fanson}}},
\bauthor{\binits{D.} \bsnm{{Schiminovich}}},
\bauthor{\binits{P.} \bsnm{{Morrissey}}},
\bauthor{\binits{P.G.} \bsnm{{Friedman}}},
\bauthor{\binits{T.A.} \bsnm{{Barlow}}},
\bauthor{\binits{T.} \bsnm{{Conrow}}},
\bauthor{\binits{R.} \bsnm{{Grange}}},
\bauthor{\binits{P.N.} \bsnm{{Jelinsky}}},
\bauthor{\binits{B.} \bsnm{{Milliard}}},
\bauthor{\binits{O.H.W.} \bsnm{{Siegmund}}},
\bauthor{\binits{L.} \bsnm{{Bianchi}}},
\bauthor{\binits{Y.-I.} \bsnm{{Byun}}},
\bauthor{\binits{J.} \bsnm{{Donas}}},
\bauthor{\binits{K.} \bsnm{{Forster}}},
\bauthor{\binits{T.M.} \bsnm{{Heckman}}},
\bauthor{\binits{Y.-W.} \bsnm{{Lee}}},
\bauthor{\binits{B.F.} \bsnm{{Madore}}},
\bauthor{\binits{R.F.} \bsnm{{Malina}}},
\bauthor{\binits{S.G.} \bsnm{{Neff}}},
\bauthor{\binits{R.M.} \bsnm{{Rich}}},
\bauthor{\binits{T.} \bsnm{{Small}}},
\bauthor{\binits{F.} \bsnm{{Surber}}},
\bauthor{\binits{A.S.} \bsnm{{Szalay}}},
\bauthor{\binits{B.} \bsnm{{Welsh}}},
\bauthor{\binits{T.K.} \bsnm{{Wyder}}},
\batitle{{The Galaxy Evolution Explorer: A Space Ultraviolet Survey Mission}}.
\bjtitle{\apjl}
\bvolume{619},
\bfpage{1}--\blpage{6}
(\byear{2005}).
doi:\doiurl{10.1086/426387}
\end{barticle}
\endbibitem

\bibitem[\protect\citeauthoryear{{Miller} et~al.}{2015}]{Miller_14li}
\begin{barticle}
\bauthor{\binits{J.M.} \bsnm{{Miller}}},
\bauthor{\binits{J.S.} \bsnm{{Kaastra}}},
\bauthor{\binits{M.C.} \bsnm{{Miller}}},
\bauthor{\binits{M.T.} \bsnm{{Reynolds}}},
\bauthor{\binits{G.} \bsnm{{Brown}}},
\bauthor{\binits{S.B.} \bsnm{{Cenko}}},
\bauthor{\binits{J.J.} \bsnm{{Drake}}},
\bauthor{\binits{S.} \bsnm{{Gezari}}},
\bauthor{\binits{J.} \bsnm{{Guillochon}}},
\bauthor{\binits{K.} \bsnm{{Gultekin}}},
\bauthor{\binits{J.} \bsnm{{Irwin}}},
\bauthor{\binits{A.} \bsnm{{Levan}}},
\bauthor{\binits{D.} \bsnm{{Maitra}}},
\bauthor{\binits{W.P.} \bsnm{{Maksym}}},
\bauthor{\binits{R.} \bsnm{{Mushotzky}}},
\bauthor{\binits{P.} \bsnm{{O'Brien}}},
\bauthor{\binits{F.} \bsnm{{Paerels}}},
\bauthor{\binits{J.} \bsnm{{de Plaa}}},
\bauthor{\binits{E.} \bsnm{{Ramirez-Ruiz}}},
\bauthor{\binits{T.} \bsnm{{Strohmayer}}},
\bauthor{\binits{N.} \bsnm{{Tanvir}}},
\batitle{{Flows of X-ray gas reveal the disruption of a star by a massive black
  hole}}.
\bjtitle{\nat}
\bvolume{526},
\bfpage{542}--\blpage{545}
(\byear{2015}).
doi:\doiurl{10.1038/nature15708}
\end{barticle}
\endbibitem

\bibitem[\protect\citeauthoryear{{Miller}}{2015}]{Miller15}
\begin{barticle}
\bauthor{\binits{M.C.} \bsnm{{Miller}}},
\batitle{{Disk Winds as an Explanation for Slowly Evolving Temperatures in
  Tidal Disruption Events}}.
\bjtitle{\apj}
\bvolume{805},
\bfpage{83}
(\byear{2015}).
doi:\doiurl{10.1088/0004-637X/805/1/83}
\end{barticle}
\endbibitem

\bibitem[\protect\citeauthoryear{{Mummery} and {Balbus}}{2020}]{Mummery20}
\begin{barticle}
\bauthor{\binits{A.} \bsnm{{Mummery}}},
\bauthor{\binits{S.A.} \bsnm{{Balbus}}},
\batitle{{The spectral evolution of disc dominated tidal disruption events}}.
\bjtitle{\mnras}
\bvolume{492}(\bissue{4}),
\bfpage{5655}--\blpage{5674}
(\byear{2020}).
doi:\doiurl{10.1093/mnras/staa192}
\end{barticle}
\endbibitem

\bibitem[\protect\citeauthoryear{{Netzer} et~al.}{1985}]{Netzer1985}
\begin{barticle}
\bauthor{\binits{H.} \bsnm{{Netzer}}},
\bauthor{\binits{M.} \bsnm{{Elitzur}}},
\bauthor{\binits{G.J.} \bsnm{{Ferland}}},
\batitle{{Bowen fluoresence and He II lines in active galaxies and gaseous
  nebulae}}.
\bjtitle{\apj}
\bvolume{299},
\bfpage{752}--\blpage{762}
(\byear{1985}).
doi:\doiurl{10.1086/163741}
\end{barticle}
\endbibitem

\bibitem[\protect\citeauthoryear{{Neustadt} et~al.}{2020}]{Neustadt20}
\begin{barticle}
\bauthor{\binits{J.M.M.} \bsnm{{Neustadt}}},
\bauthor{\binits{T.W.-S.} \bsnm{{Holoien}}},
\bauthor{\binits{C.S.} \bsnm{{Kochanek}}},
\bauthor{\binits{K.} \bsnm{{Auchettl}}},
\bauthor{\binits{J.S.} \bsnm{{Brown}}},
\bauthor{\binits{B.J.} \bsnm{{Shappee}}},
\bauthor{\binits{R.W.} \bsnm{{Pogge}}},
\bauthor{\binits{S.} \bsnm{{Dong}}},
\bauthor{\binits{K.Z.} \bsnm{{Stanek}}},
\bauthor{\binits{M.A.} \bsnm{{Tucker}}},
\bauthor{\binits{S.} \bsnm{{Bose}}},
\bauthor{\binits{P.} \bsnm{{Chen}}},
\bauthor{\binits{C.} \bsnm{{Ricci}}},
\bauthor{\binits{P.J.} \bsnm{{Vallely}}},
\bauthor{\binits{J.L.} \bsnm{{Prieto}}},
\bauthor{\binits{T.A.} \bsnm{{Thompson}}},
\bauthor{\binits{D.A.} \bsnm{{Coulter}}},
\bauthor{\binits{M.R.} \bsnm{{Drout}}},
\bauthor{\binits{R.J.} \bsnm{{Foley}}},
\bauthor{\binits{C.D.} \bsnm{{Kilpatrick}}},
\bauthor{\binits{A.L.} \bsnm{{Piro}}},
\bauthor{\binits{C.} \bsnm{{Rojas-Bravo}}},
\bauthor{\binits{D.A.H.} \bsnm{{Buckley}}},
\bauthor{\binits{M.} \bsnm{{Gromadzki}}},
\bauthor{\binits{G.} \bsnm{{Dimitriadis}}},
\bauthor{\binits{M.R.} \bsnm{{Siebert}}},
\bauthor{\binits{A.} \bsnm{{Do}}},
\bauthor{\binits{M.E.} \bsnm{{Huber}}},
\bauthor{\binits{A.V.} \bsnm{{Payne}}},
\batitle{{To TDE or not to TDE: the luminous transient ASASSN-18jd with
  TDE-like and AGN-like qualities}}.
\bjtitle{\mnras}
\bvolume{494}(\bissue{2}),
\bfpage{2538}--\blpage{2560}
(\byear{2020}).
doi:\doiurl{10.1093/mnras/staa859}
\end{barticle}
\endbibitem

\bibitem[\protect\citeauthoryear{{Nicholl} et~al.}{2019}]{Nicholl2019}
\begin{barticle}
\bauthor{\binits{M.} \bsnm{{Nicholl}}},
\bauthor{\binits{P.K.} \bsnm{{Blanchard}}},
\bauthor{\binits{E.} \bsnm{{Berger}}},
\bauthor{\binits{S.} \bsnm{{Gomez}}},
\bauthor{\binits{R.} \bsnm{{Margutti}}},
\bauthor{\binits{K.D.} \bsnm{{Alexander}}},
\bauthor{\binits{J.} \bsnm{{Guillochon}}},
\bauthor{\binits{J.} \bsnm{{Leja}}},
\bauthor{\binits{R.} \bsnm{{Chornock}}},
\bauthor{\binits{B.} \bsnm{{Snios}}},
\bauthor{\binits{K.} \bsnm{{Auchettl}}},
\bauthor{\binits{A.G.} \bsnm{{Bruce}}},
\bauthor{\binits{P.} \bsnm{{Challis}}},
\bauthor{\binits{D.J.} \bsnm{{D'Orazio}}},
\bauthor{\binits{M.R.} \bsnm{{Drout}}},
\bauthor{\binits{T.} \bsnm{{Eftekhari}}},
\bauthor{\binits{R.J.} \bsnm{{Foley}}},
\bauthor{\binits{O.} \bsnm{{Graur}}},
\bauthor{\binits{C.D.} \bsnm{{Kilpatrick}}},
\bauthor{\binits{A.} \bsnm{{Lawrence}}},
\bauthor{\binits{A.L.} \bsnm{{Piro}}},
\bauthor{\binits{C.} \bsnm{{Rojas-Bravo}}},
\bauthor{\binits{N.P.} \bsnm{{Ross}}},
\bauthor{\binits{P.} \bsnm{{Short}}},
\bauthor{\binits{S.J.} \bsnm{{Smartt}}},
\bauthor{\binits{K.W.} \bsnm{{Smith}}},
\bauthor{\binits{B.} \bsnm{{Stalder}}},
\batitle{{The tidal disruption event AT2017eqx: spectroscopic evolution from
  hydrogen rich to poor suggests an atmosphere and outflow}}.
\bjtitle{\mnras}
\bvolume{488}(\bissue{2}),
\bfpage{1878}--\blpage{1893}
(\byear{2019}).
doi:\doiurl{10.1093/mnras/stz1837}
\end{barticle}
\endbibitem

\bibitem[\protect\citeauthoryear{{Nicholl} et~al.}{2020}]{Nicholl2020}
\begin{botherref}
\oauthor{\binits{M.} \bsnm{{Nicholl}}},
\oauthor{\binits{T.} \bsnm{{Wevers}}},
\oauthor{\binits{S.R.} \bsnm{{Oates}}},
\oauthor{\binits{K.D.} \bsnm{{Alexand er}}},
\oauthor{\binits{G.} \bsnm{{Leloudas}}},
\oauthor{\binits{F.} \bsnm{{Onori}}},
\oauthor{\binits{A.} \bsnm{{Jerkstrand}}},
\oauthor{\binits{S.} \bsnm{{Gomez}}},
\oauthor{\binits{S.} \bsnm{{Campana}}},
\oauthor{\binits{I.} \bsnm{{Arcavi}}},
\oauthor{\binits{P.} \bsnm{{Charalampopoulos}}},
\oauthor{\binits{M.} \bsnm{{Gromadzki}}},
\oauthor{\binits{N.} \bsnm{{Ihanec}}},
\oauthor{\binits{P.G.} \bsnm{{Jonker}}},
\oauthor{\binits{A.} \bsnm{{Lawrence}}},
\oauthor{\binits{I.} \bsnm{{Mandel}}},
\oauthor{\binits{P.} \bsnm{{Short}}},
\oauthor{\binits{J.} \bsnm{{Burke}}},
\oauthor{\binits{D.} \bsnm{{Hiramatsu}}},
\oauthor{\binits{D.A.} \bsnm{{Howell}}},
\oauthor{\binits{C.} \bsnm{{Pellegrino}}},
\oauthor{\binits{H.} \bsnm{{Abbot}}},
\oauthor{\binits{J.P.} \bsnm{{Anderson}}},
\oauthor{\binits{E.} \bsnm{{Berger}}},
\oauthor{\binits{P.K.} \bsnm{{Blanchard}}},
\oauthor{\binits{G.} \bsnm{{Cannizzaro}}},
\oauthor{\binits{T.-W.} \bsnm{{Chen}}},
\oauthor{\binits{M.} \bsnm{{Dennefeld}}},
\oauthor{\binits{L.} \bsnm{{Galbany}}},
\oauthor{\binits{S.} \bsnm{{Gonzalez-Gaitan}}},
\oauthor{\binits{G.} \bsnm{{Hosseinzadeh}}},
\oauthor{\binits{C.} \bsnm{{Inserra}}},
\oauthor{\binits{I.} \bsnm{{Irani}}},
\oauthor{\binits{P.} \bsnm{{Kuin}}},
\oauthor{\binits{T.} \bsnm{{Muller-Bravo}}},
\oauthor{\binits{J.} \bsnm{{Pineda}}},
\oauthor{\binits{N.P.} \bsnm{{Ross}}},
\oauthor{\binits{R.} \bsnm{{Roy}}},
\oauthor{\binits{B.} \bsnm{{Tucker}}},
\oauthor{\binits{L.} \bsnm{{Wyrzykowski}}},
\oauthor{\binits{D.R.} \bsnm{{Young}}},
{An outflow powers the optical rise of the nearby, fast-evolving tidal
  disruption event AT2019qiz}.
arXiv e-prints,
2006--02454
(2020)
\end{botherref}
\endbibitem

\bibitem[\protect\citeauthoryear{{Onori} et~al.}{2019}]{Onori2019}
\begin{barticle}
\bauthor{\binits{F.} \bsnm{{Onori}}},
\bauthor{\binits{G.} \bsnm{{Cannizzaro}}},
\bauthor{\binits{P.G.} \bsnm{{Jonker}}},
\bauthor{\binits{M.} \bsnm{{Fraser}}},
\bauthor{\binits{Z.} \bsnm{{Kostrzewa-Rutkowska}}},
\bauthor{\binits{A.} \bsnm{{Martin-Carrillo}}},
\bauthor{\binits{S.} \bsnm{{Benetti}}},
\bauthor{\binits{N.} \bsnm{{Elias-Rosa}}},
\bauthor{\binits{M.} \bsnm{{Gromadzki}}},
\bauthor{\binits{J.} \bsnm{{Harmanen}}},
\bauthor{\binits{S.} \bsnm{{Mattila}}},
\bauthor{\binits{M.D.} \bsnm{{Strizinger}}},
\bauthor{\binits{G.} \bsnm{{Terreran}}},
\bauthor{\binits{T.} \bsnm{{Wevers}}},
\batitle{{Optical follow-up of the tidal disruption event iPTF16fnl: new
  insights from X-shooter observations}}.
\bjtitle{\mnras}
\bvolume{489}(\bissue{1}),
\bfpage{1463}--\blpage{1480}
(\byear{2019}).
doi:\doiurl{10.1093/mnras/stz2053}
\end{barticle}
\endbibitem

\bibitem[\protect\citeauthoryear{{Parkinson} et~al.}{2020}]{Parkinson2020}
\begin{barticle}
\bauthor{\binits{E.J.} \bsnm{{Parkinson}}},
\bauthor{\binits{C.} \bsnm{{Knigge}}},
\bauthor{\binits{K.S.} \bsnm{{Long}}},
\bauthor{\binits{J.H.} \bsnm{{Matthews}}},
\bauthor{\binits{N.} \bsnm{{Higginbottom}}},
\bauthor{\binits{S.A.} \bsnm{{Sim}}},
\bauthor{\binits{H.A.} \bsnm{{Hewitt}}},
\batitle{{Accretion disc winds in tidal disruption events: ultraviolet spectral
  lines as orientation indicators}}.
\bjtitle{\mnras}
\bvolume{494}(\bissue{4}),
\bfpage{4914}--\blpage{4929}
(\byear{2020}).
doi:\doiurl{10.1093/mnras/staa1060}
\end{barticle}
\endbibitem

\bibitem[\protect\citeauthoryear{{Phinney}}{1989}]{Phinney89}
\begin{bchapter}
\bauthor{\binits{E.S.} \bsnm{{Phinney}}},
\bctitle{{Manifestations of a Massive Black Hole in the Galactic Center}},
in \bbtitle{The Center of the Galaxy},
ed. by \beditor{\bsnm{{M.~Morris}}}
\bsertitle{IAU Symposium},
vol. \bseriesno{136}
(\bpublisher{Dordrecht: Kluwer}, \blocation{???}, \byear{1989}),
p. \bfpage{543}
\end{bchapter}
\endbibitem

\bibitem[\protect\citeauthoryear{{Piran} et~al.}{2015}]{Piran15}
\begin{barticle}
\bauthor{\binits{T.} \bsnm{{Piran}}},
\bauthor{\binits{G.} \bsnm{{Svirski}}},
\bauthor{\binits{J.} \bsnm{{Krolik}}},
\bauthor{\binits{R.M.} \bsnm{{Cheng}}},
\bauthor{\binits{H.} \bsnm{{Shiokawa}}},
\batitle{{Disk Formation Versus Disk Accretion: What Powers Tidal Disruption
  Events?}}
\bjtitle{\apj}
\bvolume{806},
\bfpage{164}
(\byear{2015}).
doi:\doiurl{10.1088/0004-637X/806/2/164}
\end{barticle}
\endbibitem

\bibitem[\protect\citeauthoryear{{Rees}}{1988}]{Rees88}
\begin{barticle}
\bauthor{\binits{M.J.} \bsnm{{Rees}}},
\batitle{{Tidal disruption of stars by black holes of 10 to the 6th-10 to the
  8th solar masses in nearby galaxies}}.
\bjtitle{\nat}
\bvolume{333},
\bfpage{523}--\blpage{528}
(\byear{1988}).
doi:\doiurl{10.1038/333523a0}
\end{barticle}
\endbibitem

\bibitem[\protect\citeauthoryear{{Roth} and {Kasen}}{2018}]{Roth18}
\begin{barticle}
\bauthor{\binits{N.} \bsnm{{Roth}}},
\bauthor{\binits{D.} \bsnm{{Kasen}}},
\batitle{{What Sets the Line Profiles in Tidal Disruption Events?}}
\bjtitle{\apj}
\bvolume{855},
\bfpage{54}
(\byear{2018}).
doi:\doiurl{10.3847/1538-4357/aaaec6}
\end{barticle}
\endbibitem

\bibitem[\protect\citeauthoryear{{Roth} et~al.}{2016}]{Roth16}
\begin{barticle}
\bauthor{\binits{N.} \bsnm{{Roth}}},
\bauthor{\binits{D.} \bsnm{{Kasen}}},
\bauthor{\binits{J.} \bsnm{{Guillochon}}},
\bauthor{\binits{E.} \bsnm{{Ramirez-Ruiz}}},
\batitle{{The X-Ray through Optical Fluxes and Line Strengths of Tidal
  Disruption Events}}.
\bjtitle{\apj}
\bvolume{827}(\bissue{1}),
\bfpage{3}
(\byear{2016}).
doi:\doiurl{10.3847/0004-637X/827/1/3}
\end{barticle}
\endbibitem

\bibitem[\protect\citeauthoryear{{Saxton} et~al.}{2012}]{Saxton12}
\begin{barticle}
\bauthor{\binits{R.D.} \bsnm{{Saxton}}},
\bauthor{\binits{A.M.} \bsnm{{Read}}},
\bauthor{\binits{P.} \bsnm{{Esquej}}},
\bauthor{\binits{S.} \bsnm{{Komossa}}},
\bauthor{\binits{S.} \bsnm{{Dougherty}}},
\bauthor{\binits{P.} \bsnm{{Rodriguez-Pascual}}},
\bauthor{\binits{D.} \bsnm{{Barrado}}},
\batitle{{A tidal disruption-like X-ray flare from the quiescent galaxy SDSS
  J120136.02+300305.5}}.
\bjtitle{\aap}
\bvolume{541},
\bfpage{106}
(\byear{2012}).
doi:\doiurl{10.1051/0004-6361/201118367}
\end{barticle}
\endbibitem

\bibitem[\protect\citeauthoryear{{Schachter} et~al.}{1989}]{Schachter1989}
\begin{barticle}
\bauthor{\binits{J.} \bsnm{{Schachter}}},
\bauthor{\binits{A.V.} \bsnm{{Filippenko}}},
\bauthor{\binits{S.M.} \bsnm{{Kahn}}},
\batitle{{Bowen Fluorescence in Scorpius X-1}}.
\bjtitle{\apj}
\bvolume{340},
\bfpage{1049}
(\byear{1989}).
doi:\doiurl{10.1086/167457}
\end{barticle}
\endbibitem

\bibitem[\protect\citeauthoryear{{Shankar} et~al.}{2004}]{Shankar04}
\begin{barticle}
\bauthor{\binits{F.} \bsnm{{Shankar}}},
\bauthor{\binits{P.} \bsnm{{Salucci}}},
\bauthor{\binits{G.L.} \bsnm{{Granato}}},
\bauthor{\binits{G.} \bsnm{{De Zotti}}},
\bauthor{\binits{L.} \bsnm{{Danese}}},
\batitle{{Supermassive black hole demography: the match between the local and
  accreted mass functions}}.
\bjtitle{\mnras}
\bvolume{354},
\bfpage{1020}--\blpage{1030}
(\byear{2004}).
doi:\doiurl{10.1111/j.1365-2966.2004.08261.x}
\end{barticle}
\endbibitem

\bibitem[\protect\citeauthoryear{{Shappee} et~al.}{2014}]{Shappee14}
\begin{barticle}
\bauthor{\binits{B.J.} \bsnm{{Shappee}}},
\bauthor{\binits{J.L.} \bsnm{{Prieto}}},
\bauthor{\binits{D.} \bsnm{{Grupe}}},
\bauthor{\binits{C.S.} \bsnm{{Kochanek}}},
\bauthor{\binits{K.Z.} \bsnm{{Stanek}}},
\bauthor{\binits{G.} \bsnm{{De Rosa}}},
\bauthor{\binits{S.} \bsnm{{Mathur}}},
\bauthor{\binits{Y.} \bsnm{{Zu}}},
\bauthor{\binits{B.M.} \bsnm{{Peterson}}},
\bauthor{\binits{R.W.} \bsnm{{Pogge}}},
\bauthor{\binits{S.} \bsnm{{Komossa}}},
\bauthor{\binits{M.} \bsnm{{Im}}},
\bauthor{\binits{J.} \bsnm{{Jencson}}},
\bauthor{\binits{T.W.-S.} \bsnm{{Holoien}}},
\bauthor{\binits{U.} \bsnm{{Basu}}},
\bauthor{\binits{J.F.} \bsnm{{Beacom}}},
\bauthor{\binits{D.M.} \bsnm{{Szczygie{\l}}}},
\bauthor{\binits{J.} \bsnm{{Brimacombe}}},
\bauthor{\binits{S.} \bsnm{{Adams}}},
\bauthor{\binits{A.} \bsnm{{Campillay}}},
\bauthor{\binits{C.} \bsnm{{Choi}}},
\bauthor{\binits{C.} \bsnm{{Contreras}}},
\bauthor{\binits{M.} \bsnm{{Dietrich}}},
\bauthor{\binits{M.} \bsnm{{Dubberley}}},
\bauthor{\binits{M.} \bsnm{{Elphick}}},
\bauthor{\binits{S.} \bsnm{{Foale}}},
\bauthor{\binits{M.} \bsnm{{Giustini}}},
\bauthor{\binits{C.} \bsnm{{Gonzalez}}},
\bauthor{\binits{E.} \bsnm{{Hawkins}}},
\bauthor{\binits{D.A.} \bsnm{{Howell}}},
\bauthor{\binits{E.Y.} \bsnm{{Hsiao}}},
\bauthor{\binits{M.} \bsnm{{Koss}}},
\bauthor{\binits{K.M.} \bsnm{{Leighly}}},
\bauthor{\binits{N.} \bsnm{{Morrell}}},
\bauthor{\binits{D.} \bsnm{{Mudd}}},
\bauthor{\binits{D.} \bsnm{{Mullins}}},
\bauthor{\binits{J.M.} \bsnm{{Nugent}}},
\bauthor{\binits{J.} \bsnm{{Parrent}}},
\bauthor{\binits{M.M.} \bsnm{{Phillips}}},
\bauthor{\binits{G.} \bsnm{{Pojmanski}}},
\bauthor{\binits{W.} \bsnm{{Rosing}}},
\bauthor{\binits{R.} \bsnm{{Ross}}},
\bauthor{\binits{D.} \bsnm{{Sand}}},
\bauthor{\binits{D.M.} \bsnm{{Terndrup}}},
\bauthor{\binits{S.} \bsnm{{Valenti}}},
\bauthor{\binits{Z.} \bsnm{{Walker}}},
\bauthor{\binits{Y.} \bsnm{{Yoon}}},
\batitle{{The Man behind the Curtain: X-Rays Drive the UV through NIR
  Variability in the 2013 Active Galactic Nucleus Outburst in NGC 2617}}.
\bjtitle{\apj}
\bvolume{788},
\bfpage{48}
(\byear{2014}).
doi:\doiurl{10.1088/0004-637X/788/1/48}
\end{barticle}
\endbibitem

\bibitem[\protect\citeauthoryear{{Short} et~al.}{2020}]{Short2020}
\begin{botherref}
\oauthor{\binits{P.} \bsnm{{Short}}},
\oauthor{\binits{M.} \bsnm{{Nicholl}}},
\oauthor{\binits{A.} \bsnm{{Lawrence}}},
\oauthor{\binits{S.} \bsnm{{Gomez}}},
\oauthor{\binits{I.} \bsnm{{Arcavi}}},
\oauthor{\binits{T.} \bsnm{{Wevers}}},
\oauthor{\binits{G.} \bsnm{{Leloudas}}},
\oauthor{\binits{S.} \bsnm{{Schulze}}},
\oauthor{\binits{J.P.} \bsnm{{Anderson}}},
\oauthor{\binits{E.} \bsnm{{Berger}}},
\oauthor{\binits{P.K.} \bsnm{{Blanchard}}},
\oauthor{\binits{J.} \bsnm{{Burke}}},
\oauthor{\binits{P.} \bsnm{{Charalampopoulos}}},
\oauthor{\binits{R.} \bsnm{{Chornock}}},
\oauthor{\binits{L.} \bsnm{{Galbany}}},
\oauthor{\binits{M.} \bsnm{{Gromadzki}}},
\oauthor{\binits{L.J.} \bsnm{{Herzog}}},
\oauthor{\binits{D.} \bsnm{{Hiramatsu}}},
\oauthor{\binits{K.} \bsnm{{Horne}}},
\oauthor{\binits{G.} \bsnm{{Hosseinzadeh}}},
\oauthor{\binits{D.A.} \bsnm{{Howell}}},
\oauthor{\binits{N.} \bsnm{{Ihanec}}},
\oauthor{\binits{C.} \bsnm{{Inserra}}},
\oauthor{\binits{E.} \bsnm{{Kankare}}},
\oauthor{\binits{K.} \bsnm{{Maguire}}},
\oauthor{\binits{C.} \bsnm{{McCully}}},
\oauthor{\binits{T.E.} \bsnm{{M{\"u}ller Bravo}}},
\oauthor{\binits{F.} \bsnm{{Onori}}},
\oauthor{\binits{J.} \bsnm{{Sollerman}}},
\oauthor{\binits{D.R.} \bsnm{{Young}}},
{The Tidal Disruption Event AT 2018hyz I: Double-peaked emission lines and a
  flat Balmer decrement}.
arXiv e-prints,
2003--05470
(2020)
\end{botherref}
\endbibitem

\bibitem[\protect\citeauthoryear{{Stern} et~al.}{2012}]{Stern12}
\begin{barticle}
\bauthor{\binits{D.} \bsnm{{Stern}}},
\bauthor{\binits{R.J.} \bsnm{{Assef}}},
\bauthor{\binits{D.J.} \bsnm{{Benford}}},
\bauthor{\binits{A.} \bsnm{{Blain}}},
\bauthor{\binits{R.} \bsnm{{Cutri}}},
\bauthor{\binits{A.} \bsnm{{Dey}}},
\bauthor{\binits{P.} \bsnm{{Eisenhardt}}},
\bauthor{\binits{R.L.} \bsnm{{Griffith}}},
\bauthor{\binits{T.H.} \bsnm{{Jarrett}}},
\bauthor{\binits{S.} \bsnm{{Lake}}},
\bauthor{\binits{F.} \bsnm{{Masci}}},
\bauthor{\binits{S.} \bsnm{{Petty}}},
\bauthor{\binits{S.A.} \bsnm{{Stanford}}},
\bauthor{\binits{C.-W.} \bsnm{{Tsai}}},
\bauthor{\binits{E.L.} \bsnm{{Wright}}},
\bauthor{\binits{L.} \bsnm{{Yan}}},
\bauthor{\binits{F.} \bsnm{{Harrison}}},
\bauthor{\binits{K.} \bsnm{{Madsen}}},
\batitle{{Mid-infrared Selection of Active Galactic Nuclei with the Wide-Field
  Infrared Survey Explorer. I. Characterizing WISE-selected Active Galactic
  Nuclei in COSMOS}}.
\bjtitle{\apj}
\bvolume{753},
\bfpage{30}
(\byear{2012}).
doi:\doiurl{10.1088/0004-637X/753/1/30}
\end{barticle}
\endbibitem

\bibitem[\protect\citeauthoryear{{Stone} et~al.}{2013}]{Stone13}
\begin{barticle}
\bauthor{\binits{N.} \bsnm{{Stone}}},
\bauthor{\binits{R.} \bsnm{{Sari}}},
\bauthor{\binits{A.} \bsnm{{Loeb}}},
\batitle{{Consequences of strong compression in tidal disruption events}}.
\bjtitle{\mnras}
\bvolume{435},
\bfpage{1809}--\blpage{1824}
(\byear{2013}).
doi:\doiurl{10.1093/mnras/stt1270}
\end{barticle}
\endbibitem

\bibitem[\protect\citeauthoryear{{Stone} et~al.}{2018}]{Stone18a}
\begin{barticle}
\bauthor{\binits{N.C.} \bsnm{{Stone}}},
\bauthor{\binits{A.} \bsnm{{Generozov}}},
\bauthor{\binits{E.} \bsnm{{Vasiliev}}},
\bauthor{\binits{B.D.} \bsnm{{Metzger}}},
\batitle{{The delay time distribution of tidal disruption flares}}.
\bjtitle{\mnras}
\bvolume{480},
\bfpage{5060}--\blpage{5077}
(\byear{2018}).
doi:\doiurl{10.1093/mnras/sty2045}
\end{barticle}
\endbibitem

\bibitem[\protect\citeauthoryear{{Strauss} et~al.}{2002}]{strauss02}
\begin{barticle}
\bauthor{\binits{M.A.} \bsnm{{Strauss}}},
\bauthor{\binits{D.H.} \bsnm{{Weinberg}}},
\bauthor{\binits{R.H.} \bsnm{{Lupton}}},
\bauthor{\binits{V.K.} \bsnm{{Narayanan}}},
\bauthor{\binits{J.} \bsnm{{Annis}}},
\bauthor{\binits{M.} \bsnm{{Bernardi}}},
\bauthor{\binits{M.} \bsnm{{Blanton}}},
\bauthor{\binits{S.} \bsnm{{Burles}}},
\bauthor{\binits{A.J.} \bsnm{{Connolly}}},
\bauthor{\binits{J.} \bsnm{{Dalcanton}}},
\bauthor{\binits{M.} \bsnm{{Doi}}},
\bauthor{\binits{D.} \bsnm{{Eisenstein}}},
\bauthor{\binits{J.A.} \bsnm{{Frieman}}},
\bauthor{\binits{M.} \bsnm{{Fukugita}}},
\bauthor{\binits{J.E.} \bsnm{{Gunn}}},
\bauthor{\binits{{\v Z}.} \bsnm{{Ivezi{\'c}}}},
\bauthor{\binits{S.} \bsnm{{Kent}}},
\bauthor{\binits{R.S.J.} \bsnm{{Kim}}},
\bauthor{\binits{G.R.} \bsnm{{Knapp}}},
\bauthor{\binits{R.G.} \bsnm{{Kron}}},
\bauthor{\binits{J.A.} \bsnm{{Munn}}},
\bauthor{\binits{H.J.} \bsnm{{Newberg}}},
\bauthor{\binits{R.C.} \bsnm{{Nichol}}},
\bauthor{\binits{S.} \bsnm{{Okamura}}},
\bauthor{\binits{T.R.} \bsnm{{Quinn}}},
\bauthor{\binits{M.W.} \bsnm{{Richmond}}},
\bauthor{\binits{D.J.} \bsnm{{Schlegel}}},
\bauthor{\binits{K.} \bsnm{{Shimasaku}}},
\bauthor{\binits{M.} \bsnm{{SubbaRao}}},
\bauthor{\binits{A.S.} \bsnm{{Szalay}}},
\bauthor{\binits{D.} \bsnm{{Vanden Berk}}},
\bauthor{\binits{M.S.} \bsnm{{Vogeley}}},
\bauthor{\binits{B.} \bsnm{{Yanny}}},
\bauthor{\binits{N.} \bsnm{{Yasuda}}},
\bauthor{\binits{D.G.} \bsnm{{York}}},
\bauthor{\binits{I.} \bsnm{{Zehavi}}},
\batitle{{Spectroscopic Target Selection in the Sloan Digital Sky Survey: The
  Main Galaxy Sample}}.
\bjtitle{\aj}
\bvolume{124},
\bfpage{1810}--\blpage{1824}
(\byear{2002}).
doi:\doiurl{10.1086/342343}
\end{barticle}
\endbibitem

\bibitem[\protect\citeauthoryear{{Strubbe} and {Murray}}{2015}]{strubbe15}
\begin{barticle}
\bauthor{\binits{L.E.} \bsnm{{Strubbe}}},
\bauthor{\binits{N.} \bsnm{{Murray}}},
\batitle{{Insights into tidal disruption of stars from PS1-10jh}}.
\bjtitle{\mnras}
\bvolume{454},
\bfpage{2321}--\blpage{2343}
(\byear{2015}).
doi:\doiurl{10.1093/mnras/stv2081}
\end{barticle}
\endbibitem

\bibitem[\protect\citeauthoryear{{Strubbe} and
  {Quataert}}{2009}]{strubbe_quataert09}
\begin{barticle}
\bauthor{\binits{L.E.} \bsnm{{Strubbe}}},
\bauthor{\binits{E.} \bsnm{{Quataert}}},
\batitle{{Optical flares from the tidal disruption of stars by massive black
  holes}}.
\bjtitle{\mnras}
\bvolume{400},
\bfpage{2070}--\blpage{2084}
(\byear{2009}).
doi:\doiurl{10.1111/j.1365-2966.2009.15599.x}
\end{barticle}
\endbibitem

\bibitem[\protect\citeauthoryear{{Strubbe} and {Quataert}}{2011}]{Strubbe11}
\begin{barticle}
\bauthor{\binits{L.E.} \bsnm{{Strubbe}}},
\bauthor{\binits{E.} \bsnm{{Quataert}}},
\batitle{{Spectroscopic signatures of the tidal disruption of stars by massive
  black holes}}.
\bjtitle{\mnras}
\bvolume{415},
\bfpage{168}--\blpage{180}
(\byear{2011}).
doi:\doiurl{10.1111/j.1365-2966.2011.18686.x}
\end{barticle}
\endbibitem

\bibitem[\protect\citeauthoryear{{Thomas} et~al.}{2013}]{Thomas13}
\begin{barticle}
\bauthor{\binits{D.} \bsnm{{Thomas}}},
\bauthor{\binits{O.} \bsnm{{Steele}}},
\bauthor{\binits{C.} \bsnm{{Maraston}}},
\bauthor{\binits{J.} \bsnm{{Johansson}}},
\bauthor{\binits{A.} \bsnm{{Beifiori}}},
\bauthor{\binits{J.} \bsnm{{Pforr}}},
\bauthor{\binits{G.} \bsnm{{Str{\"o}mb{\"a}ck}}},
\bauthor{\binits{C.A.} \bsnm{{Tremonti}}},
\bauthor{\binits{D.} \bsnm{{Wake}}},
\bauthor{\binits{D.} \bsnm{{Bizyaev}}},
\bauthor{\binits{A.} \bsnm{{Bolton}}},
\bauthor{\binits{H.} \bsnm{{Brewington}}},
\bauthor{\binits{J.R.} \bsnm{{Brownstein}}},
\bauthor{\binits{J.} \bsnm{{Comparat}}},
\bauthor{\binits{J.-P.} \bsnm{{Kneib}}},
\bauthor{\binits{E.} \bsnm{{Malanushenko}}},
\bauthor{\binits{V.} \bsnm{{Malanushenko}}},
\bauthor{\binits{D.} \bsnm{{Oravetz}}},
\bauthor{\binits{K.} \bsnm{{Pan}}},
\bauthor{\binits{J.K.} \bsnm{{Parejko}}},
\bauthor{\binits{D.P.} \bsnm{{Schneider}}},
\bauthor{\binits{A.} \bsnm{{Shelden}}},
\bauthor{\binits{A.} \bsnm{{Simmons}}},
\bauthor{\binits{S.} \bsnm{{Snedden}}},
\bauthor{\binits{M.} \bsnm{{Tanaka}}},
\bauthor{\binits{B.A.} \bsnm{{Weaver}}},
\bauthor{\binits{R.} \bsnm{{Yan}}},
\batitle{{Stellar velocity dispersions and emission line properties of
  SDSS-III/BOSS galaxies}}.
\bjtitle{\mnras}
\bvolume{431}(\bissue{2}),
\bfpage{1383}--\blpage{1397}
(\byear{2013}).
doi:\doiurl{10.1093/mnras/stt261}
\end{barticle}
\endbibitem

\bibitem[\protect\citeauthoryear{{Tonry} et~al.}{2018}]{Tonry2018}
\begin{barticle}
\bauthor{\binits{J.L.} \bsnm{{Tonry}}},
\bauthor{\binits{L.} \bsnm{{Denneau}}},
\bauthor{\binits{A.N.} \bsnm{{Heinze}}},
\bauthor{\binits{B.} \bsnm{{Stalder}}},
\bauthor{\binits{K.W.} \bsnm{{Smith}}},
\bauthor{\binits{S.J.} \bsnm{{Smartt}}},
\bauthor{\binits{C.W.} \bsnm{{Stubbs}}},
\bauthor{\binits{H.J.} \bsnm{{Weiland }}},
\bauthor{\binits{A.} \bsnm{{Rest}}},
\batitle{{ATLAS: A High-cadence All-sky Survey System}}.
\bjtitle{\pasp}
\bvolume{130}(\bissue{988}),
\bfpage{064505}
(\byear{2018}).
doi:\doiurl{10.1088/1538-3873/aabadf}
\end{barticle}
\endbibitem

\bibitem[\protect\citeauthoryear{{Trakhtenbrot}
  et~al.}{2019a}]{Trakhtenbrot2019_CLAGN}
\begin{barticle}
\bauthor{\binits{B.} \bsnm{{Trakhtenbrot}}},
\bauthor{\binits{I.} \bsnm{{Arcavi}}},
\bauthor{\binits{C.L.} \bsnm{{MacLeod}}},
\bauthor{\binits{C.} \bsnm{{Ricci}}},
\bauthor{\binits{E.} \bsnm{{Kara}}},
\bauthor{\binits{M.L.} \bsnm{{Graham}}},
\bauthor{\binits{D.} \bsnm{{Stern}}},
\bauthor{\binits{F.A.} \bsnm{{Harrison}}},
\bauthor{\binits{J.} \bsnm{{Burke}}},
\bauthor{\binits{D.} \bsnm{{Hiramatsu}}},
\bauthor{\binits{G.} \bsnm{{Hosseinzadeh}}},
\bauthor{\binits{D.A.} \bsnm{{Howell}}},
\bauthor{\binits{S.J.} \bsnm{{Smartt}}},
\bauthor{\binits{A.} \bsnm{{Rest}}},
\bauthor{\binits{J.L.} \bsnm{{Prieto}}},
\bauthor{\binits{B.J.} \bsnm{{Shappee}}},
\bauthor{\binits{T.W.-S.} \bsnm{{Holoien}}},
\bauthor{\binits{D.} \bsnm{{Bersier}}},
\bauthor{\binits{A.V.} \bsnm{{Filippenko}}},
\bauthor{\binits{T.G.} \bsnm{{Brink}}},
\bauthor{\binits{W.} \bsnm{{Zheng}}},
\bauthor{\binits{R.} \bsnm{{Li}}},
\bauthor{\binits{R.A.} \bsnm{{Remillard}}},
\bauthor{\binits{M.} \bsnm{{Loewenstein}}},
\batitle{{1ES 1927+654: An AGN Caught Changing Look on a Timescale of Months}}.
\bjtitle{\apj}
\bvolume{883}(\bissue{1}),
\bfpage{94}
(\byear{2019}a).
doi:\doiurl{10.3847/1538-4357/ab39e4}
\end{barticle}
\endbibitem

\bibitem[\protect\citeauthoryear{{Trakhtenbrot}
  et~al.}{2019b}]{Trakhtenbrot2019_Bowen}
\begin{barticle}
\bauthor{\binits{B.} \bsnm{{Trakhtenbrot}}},
\bauthor{\binits{I.} \bsnm{{Arcavi}}},
\bauthor{\binits{C.} \bsnm{{Ricci}}},
\bauthor{\binits{S.} \bsnm{{Tacchella}}},
\bauthor{\binits{D.} \bsnm{{Stern}}},
\bauthor{\binits{H.} \bsnm{{Netzer}}},
\bauthor{\binits{P.G.} \bsnm{{Jonker}}},
\bauthor{\binits{A.} \bsnm{{Horesh}}},
\bauthor{\binits{J.E.} \bsnm{{Mej{\'\i}a-Restrepo}}},
\bauthor{\binits{G.} \bsnm{{Hosseinzadeh}}},
\bauthor{\binits{V.} \bsnm{{Hallefors}}},
\bauthor{\binits{D.A.} \bsnm{{Howell}}},
\bauthor{\binits{C.} \bsnm{{McCully}}},
\bauthor{\binits{M.} \bsnm{{Balokovi{\'c}}}},
\bauthor{\binits{M.} \bsnm{{Heida}}},
\bauthor{\binits{N.} \bsnm{{Kamraj}}},
\bauthor{\binits{G.B.} \bsnm{{Lansbury}}},
\bauthor{\binits{{\L}.} \bsnm{{Wyrzykowski}}},
\bauthor{\binits{M.} \bsnm{{Gromadzki}}},
\bauthor{\binits{A.} \bsnm{{Hamanowicz}}},
\bauthor{\binits{S.B.} \bsnm{{Cenko}}},
\bauthor{\binits{D.J.} \bsnm{{Sand}}},
\bauthor{\binits{E.Y.} \bsnm{{Hsiao}}},
\bauthor{\binits{M.M.} \bsnm{{Phillips}}},
\bauthor{\binits{T.R.} \bsnm{{Diamond}}},
\bauthor{\binits{E.} \bsnm{{Kara}}},
\bauthor{\binits{K.C.} \bsnm{{Gendreau}}},
\bauthor{\binits{Z.} \bsnm{{Arzoumanian}}},
\bauthor{\binits{R.} \bsnm{{Remillard}}},
\batitle{{A new class of flares from accreting supermassive black holes}}.
\bjtitle{Nature Astronomy}
\bvolume{3},
\bfpage{242}--\blpage{250}
(\byear{2019}b).
doi:\doiurl{10.1038/s41550-018-0661-3}
\end{barticle}
\endbibitem

\bibitem[\protect\citeauthoryear{{Udalski} et~al.}{2015}]{Udalski2015}
\begin{barticle}
\bauthor{\binits{A.} \bsnm{{Udalski}}},
\bauthor{\binits{M.K.} \bsnm{{Szyma{\'n}ski}}},
\bauthor{\binits{G.} \bsnm{{Szyma{\'n}ski}}},
\batitle{{OGLE-IV: Fourth Phase of the Optical Gravitational Lensing
  Experiment}}.
\bjtitle{Acta Astronomica}
\bvolume{65}(\bissue{1}),
\bfpage{1}--\blpage{38}
(\byear{2015})
\end{barticle}
\endbibitem

\bibitem[\protect\citeauthoryear{{van Velzen}}{2018}]{vanVelzen18}
\begin{barticle}
\bauthor{\binits{S.} \bsnm{{van Velzen}}},
\batitle{{On the Mass and Luminosity Functions of Tidal Disruption Flares: Rate
  Suppression due to Black Hole Event Horizons}}.
\bjtitle{\apj}
\bvolume{852},
\bfpage{72}
(\byear{2018}).
doi:\doiurl{10.3847/1538-4357/aa998e}
\end{barticle}
\endbibitem

\bibitem[\protect\citeauthoryear{{van Velzen} and {Farrar}}{2014}]{vanVelzen14}
\begin{barticle}
\bauthor{\binits{S.} \bsnm{{van Velzen}}},
\bauthor{\binits{G.R.} \bsnm{{Farrar}}},
\batitle{{Measurement of the Rate of Stellar Tidal Disruption Flares}}.
\bjtitle{\apj}
\bvolume{792},
\bfpage{53}
(\byear{2014}).
doi:\doiurl{10.1088/0004-637X/792/1/53}
\end{barticle}
\endbibitem

\bibitem[\protect\citeauthoryear{{van Velzen} et~al.}{2011}]{vanVelzen10}
\begin{barticle}
\bauthor{\binits{S.} \bsnm{{van Velzen}}},
\bauthor{\binits{G.R.} \bsnm{{Farrar}}},
\bauthor{\binits{S.} \bsnm{{Gezari}}},
\bauthor{\binits{N.} \bsnm{{Morrell}}},
\bauthor{\binits{D.} \bsnm{{Zaritsky}}},
\bauthor{\binits{L.} \bsnm{{{\"O}stman}}},
\bauthor{\binits{M.} \bsnm{{Smith}}},
\bauthor{\binits{J.} \bsnm{{Gelfand}}},
\bauthor{\binits{A.J.} \bsnm{{Drake}}},
\batitle{{Optical Discovery of Probable Stellar Tidal Disruption Flares}}.
\bjtitle{\apj}
\bvolume{741},
\bfpage{73}
(\byear{2011}).
doi:\doiurl{10.1088/0004-637X/741/2/73}
\end{barticle}
\endbibitem

\bibitem[\protect\citeauthoryear{{van Velzen} et~al.}{2016}]{vanVelzen16}
\begin{barticle}
\bauthor{\binits{S.} \bsnm{{van Velzen}}},
\bauthor{\binits{G.E.} \bsnm{{Anderson}}},
\bauthor{\binits{N.C.} \bsnm{{Stone}}},
\bauthor{\binits{M.} \bsnm{{Fraser}}},
\bauthor{\binits{T.} \bsnm{{Wevers}}},
\bauthor{\binits{B.D.} \bsnm{{Metzger}}},
\bauthor{\binits{P.G.} \bsnm{{Jonker}}},
\bauthor{\binits{A.J.} \bsnm{{van der Horst}}},
\bauthor{\binits{T.D.} \bsnm{{Staley}}},
\bauthor{\binits{A.J.} \bsnm{{Mendez}}},
\bauthor{\binits{J.C.A.} \bsnm{{Miller-Jones}}},
\bauthor{\binits{S.T.} \bsnm{{Hodgkin}}},
\bauthor{\binits{H.C.} \bsnm{{Campbell}}},
\bauthor{\binits{R.P.} \bsnm{{Fender}}},
\batitle{{A radio jet from the optical and x-ray bright stellar tidal
  disruption flare ASASSN-14li}}.
\bjtitle{Science}
\bvolume{351},
\bfpage{62}--\blpage{65}
(\byear{2016}).
doi:\doiurl{10.1126/science.aad1182}
\end{barticle}
\endbibitem

\bibitem[\protect\citeauthoryear{{van Velzen} et~al.}{2019a}]{vanVelzen18_FUV}
\begin{barticle}
\bauthor{\binits{S.} \bsnm{{van Velzen}}},
\bauthor{\binits{N.C.} \bsnm{{Stone}}},
\bauthor{\binits{B.D.} \bsnm{{Metzger}}},
\bauthor{\binits{S.} \bsnm{{Gezari}}},
\bauthor{\binits{T.M.} \bsnm{{Brown}}},
\bauthor{\binits{A.S.} \bsnm{{Fruchter}}},
\batitle{{Late-time UV Observations of Tidal Disruption Flares Reveal
  Unobscured, Compact Accretion Disks}}.
\bjtitle{\apj}
\bvolume{878}(\bissue{2}),
\bfpage{82}
(\byear{2019}a).
doi:\doiurl{10.3847/1538-4357/ab1844}
\end{barticle}
\endbibitem

\bibitem[\protect\citeauthoryear{{van Velzen}
  et~al.}{2019b}]{vanVelzen18_NedStark}
\begin{barticle}
\bauthor{\binits{S.} \bsnm{{van Velzen}}},
\bauthor{\binits{S.} \bsnm{{Gezari}}},
\bauthor{\binits{S.B.} \bsnm{{Cenko}}},
\bauthor{\binits{E.} \bsnm{{Kara}}},
\bauthor{\binits{J.C.A.} \bsnm{{Miller-Jones}}},
\bauthor{\binits{T.} \bsnm{{Hung}}},
\bauthor{\binits{J.} \bsnm{{Bright}}},
\bauthor{\binits{N.} \bsnm{{Roth}}},
\bauthor{\binits{N.} \bsnm{{Blagorodnova}}},
\bauthor{\binits{D.} \bsnm{{Huppenkothen}}},
\bauthor{\binits{L.} \bsnm{{Yan}}},
\bauthor{\binits{E.} \bsnm{{Ofek}}},
\bauthor{\binits{J.} \bsnm{{Sollerman}}},
\bauthor{\binits{S.} \bsnm{{Frederick}}},
\bauthor{\binits{C.} \bsnm{{Ward}}},
\bauthor{\binits{M.J.} \bsnm{{Graham}}},
\bauthor{\binits{R.} \bsnm{{Fender}}},
\bauthor{\binits{M.M.} \bsnm{{Kasliwal}}},
\bauthor{\binits{C.} \bsnm{{Canella}}},
\bauthor{\binits{R.} \bsnm{{Stein}}},
\bauthor{\binits{M.} \bsnm{{Giomi}}},
\bauthor{\binits{V.} \bsnm{{Brinnel}}},
\bauthor{\binits{J.} \bsnm{{van Santen}}},
\bauthor{\binits{J.} \bsnm{{Nordin}}},
\bauthor{\binits{E.C.} \bsnm{{Bellm}}},
\bauthor{\binits{R.} \bsnm{{Dekany}}},
\bauthor{\binits{C.} \bsnm{{Fremling}}},
\bauthor{\binits{V.Z.} \bsnm{{Golkhou}}},
\bauthor{\binits{T.} \bsnm{{Kupfer}}},
\bauthor{\binits{S.R.} \bsnm{{Kulkarni}}},
\bauthor{\binits{R.R.} \bsnm{{Laher}}},
\bauthor{\binits{A.} \bsnm{{Mahabal}}},
\bauthor{\binits{F.J.} \bsnm{{Masci}}},
\bauthor{\binits{A.A.} \bsnm{{Miller}}},
\bauthor{\binits{J.D.} \bsnm{{Neill}}},
\bauthor{\binits{R.} \bsnm{{Riddle}}},
\bauthor{\binits{M.} \bsnm{{Rigault}}},
\bauthor{\binits{B.} \bsnm{{Rusholme}}},
\bauthor{\binits{M.T.} \bsnm{{Soumagnac}}},
\bauthor{\binits{Y.} \bsnm{{Tachibana}}},
\batitle{{The First Tidal Disruption Flare in ZTF: From Photometric Selection
  to Multi-wavelength Characterization}}.
\bjtitle{\apj}
\bvolume{872},
\bfpage{198}
(\byear{2019}b).
doi:\doiurl{10.3847/1538-4357/aafe0c}
\end{barticle}
\endbibitem

\bibitem[\protect\citeauthoryear{{van Velzen} et~al.}{2020a}]{vanvelzen2020ZTF}
\begin{botherref}
\oauthor{\binits{S.} \bsnm{{van Velzen}}},
\oauthor{\binits{S.} \bsnm{{Gezari}}},
\oauthor{\binits{E.} \bsnm{{Hammerstein}}},
\oauthor{\binits{N.} \bsnm{{Roth}}},
\oauthor{\binits{S.} \bsnm{{Frederick}}},
\oauthor{\binits{C.} \bsnm{{Ward}}},
\oauthor{\binits{T.} \bsnm{{Hung}}},
\oauthor{\binits{S.B.} \bsnm{{Cenko}}},
\oauthor{\binits{R.} \bsnm{{Stein}}},
\oauthor{\binits{D.A.} \bsnm{{Perley}}},
\oauthor{\binits{K.} \bsnm{{Taggart}}},
\oauthor{\binits{J.} \bsnm{{Sollerman}}},
\oauthor{\binits{I.} \bsnm{{Andreoni}}},
\oauthor{\binits{E.C.} \bsnm{{Bellm}}},
\oauthor{\binits{V.} \bsnm{{Brinnel}}},
\oauthor{\binits{K.} \bsnm{{De}}},
\oauthor{\binits{R.} \bsnm{{Dekany}}},
\oauthor{\binits{M.} \bsnm{{Feeney}}},
\oauthor{\binits{R.J.} \bsnm{{Foley}}},
\oauthor{\binits{C.} \bsnm{{Fremling}}},
\oauthor{\binits{M.} \bsnm{{Giomi}}},
\oauthor{\binits{V.Z.} \bsnm{{Golkhou}}},
\oauthor{\binits{A.Y.Q.} \bsnm{{Ho}}},
\oauthor{\binits{M.M.} \bsnm{{Kasliwal}}},
\oauthor{\binits{C.D.} \bsnm{{Kilpatrick}}},
\oauthor{\binits{S.R.} \bsnm{{Kulkarni}}},
\oauthor{\binits{T.} \bsnm{{Kupfer}}},
\oauthor{\binits{R.R.} \bsnm{{Laher}}},
\oauthor{\binits{A.} \bsnm{{Mahabal}}},
\oauthor{\binits{F.J.} \bsnm{{Masci}}},
\oauthor{\binits{J.} \bsnm{{Nordin}}},
\oauthor{\binits{R.} \bsnm{{Riddle}}},
\oauthor{\binits{B.} \bsnm{{Rusholme}}},
\oauthor{\binits{Y.} \bsnm{{Sharma}}},
\oauthor{\binits{J.} \bsnm{{van Santen}}},
\oauthor{\binits{D.L.} \bsnm{{Shupe}}},
\oauthor{\binits{M.T.} \bsnm{{Soumagnac}}},
{Seventeen Tidal Disruption Events from the First Half of ZTF Survey
  Observations: Entering a New Era of Population Studies}.
arXiv e-prints,
2001--01409
(2020a)
\end{botherref}
\endbibitem

\bibitem[\protect\citeauthoryear{{van Velzen} et~al.}{2020b}]{vanVelzen20}
\begin{botherref}
\oauthor{\binits{S.} \bsnm{{van Velzen}}},
\oauthor{\binits{S.} \bsnm{{Gezari}}},
\oauthor{\binits{E.} \bsnm{{Hammerstein}}},
\oauthor{\binits{N.} \bsnm{{Roth}}},
\oauthor{\binits{S.} \bsnm{{Frederick}}},
\oauthor{\binits{C.} \bsnm{{Ward}}},
\oauthor{\binits{T.} \bsnm{{Hung}}},
\oauthor{\binits{S.B.} \bsnm{{Cenko}}},
\oauthor{\binits{R.} \bsnm{{Stein}}},
\oauthor{\binits{D.A.} \bsnm{{Perley}}},
\oauthor{\binits{K.} \bsnm{{Taggart}}},
\oauthor{\binits{J.} \bsnm{{Sollerman}}},
\oauthor{\binits{I.} \bsnm{{Andreoni}}},
\oauthor{\binits{E.C.} \bsnm{{Bellm}}},
\oauthor{\binits{V.} \bsnm{{Brinnel}}},
\oauthor{\binits{K.} \bsnm{{De}}},
\oauthor{\binits{R.} \bsnm{{Dekany}}},
\oauthor{\binits{M.} \bsnm{{Feeney}}},
\oauthor{\binits{R.J.} \bsnm{{Foley}}},
\oauthor{\binits{C.} \bsnm{{Fremling}}},
\oauthor{\binits{M.} \bsnm{{Giomi}}},
\oauthor{\binits{V.Z.} \bsnm{{Golkhou}}},
\oauthor{\binits{A.Y.Q.} \bsnm{{Ho}}},
\oauthor{\binits{M.M.} \bsnm{{Kasliwal}}},
\oauthor{\binits{C.D.} \bsnm{{Kilpatrick}}},
\oauthor{\binits{S.R.} \bsnm{{Kulkarni}}},
\oauthor{\binits{T.} \bsnm{{Kupfer}}},
\oauthor{\binits{R.R.} \bsnm{{Laher}}},
\oauthor{\binits{A.} \bsnm{{Mahabal}}},
\oauthor{\binits{F.J.} \bsnm{{Masci}}},
\oauthor{\binits{J.} \bsnm{{Nordin}}},
\oauthor{\binits{R.} \bsnm{{Riddle}}},
\oauthor{\binits{B.} \bsnm{{Rusholme}}},
\oauthor{\binits{Y.} \bsnm{{Sharma}}},
\oauthor{\binits{J.} \bsnm{{van Santen}}},
\oauthor{\binits{D.L.} \bsnm{{Shupe}}},
\oauthor{\binits{M.T.} \bsnm{{Soumagnac}}},
{Seventeen Tidal Disruption Events from the First Half of ZTF Survey
  Observations: Entering a New Era of Population Studies}.
arXiv e-prints,
2001--01409
(2020b)
\end{botherref}
\endbibitem

\bibitem[\protect\citeauthoryear{{Wang} et~al.}{2011}]{Wang2011}
\begin{barticle}
\bauthor{\binits{T.-G.} \bsnm{{Wang}}},
\bauthor{\binits{H.-Y.} \bsnm{{Zhou}}},
\bauthor{\binits{L.-F.} \bsnm{{Wang}}},
\bauthor{\binits{H.-L.} \bsnm{{Lu}}},
\bauthor{\binits{D.} \bsnm{{Xu}}},
\batitle{{Transient Superstrong Coronal Lines and Broad Bumps in the Galaxy
  SDSS J074820.67+471214.3}}.
\bjtitle{\apj}
\bvolume{740}(\bissue{2}),
\bfpage{85}
(\byear{2011}).
doi:\doiurl{10.1088/0004-637X/740/2/85}
\end{barticle}
\endbibitem

\bibitem[\protect\citeauthoryear{{Wang} et~al.}{2012}]{Wang2012}
\begin{barticle}
\bauthor{\binits{T.-G.} \bsnm{{Wang}}},
\bauthor{\binits{H.-Y.} \bsnm{{Zhou}}},
\bauthor{\binits{S.} \bsnm{{Komossa}}},
\bauthor{\binits{H.-Y.} \bsnm{{Wang}}},
\bauthor{\binits{W.} \bsnm{{Yuan}}},
\bauthor{\binits{C.} \bsnm{{Yang}}},
\batitle{{Extreme Coronal Line Emitters: Tidal Disruption of Stars by Massive
  Black Holes in Galactic Nuclei?}}
\bjtitle{\apj}
\bvolume{749}(\bissue{2}),
\bfpage{115}
(\byear{2012}).
doi:\doiurl{10.1088/0004-637X/749/2/115}
\end{barticle}
\endbibitem

\bibitem[\protect\citeauthoryear{{Wevers} et~al.}{2017}]{Wevers17}
\begin{barticle}
\bauthor{\binits{T.} \bsnm{{Wevers}}},
\bauthor{\binits{S.} \bsnm{{van Velzen}}},
\bauthor{\binits{P.G.} \bsnm{{Jonker}}},
\bauthor{\binits{N.C.} \bsnm{{Stone}}},
\bauthor{\binits{T.} \bsnm{{Hung}}},
\bauthor{\binits{F.} \bsnm{{Onori}}},
\bauthor{\binits{S.} \bsnm{{Gezari}}},
\bauthor{\binits{N.} \bsnm{{Blagorodnova}}},
\batitle{{Black hole masses of tidal disruption event host galaxies}}.
\bjtitle{\mnras}
\bvolume{471},
\bfpage{1694}--\blpage{1708}
(\byear{2017}).
doi:\doiurl{10.1093/mnras/stx1703}
\end{barticle}
\endbibitem

\bibitem[\protect\citeauthoryear{{Wevers} et~al.}{2019a}]{Wevers19a}
\begin{barticle}
\bauthor{\binits{T.} \bsnm{{Wevers}}},
\bauthor{\binits{D.R.} \bsnm{{Pasham}}},
\bauthor{\binits{S.} \bsnm{{van Velzen}}},
\bauthor{\binits{G.} \bsnm{{Leloudas}}},
\bauthor{\binits{S.} \bsnm{{Schulze}}},
\bauthor{\binits{J.C.A.} \bsnm{{Miller-Jones}}},
\bauthor{\binits{P.G.} \bsnm{{Jonker}}},
\bauthor{\binits{M.} \bsnm{{Gromadzki}}},
\bauthor{\binits{E.} \bsnm{{Kankare}}},
\bauthor{\binits{S.T.} \bsnm{{Hodgkin}}},
\bauthor{\binits{{\L}.} \bsnm{{Wyrzykowski}}},
\bauthor{\binits{Z.} \bsnm{{Kostrzewa-Rutkowska}}},
\bauthor{\binits{S.} \bsnm{{Moran}}},
\bauthor{\binits{M.} \bsnm{{Berton}}},
\bauthor{\binits{K.} \bsnm{{Maguire}}},
\bauthor{\binits{F.} \bsnm{{Onori}}},
\bauthor{\binits{S.} \bsnm{{Mattila}}},
\bauthor{\binits{M.} \bsnm{{Nicholl}}},
\batitle{{Evidence for rapid disc formation and reprocessing in the X-ray
  bright tidal disruption event candidate AT 2018fyk}}.
\bjtitle{\mnras}
\bvolume{488}(\bissue{4}),
\bfpage{4816}--\blpage{4830}
(\byear{2019}a).
doi:\doiurl{10.1093/mnras/stz1976}
\end{barticle}
\endbibitem

\bibitem[\protect\citeauthoryear{{Wevers} et~al.}{2019b}]{wever19}
\begin{botherref}
\oauthor{\binits{T.} \bsnm{{Wevers}}},
\oauthor{\binits{D.R.} \bsnm{{Pasham}}},
\oauthor{\binits{S.} \bsnm{{van Velzen}}},
\oauthor{\binits{G.} \bsnm{{Leloudas}}},
\oauthor{\binits{S.} \bsnm{{Schulze}}},
\oauthor{\binits{J.C.A.} \bsnm{{Miller-Jones}}},
\oauthor{\binits{P.G.} \bsnm{{Jonker}}},
\oauthor{\binits{M.} \bsnm{{Gromadzki}}},
\oauthor{\binits{E.} \bsnm{{Kankare}}},
\oauthor{\binits{S.T.} \bsnm{{Hodgkin}}},
\oauthor{\binits{L.} \bsnm{{. Wyrzykowski}}},
\oauthor{\binits{Z.} \bsnm{{Kostrzewa-Rutkowska}}},
\oauthor{\binits{S.} \bsnm{{Moran}}},
\oauthor{\binits{M.} \bsnm{{Berton}}},
\oauthor{\binits{K.} \bsnm{{Maguire}}},
\oauthor{\binits{F.} \bsnm{{Onori}}},
\oauthor{\binits{S.} \bsnm{{Matilla}}},
\oauthor{\binits{M.} \bsnm{{Nicholl}}},
{Evidence for rapid disk formation and reprocessing in the X-ray bright tidal
  disruption event AT 2018fyk}.
arXiv e-prints,
1903--12203
(2019b)
\end{botherref}
\endbibitem

\bibitem[\protect\citeauthoryear{{Wevers} et~al.}{2019c}]{Wevers19}
\begin{barticle}
\bauthor{\binits{T.} \bsnm{{Wevers}}},
\bauthor{\binits{N.C.} \bsnm{{Stone}}},
\bauthor{\binits{S.} \bsnm{{van Velzen}}},
\bauthor{\binits{P.G.} \bsnm{{Jonker}}},
\bauthor{\binits{T.} \bsnm{{Hung}}},
\bauthor{\binits{K.} \bsnm{{Auchettl}}},
\bauthor{\binits{S.} \bsnm{{Gezari}}},
\bauthor{\binits{F.} \bsnm{{Onori}}},
\bauthor{\binits{D.} \bsnm{{Mata S{\'a}nchez}}},
\bauthor{\binits{Z.} \bsnm{{Kostrzewa-Rutkowska}}},
\bauthor{\binits{J.} \bsnm{{Casares}}},
\batitle{{Black hole masses of tidal disruption event host galaxies II}}.
\bjtitle{\mnras}
\bvolume{487}(\bissue{3}),
\bfpage{4136}--\blpage{4152}
(\byear{2019}c).
doi:\doiurl{10.1093/mnras/stz1602}
\end{barticle}
\endbibitem

\bibitem[\protect\citeauthoryear{{Weymann} and {Williams}}{1969}]{Weyman1969}
\begin{barticle}
\bauthor{\binits{R.J.} \bsnm{{Weymann}}},
\bauthor{\binits{R.E.} \bsnm{{Williams}}},
\batitle{{The Bowen Fluorescence Mechanism in Planetary Nebulae and the Nuclei
  of Seyfert Galaxies}}.
\bjtitle{\apj}
\bvolume{157},
\bfpage{1201}
(\byear{1969}).
doi:\doiurl{10.1086/150147}
\end{barticle}
\endbibitem

\bibitem[\protect\citeauthoryear{{Wyrzykowski} et~al.}{2014}]{Wyrzykowski2014}
\begin{barticle}
\bauthor{\binits{{\L}.} \bsnm{{Wyrzykowski}}},
\bauthor{\binits{Z.} \bsnm{{Kostrzewa-Rutkowska}}},
\bauthor{\binits{S.} \bsnm{{Koz{\l}owski}}},
\bauthor{\binits{A.} \bsnm{{Udalski}}},
\bauthor{\binits{R.} \bsnm{{Poleski}}},
\bauthor{\binits{J.} \bsnm{{Skowron}}},
\bauthor{\binits{N.} \bsnm{{Blagorodnova}}},
\bauthor{\binits{M.} \bsnm{{Kubiak}}},
\bauthor{\binits{M.K.} \bsnm{{Szyma{\'n}ski}}},
\bauthor{\binits{G.} \bsnm{{Pietrzy{\'n}ski}}},
\bauthor{\binits{I.} \bsnm{{Soszy{\'n}ski}}},
\bauthor{\binits{K.} \bsnm{{Ulaczyk}}},
\bauthor{\binits{P.} \bsnm{{Pietrukowicz}}},
\bauthor{\binits{P.} \bsnm{{Mr{\'o}z}}},
\batitle{{OGLE-IV Real-Time Transient Search}}.
\bjtitle{Acta Astronomica}
\bvolume{64}(\bissue{3}),
\bfpage{197}--\blpage{232}
(\byear{2014})
\end{barticle}
\endbibitem

\bibitem[\protect\citeauthoryear{{Wyrzykowski} et~al.}{2017}]{Wyrzykowski17}
\begin{barticle}
\bauthor{\binits{{\L}.} \bsnm{{Wyrzykowski}}},
\bauthor{\binits{M.} \bsnm{{Zieli{\'n}ski}}},
\bauthor{\binits{Z.} \bsnm{{Kostrzewa-Rutkowska}}},
\bauthor{\binits{A.} \bsnm{{Hamanowicz}}},
\bauthor{\binits{P.G.} \bsnm{{Jonker}}},
\bauthor{\binits{I.} \bsnm{{Arcavi}}},
\bauthor{\binits{J.} \bsnm{{Guillochon}}},
\bauthor{\binits{P.J.} \bsnm{{Brown}}},
\bauthor{\binits{S.} \bsnm{{Koz{\l}owski}}},
\bauthor{\binits{A.} \bsnm{{Udalski}}},
\bauthor{\binits{M.K.} \bsnm{{Szyma{\'n}ski}}},
\bauthor{\binits{I.} \bsnm{{Soszy{\'n}ski}}},
\bauthor{\binits{R.} \bsnm{{Poleski}}},
\bauthor{\binits{P.} \bsnm{{Pietrukowicz}}},
\bauthor{\binits{J.} \bsnm{{Skowron}}},
\bauthor{\binits{P.} \bsnm{{Mr{\'o}z}}},
\bauthor{\binits{K.} \bsnm{{Ulaczyk}}},
\bauthor{\binits{M.} \bsnm{{Pawlak}}},
\bauthor{\binits{K.A.} \bsnm{{Rybicki}}},
\bauthor{\binits{J.} \bsnm{{Greiner}}},
\bauthor{\binits{T.} \bsnm{{Kr{\"u}hler}}},
\bauthor{\binits{J.} \bsnm{{Bolmer}}},
\bauthor{\binits{S.J.} \bsnm{{Smartt}}},
\bauthor{\binits{K.} \bsnm{{Maguire}}},
\bauthor{\binits{K.} \bsnm{{Smith}}},
\batitle{{OGLE16aaa - a signature of a hungry supermassive black hole}}.
\bjtitle{\mnras}
\bvolume{465},
\bfpage{114}--\blpage{118}
(\byear{2017}).
doi:\doiurl{10.1093/mnrasl/slw213}
\end{barticle}
\endbibitem

\bibitem[\protect\citeauthoryear{{Yang} et~al.}{2013}]{Yang2013}
\begin{barticle}
\bauthor{\binits{C.-W.} \bsnm{{Yang}}},
\bauthor{\binits{T.-G.} \bsnm{{Wang}}},
\bauthor{\binits{G.} \bsnm{{Ferland}}},
\bauthor{\binits{W.} \bsnm{{Yuan}}},
\bauthor{\binits{H.-Y.} \bsnm{{Zhou}}},
\bauthor{\binits{P.} \bsnm{{Jiang}}},
\batitle{{Long-term Spectral Evolution of Tidal Disruption Candidates Selected
  by Strong Coronal Lines}}.
\bjtitle{\apj}
\bvolume{774}(\bissue{1}),
\bfpage{46}
(\byear{2013}).
doi:\doiurl{10.1088/0004-637X/774/1/46}
\end{barticle}
\endbibitem

\bibitem[\protect\citeauthoryear{{Yang} et~al.}{2017}]{Yang17}
\begin{barticle}
\bauthor{\binits{C.} \bsnm{{Yang}}},
\bauthor{\binits{T.} \bsnm{{Wang}}},
\bauthor{\binits{G.J.} \bsnm{{Ferland}}},
\bauthor{\binits{L.} \bsnm{{Dou}}},
\bauthor{\binits{H.} \bsnm{{Zhou}}},
\bauthor{\binits{N.} \bsnm{{Jiang}}},
\bauthor{\binits{Z.} \bsnm{{Sheng}}},
\batitle{{The Carbon and Nitrogen Abundance Ratio in the Broad Line Region of
  Tidal Disruption Events}}.
\bjtitle{\apj}
\bvolume{846}(\bissue{2}),
\bfpage{150}
(\byear{2017}).
doi:\doiurl{10.3847/1538-4357/aa8598}
\end{barticle}
\endbibitem

\bibitem[\protect\citeauthoryear{{York} et~al.}{2000}]{york02}
\begin{barticle}
\bauthor{\binits{D.G.} \bsnm{{York}}},
\bauthor{\binits{J.} \bsnm{{Adelman}}},
\bauthor{\binits{J.E.} \bsnm{{Anderson}} \bsuffix{Jr.}},
\bauthor{\binits{S.F.} \bsnm{{Anderson}}},
\bauthor{\binits{J.} \bsnm{{Annis}}},
\bauthor{\binits{N.A.} \bsnm{{Bahcall}}},
\bauthor{\binits{J.A.} \bsnm{{Bakken}}},
\bauthor{\binits{R.} \bsnm{{Barkhouser}}},
\bauthor{\binits{S.} \bsnm{{Bastian}}},
\bauthor{\binits{E.} \bsnm{{Berman}}},
\bauthor{\binits{W.N.} \bsnm{{Boroski}}},
\bauthor{\binits{S.} \bsnm{{Bracker}}},
\bauthor{\binits{C.} \bsnm{{Briegel}}},
\bauthor{\binits{J.W.} \bsnm{{Briggs}}},
\bauthor{\binits{J.} \bsnm{{Brinkmann}}},
\bauthor{\binits{R.} \bsnm{{Brunner}}},
\bauthor{\binits{S.} \bsnm{{Burles}}},
\bauthor{\binits{L.} \bsnm{{Carey}}},
\bauthor{\binits{M.A.} \bsnm{{Carr}}},
\bauthor{\binits{F.J.} \bsnm{{Castander}}},
\bauthor{\binits{B.} \bsnm{{Chen}}},
\bauthor{\binits{P.L.} \bsnm{{Colestock}}},
\bauthor{\binits{A.J.} \bsnm{{Connolly}}},
\bauthor{\binits{J.H.} \bsnm{{Crocker}}},
\bauthor{\binits{I.} \bsnm{{Csabai}}},
\bauthor{\binits{P.C.} \bsnm{{Czarapata}}},
\bauthor{\binits{J.E.} \bsnm{{Davis}}},
\bauthor{\binits{M.} \bsnm{{Doi}}},
\bauthor{\binits{T.} \bsnm{{Dombeck}}},
\bauthor{\binits{D.} \bsnm{{Eisenstein}}},
\bauthor{\binits{N.} \bsnm{{Ellman}}},
\bauthor{\binits{B.R.} \bsnm{{Elms}}},
\bauthor{\binits{M.L.} \bsnm{{Evans}}},
\bauthor{\binits{X.} \bsnm{{Fan}}},
\bauthor{\binits{G.R.} \bsnm{{Federwitz}}},
\bauthor{\binits{L.} \bsnm{{Fiscelli}}},
\bauthor{\binits{S.} \bsnm{{Friedman}}},
\bauthor{\binits{J.A.} \bsnm{{Frieman}}},
\bauthor{\binits{M.} \bsnm{{Fukugita}}},
\bauthor{\binits{B.} \bsnm{{Gillespie}}},
\bauthor{\binits{J.E.} \bsnm{{Gunn}}},
\bauthor{\binits{V.K.} \bsnm{{Gurbani}}},
\bauthor{\binits{E.} \bsnm{{de Haas}}},
\bauthor{\binits{M.} \bsnm{{Haldeman}}},
\bauthor{\binits{F.H.} \bsnm{{Harris}}},
\bauthor{\binits{J.} \bsnm{{Hayes}}},
\bauthor{\binits{T.M.} \bsnm{{Heckman}}},
\bauthor{\binits{G.S.} \bsnm{{Hennessy}}},
\bauthor{\binits{R.B.} \bsnm{{Hindsley}}},
\bauthor{\binits{S.} \bsnm{{Holm}}},
\bauthor{\binits{D.J.} \bsnm{{Holmgren}}},
\bauthor{\binits{C.} \bsnm{{Huang}}},
\bauthor{\binits{C.} \bsnm{{Hull}}},
\bauthor{\binits{D.} \bsnm{{Husby}}},
\bauthor{\binits{S.} \bsnm{{Ichikawa}}},
\bauthor{\binits{T.} \bsnm{{Ichikawa}}},
\bauthor{\binits{{\v Z}.} \bsnm{{Ivezi{\'c}}}},
\bauthor{\binits{S.} \bsnm{{Kent}}},
\bauthor{\binits{R.S.J.} \bsnm{{Kim}}},
\bauthor{\binits{E.} \bsnm{{Kinney}}},
\bauthor{\binits{M.} \bsnm{{Klaene}}},
\bauthor{\binits{A.N.} \bsnm{{Kleinman}}},
\bauthor{\binits{S.} \bsnm{{Kleinman}}},
\bauthor{\binits{G.R.} \bsnm{{Knapp}}},
\bauthor{\binits{J.} \bsnm{{Korienek}}},
\bauthor{\binits{R.G.} \bsnm{{Kron}}},
\bauthor{\binits{P.Z.} \bsnm{{Kunszt}}},
\bauthor{\binits{D.Q.} \bsnm{{Lamb}}},
\bauthor{\binits{B.} \bsnm{{Lee}}},
\bauthor{\binits{R.F.} \bsnm{{Leger}}},
\bauthor{\binits{S.} \bsnm{{Limmongkol}}},
\bauthor{\binits{C.} \bsnm{{Lindenmeyer}}},
\bauthor{\binits{D.C.} \bsnm{{Long}}},
\bauthor{\binits{C.} \bsnm{{Loomis}}},
\bauthor{\binits{J.} \bsnm{{Loveday}}},
\bauthor{\binits{R.} \bsnm{{Lucinio}}},
\bauthor{\binits{R.H.} \bsnm{{Lupton}}},
\bauthor{\binits{B.} \bsnm{{MacKinnon}}},
\bauthor{\binits{E.J.} \bsnm{{Mannery}}},
\bauthor{\binits{P.M.} \bsnm{{Mantsch}}},
\bauthor{\binits{B.} \bsnm{{Margon}}},
\bauthor{\binits{P.} \bsnm{{McGehee}}},
\bauthor{\binits{T.A.} \bsnm{{McKay}}},
\bauthor{\binits{A.} \bsnm{{Meiksin}}},
\bauthor{\binits{A.} \bsnm{{Merelli}}},
\bauthor{\binits{D.G.} \bsnm{{Monet}}},
\bauthor{\binits{J.A.} \bsnm{{Munn}}},
\bauthor{\binits{V.K.} \bsnm{{Narayanan}}},
\bauthor{\binits{T.} \bsnm{{Nash}}},
\bauthor{\binits{E.} \bsnm{{Neilsen}}},
\bauthor{\binits{R.} \bsnm{{Neswold}}},
\bauthor{\binits{H.J.} \bsnm{{Newberg}}},
\bauthor{\binits{R.C.} \bsnm{{Nichol}}},
\bauthor{\binits{T.} \bsnm{{Nicinski}}},
\bauthor{\binits{M.} \bsnm{{Nonino}}},
\bauthor{\binits{N.} \bsnm{{Okada}}},
\bauthor{\binits{S.} \bsnm{{Okamura}}},
\bauthor{\binits{J.P.} \bsnm{{Ostriker}}},
\bauthor{\binits{R.} \bsnm{{Owen}}},
\bauthor{\binits{A.G.} \bsnm{{Pauls}}},
\bauthor{\binits{J.} \bsnm{{Peoples}}},
\bauthor{\binits{R.L.} \bsnm{{Peterson}}},
\bauthor{\binits{D.} \bsnm{{Petravick}}},
\bauthor{\binits{J.R.} \bsnm{{Pier}}},
\bauthor{\binits{A.} \bsnm{{Pope}}},
\bauthor{\binits{R.} \bsnm{{Pordes}}},
\bauthor{\binits{A.} \bsnm{{Prosapio}}},
\bauthor{\binits{R.} \bsnm{{Rechenmacher}}},
\bauthor{\binits{T.R.} \bsnm{{Quinn}}},
\bauthor{\binits{G.T.} \bsnm{{Richards}}},
\bauthor{\binits{M.W.} \bsnm{{Richmond}}},
\bauthor{\binits{C.H.} \bsnm{{Rivetta}}},
\bauthor{\binits{C.M.} \bsnm{{Rockosi}}},
\bauthor{\binits{K.} \bsnm{{Ruthmansdorfer}}},
\bauthor{\binits{D.} \bsnm{{Sandford}}},
\bauthor{\binits{D.J.} \bsnm{{Schlegel}}},
\bauthor{\binits{D.P.} \bsnm{{Schneider}}},
\bauthor{\binits{M.} \bsnm{{Sekiguchi}}},
\bauthor{\binits{G.} \bsnm{{Sergey}}},
\bauthor{\binits{K.} \bsnm{{Shimasaku}}},
\bauthor{\binits{W.A.} \bsnm{{Siegmund}}},
\bauthor{\binits{S.} \bsnm{{Smee}}},
\bauthor{\binits{J.A.} \bsnm{{Smith}}},
\bauthor{\binits{S.} \bsnm{{Snedden}}},
\bauthor{\binits{R.} \bsnm{{Stone}}},
\bauthor{\binits{C.} \bsnm{{Stoughton}}},
\bauthor{\binits{M.A.} \bsnm{{Strauss}}},
\bauthor{\binits{C.} \bsnm{{Stubbs}}},
\bauthor{\binits{M.} \bsnm{{SubbaRao}}},
\bauthor{\binits{A.S.} \bsnm{{Szalay}}},
\bauthor{\binits{I.} \bsnm{{Szapudi}}},
\bauthor{\binits{G.P.} \bsnm{{Szokoly}}},
\bauthor{\binits{A.R.} \bsnm{{Thakar}}},
\bauthor{\binits{C.} \bsnm{{Tremonti}}},
\bauthor{\binits{D.L.} \bsnm{{Tucker}}},
\bauthor{\binits{A.} \bsnm{{Uomoto}}},
\bauthor{\binits{D.} \bsnm{{Vanden Berk}}},
\bauthor{\binits{M.S.} \bsnm{{Vogeley}}},
\bauthor{\binits{P.} \bsnm{{Waddell}}},
\bauthor{\binits{S.} \bsnm{{Wang}}},
\bauthor{\binits{M.} \bsnm{{Watanabe}}},
\bauthor{\binits{D.H.} \bsnm{{Weinberg}}},
\bauthor{\binits{B.} \bsnm{{Yanny}}},
\bauthor{\binits{N.} \bsnm{{Yasuda}}},
\batitle{{The Sloan Digital Sky Survey: Technical Summary}}.
\bjtitle{\aj}
\bvolume{120},
\bfpage{1579}--\blpage{1587}
(\byear{2000}).
doi:\doiurl{10.1086/301513}
\end{barticle}
\endbibitem

\end{thebibliography}
\end{document}